\documentclass[prd,preprint,tightenlines,a4paper,nofootinbib]{revtex4-1}

\usepackage{amsbsy}     % Bolds (vectors, tensors, etc.)
\usepackage{amsfonts}       % AMS fonts
\usepackage{amsmath}        % AMS maths
\usepackage{amssymb}        % AMS symbols
\usepackage{wasysym}

\usepackage{graphicx}

\begin{document}

 \begin{flushright}
 LU TP 12-24\\
 June 2012
 \end{flushright}

\title{Closing the Window on \\ Light Charged Higgs Bosons in the NMSSM}

\author{Johan Rathsman}
 \email{Johan.Rathsman@thep.lu.se}
\author{Thomas R{\"o}ssler}
 \email{tr@thep.lu.se}
\affiliation{Department of Astronomy and Theoretical Physics, Lund
University, S{\"o}lvegatan 14A, SE-223 62 Lund, Sweden}

\begin{abstract}
In the Next-to-Minimal SuperSymmetric Model (NMSSM) the lightest CP-odd Higgs bosons ($a_1$) can be very light. As a consequence, in addition to the standard charged Higgs boson ($h^\pm$) decays considered in the MSSM for a light charged Higgs ($m_{h^\pm} < m_t$), the branching fraction for $h^\pm \to a_1 W$ can be dominant.  We investigate how this signal can be searched for in $t\bar{t}$ production at the Large Hadron Collider (LHC) in the case that  ($m_{a_1} \gtrsim 2m_B$) with the $a_1$ giving rise to a single $b\bar{b}$--jet and discuss to what extent the LHC experiments are able to discover such a scenario with an integrated luminosity $\sim 20$ fb$^{-1}$. 
We also discuss the implications of the possible Higgs-signal observed at the LHC.
\end{abstract}

\maketitle

\section{Introduction}
With the successful start-up of the LHC and the intriguing results from the first year of data-taking at the center of mass energy of 7 TeV  and a integrated luminosity close to  $\approx 5$ fb$^{-1}$ for each of the ATLAS and CMS experiments, the ongoing run in 2012 is set to be a milestone in particle physics. The possible signal for a Higgs boson around $125$ GeV may or may not be confirmed and the search for physics Beyond the Standard Model (BSM) will continue. Contrary to the neutral Higgs boson, which if discovered may need to be analyzed in detail regarding its branching fraction into various channels in order to determine whether it is the Standard Model (SM) Higgs boson or not, the discovery of a charged Higgs boson would be an unmistakable sign of BSM physics.

The charged Higgs boson arises in theories with more than one Higgs doublet. The prime example is the MSSM \cite{Nilles:1983ge,Haber:1984rc} which has two Higgs doublets, leading to two CP-even Higgs bosons ($h,H$) and one CP-odd ($A$) in the case of CP-conservation and two charged states ($H^\pm$) after electroweak symmetry breaking. In this case the two Higgs doublets are required by supersymmetry with one of the giving masses to the up-type fermions and one to the down-type ones. For a complete introduction to the Higgs sector in the MSSM we refer to \cite{Djouadi:2005gj}. 

The main reason for introducing supersymmetry (for a general introduction to supersymmetric theories we refer to \cite{Martin:1997ns}) is to solve the so called hierarchy problem, {\it i.e.\ }why the scale of the electroweak interaction is so much smaller than the cut-off of the SM, normally taken to be the Planck mass where gravity becomes the dominant force and the SM breaks down. With the Higgs boson being a scalar particle, the higher order corrections to its mass are proportional to this cut-off and with the SM only being an effective theory this cut-off dependence cannot be renormalized away. In a supersymmetric theory this problem is essentially solved by the introduction of the fermionic partners of the Higgs fields, which only get logarithmic corrections to their masses and thereby avoids fine-tuning. In turn this means that the Higgs boson masses are also protected from receiving quadratic corrections as long as supersymmetry is not broken or only softly broken.

In addition to solving the hierarchy problem supersymmetry also offers a candidate for cold dark matter \cite{Goldberg:1983nd,Ellis:1983ew} in the case that $R$-symmetry is preserved, which in turn is introduced to avoid terms in the Lagrangian that otherwise would mediate proton decay. In this case the lightest supersymmetric particle (LSP) is stable and generically it has the right mass and cross-section to constitute the observed dark matter. Supersymmetry also improves the unification of gauge forces at an hypothesized grand unification scale although the unification is not exact and it does depend on the details of the spectrum of SUSY-particles. 

As is well known, the MSSM by itself is not without problems. Leaving the question of the precise mechanism for supersymmetry breaking aside, the MSSM faces the so called $\mu$-problem. This relates to the magnitude of the dimensionful
$\mu$-parameter which couples the two Higgs doublets to each other in the superpotential. In order to avoid large cancellations between this contribution to the Higgs masses and the soft supersymmetry breaking terms as well as having a phenomenologically 
viable supersymmetric theory (mainly having a large enough chargino mass), the magnitude of $\mu$ should be of order the electroweak or supersymmetry breaking scales. The problem is then that there is no a priori reason for this parameter to have any particular value, in principle it could be anything up to the Planck scale, so why is it similar to the electroweak or supersymmetry breaking scales?

In the NMSSM the $\mu$-problem is solved by introducing an additional Higgs singlet into the theory. After supersymmetry breaking this field gets a vacuum expectation value (vev) that effectively acts as a $\mu$-term. The original $\mu$-term
in the superpotential can then be set to zero without spoiling the viability of the theory. For a more detailed review of the NMSSM we refer to \cite{Maniatis:2009re,Ellwanger:2009dp}.

The additional Higgs singlet has important consequences for the phenomenology of the Higgs sector. In the MSSM, the masses of the heavy Higgs bosons ($H, A, H^\pm$) are closely related to each other as they originate from the same (second) Higgs doublet if viewed in the Higgs basis where only one (the first) of the Higgs doublets has a vev.
For example, at tree-level the masses of the CP-odd and charged Higgs bosons are related by $m_{H^\pm}^2=m_A^2+m_W^2$. As a consequence the decay $H^\pm \to A W $ is typically not open. 

In the NMSSM the additional Higgs singlet means that there is one more CP-even and one more CP-odd field with a separate mass 
scale introduced into the Higgs sector. As a consequence, the by now three CP-even and two CP-odd electroweak states will   
mix into the respective mass eigenstates. Thus the mass-relations from the MSSM will be altered. This is particularly evident in the CP-odd sector where the lightest state $a_1$ may now be much lighter than the charged Higgs boson -- even after taking experimental constraints into account as discussed below -- opening up the possibility for the $h^\pm \to a_1 W$ decay to be dominant. In turn this means that the search for charged Higgs bosons, in $t$-quark decays for example, has to be widened also to include this decay channel.

The decay  $h^\pm \to a_1 W$  has already been considered to different levels of detail \cite{Drees:1998pw,Drees:1999sb,Akeroyd:2007yj} in the literature
and there are constraints from the DELPHI experiment for $m_{a_1} > 12 $ GeV~\cite{Abdallah:2003wd}, as well as the CDF experiment for the case $a_1 \to \tau^+ \tau^-$~\cite{CDF-10104}. 
In this paper we want to focus on the  region in parameter space where the $a_1$ mass is above the $b\bar{b}$ threshold but still so close to it that the two $b$-quarks will fragment into a single  
$b\bar{b}$--jet. The viability of scenarios with light $a_1$'s have also been considered by \cite{Dermisek:2008uu,Dermisek:2009fd,Dermisek:2010mg,Mahmoudi:2010xp,Stal:2011cz}.
 
Our paper is organized as follows. In the next section we give some basic properties of the Higgs sector in the NMSSM that are relevant to our discussion. We then discuss the constraints on the parameter space in section 3, including the latest results from LHC. In section 4 we illustrate how the signal $h^\pm \to a_1 W$ can be searched for in $t\bar{t}$-production taking into account the appropriate backgrounds. Section 5 contains a discussion of the implications of the possible Higgs signal from the ATLAS and CMS experiments and in section 6 we summarize and conclude.

\section{Basic properties of the NMSSM}

We consider the $Z_3$-symmetric version of the NMSSM with the superpotential given by 
\begin{equation}
W_{\rm NMSSM} = W_{\rm MSSM} + \lambda \hat{S} \hat{H}_u \cdot \hat{H}_d + \kappa \hat{S}^3
\end{equation}
where $W_{\rm MSSM}$ is the superpotential of the MSSM with $\mu$ set to zero.
The soft supersymmetry breaking potential relative to the MSSM is then given by 
\begin{equation}
V^{\rm NMSSM}_{\rm soft} = V^{\rm MSSM}_{\rm soft} + m_S^2|S|^2 + \left( \lambda A_{\lambda}  {S}  {H}_u \cdot  {H}_d + \frac{1}{3}\kappa A_{\kappa}  {S}^3 + h.c. \right)
\end{equation}
where the part of $V^{\rm MSSM}_{\rm soft}$ only depending on the Higgs fields is given by,
\begin{equation}
V^{\rm MSSM}_{\rm soft, higgs} = m_{H_u}^2 |H_u|^2 +  m_{H_d}^2 |H_d|^2 \quad .
\end{equation}
In addition $V^{\rm MSSM}_{\rm soft}$ contains all the dependence on the other soft supersymmetry breaking parameters: the gaugino masses $M_1, M_2, M_3$, the tri-linear couplings $\mathbf{a}_u, \mathbf{a}_d, \mathbf{a}_e$, the squark masses $\mathbf{M}_Q, \mathbf{M}_u, \mathbf{M}_d$, and finally the slepton masses $\mathbf{M}_L, \mathbf{M}_e$. In the following we will assume minimal flavour violation so that the sfermion mass matrices are diagonal and the tri-linear couplings are proportional to the corresponding Yukawa coupling matrices $\mathbf{a}_u = A_u \mathbf{y}_u$ etc.

After electroweak symmetry breaking, and assuming that CP is conserved, the Higgs sector will contain three CP-even Higgs bosons ($h_1, h_2, h_3$), two CP-odd ($a_1, a_2$) and one charged ($h^\pm$), where the states are ordered in terms of increasing mass. In the same way as in the MSSM the minimization conditions for the Higgs potential allows one to trade the $m_{H_u},  m_{H_d}$ parameters
for the doublet vev $v\approx 174$ GeV and $\tan\beta = v_u/v_d$. Similarly $m_S$ can be expressed in terms of the singlet vev $v_s$ which in turn gives rise to the effective $\mu$-parameter, $\mu=\lambda v_s$. All in all this leaves us with 6 unknown parameters describing the Higgs sector of NMSSM at tree-level: $ \tan\beta, \mu, \lambda, \kappa, A_\lambda, A_\kappa $. Below we will trade the latter two parameters for the masses $m_{h^\pm}$ and $m_{a_1}$.

As already alluded to the mass-eigenstates are mixtures of the electroweak eigenstates. More specifically, writing $S^{\rm weak} = (\mbox{Re}(H_u),\mbox{Re}(H_d),\mbox{Re}(S))$ we have $h_i = \mathbf{S}_{ij}S^{\rm weak}_j$.  (In the MSSM limit this means that $ \mathbf{S}_{12}=-\sin\alpha$.)
Similarly for the CP-odd states we have $a_i = \mathbf{A}_{ij}A^{\rm weak}_j$ with $A^{\rm weak}=(\mbox{Im}(\cos\beta H_u + \sin\beta H_d),\mbox{Im}(S))$. Here the mixing matrix is simply 
$\mathbf{A}=
\left( \begin{array}{cc}
\cos\theta_A & \sin\theta_A \\
-\sin\theta_A & \cos\theta_A
\end{array}
\right)
$.
Together with the ratio of the two doublet vevs (or equivalently the rotation angle needed to go to the Higgs basis where only one of the doublets have a vev) $\tan\beta$, the mixing matrices $\mathbf{S}$ and $\mathbf{A}$ specify the reduced couplings to fermions and gauge bosons as given in table \ref{tab:couplings}. 

\begin{table}[ht]
\centering
\begin{tabular}{cccc}
\hline
\hspace{10pt}Vertex\hspace{10pt} & \hspace{10pt}NMSSM\hspace{10pt} & \hspace{10pt}MSSM\hspace{10pt} &
\hspace{10pt}SM\hspace{10pt}   \\
\hline
$h_1tt$ & $\dfrac{\mathbf{S}_{11}}{\sin\beta}$ & $\dfrac{\cos\alpha}{\sin\beta}$ & 1 \\
$h_1bb$ & $\dfrac{\mathbf{S}_{12}}{\cos\beta}$ & $\dfrac{\sin\alpha}{\cos\beta}$ & 1 \\
$h_2tt$ & $\dfrac{\mathbf{S}_{21}}{\sin\beta}$ & $\dfrac{\sin\alpha}{\sin\beta}$ & {\rm n.a.} \\
$h_2bb$ & $\dfrac{\mathbf{S}_{22}}{\cos\beta}$ & $\dfrac{\cos\alpha}{\cos\beta}$ & {\rm n.a.} \\
\hline
$a_1tt$ & $\cot\beta\, \cos\theta_A$ & $\cot\beta$ & {\rm n.a.} \\
$a_1bb$ & $\tan\beta\, \cos\theta_A$ & $\tan\beta$ & {\rm n.a.} \\
%$a_2tt$ & $\cot\beta\, \sin\theta_A$ & ${\rm n.a.}$ & {\rm n.a.} \\
%$a_2bb$ & $\tan\beta\, \sin\theta_A$ & ${\rm n.a.}$ &{\rm n.a.} \\ 
\hline
$h_1VV$ & $ \sin\beta \, \mathbf{S}_{11} + \cos\beta \, \mathbf{S}_{12}$ & $\sin(\beta-\alpha)$ & 1 \\
$h_2VV$ & $ \sin\beta \, \mathbf{S}_{21} + \cos\beta \, \mathbf{S}_{22}$ & $\cos(\beta-\alpha)$ & {\rm n.a.} \\
$a_1h_1Z$ & $ (\cos\beta \, \mathbf{S}_{11} - \sin\beta \, \mathbf{S}_{12})\cos\theta_A$   & $\cos(\beta-\alpha)$ &  {\rm n.a.} \\
$a_1h_2Z$ & $ (\cos\beta \, \mathbf{S}_{21} - \sin\beta \, \mathbf{S}_{22})\cos\theta_A$   & $\sin(\beta-\alpha)$ &  {\rm n.a.} \\
\hline
$h_1h^+W^-$ & $ \cos\beta \,  \mathbf{S}_{11} - \sin\beta \, \mathbf{S}_{12}$ & $\cos(\beta-\alpha)$ & {\rm n.a.} \\
$a_1h^+W^-$ & $ \cos\theta_A$   & $1$ &  {\rm n.a.} \\
\hline
\end{tabular}
\caption{Reduced Higgs couplings in the NMSSM compared to the MSSM and the SM (when applicable). Note that the reduced couplings to fermions are identical for all three generations, even if only the third generation is displayed here. The couplings to $h_3$ can be obtained from the $h_1$ ones by the replacements 
$\mathbf{S}_{11} \to \mathbf{S}_{31}$ and $\mathbf{S}_{12} \to \mathbf{S}_{32}$ wheras the couplings to $a_2$ can be obtained from the $a_1$ ones by the replacement $\cos\theta_A \to \sin\theta_A$.}
\label{tab:couplings}
\end{table}

\section{Experimental constraints}
\label{sec:expcon}

In this section we will explore to what extent the process we are interested in is constrained by existing experimental data. Since we are interested in a light $h^\pm$ (with $m_{h^\pm}< m_t$) and a light $a_1$ there are constraints both from collider experiments as well as low-energy flavour experiments. However, before going in to the various constraints we will specify the scenarios that we have considered and then come back to the question of experimental constraints.

\subsection{Specification of SUSY scenario considered}

In the following we will consider a variant of the well motivated $m_h^{\max}$-scenario in the MSSM \cite{Carena:2002qg}, similarly to what was done in \cite{Mahmoudi:2010xp}. Thus we will consider a universal scale $M_{\rm SUSY}$ for the sfermion masses at the supersymmetry breaking scale. In other words we assume, as already stated, that the sfermion mass matrices ($\mathbf{M}_Q$ etc) are diagonal, and furthermore we assume that all diagonal entries are equal to $M_{\rm SUSY}$ which we keep fixed at 1 TeV. In addition we assume that the gaugino masses are related as in the constrained MSSM, where supersymmetry breaking is assumed to be mediated by gravity, namely $M_1=100$ GeV, $M_2=200$ GeV, $M_3=800$ GeV. 
Finally we will assume that $A_t=A_b=A_\tau$ but contrary to what was done in \cite{Mahmoudi:2010xp} we will let them vary in the range $A_t \in [-5000,5000]$ GeV so that the amount of mixing between the $\tilde{t}_L$ and $\tilde{t}_R$ is unconstrained. 

For the Higgs sector we will let all six parameters vary freely. However, as was done in \cite{Mahmoudi:2010xp}, we will trade the $A_\kappa$ parameter for $m_{a_1}$ and $A_\lambda$ parameter for $m_{h^\pm}$ using an iterative procedure starting from the tree-level relations:
\begin{eqnarray}
m_{h^\pm}^2 &  = &\frac{2\mu}{\sin2\beta}\left(A_\lambda+\dfrac{\kappa}{\lambda}\mu\right)+ m_W^2 - \lambda^2 v^2\\
{\cal M}_P^2 & = & 
\left ( 
\begin{array}{cc}
\displaystyle \frac{2\mu}{\sin2\beta}\left(A_\lambda+\dfrac{\kappa}{\lambda}\mu\right) & \lambda v \left(A_\lambda-2\dfrac{\kappa}{\lambda} \mu\right) \\
\lambda v\left( A_\lambda-2\dfrac{\kappa}{\lambda} \mu\right) & \dfrac{\lambda^2 v^2\sin2\beta}{2\mu}\left({A_\lambda}+4 \dfrac{\kappa}{\lambda}\mu \right)-3\dfrac{\kappa}{\lambda} A_\kappa \mu
\end{array}
\right),
\end{eqnarray}
where the latter gives the masses of the mass-eigenstates $a_1,a_2$ after diagonalisation.
 Thus the parameters we consider with their respective ranges are:
\begin{eqnarray*}
‪ \tan\beta & \in & [1,60], \\
 \lambda & \in & [0,0.7], \\
 \kappa & \in & [-0.7,0.7],  \\
 \mu & \in & [125,1000] \mbox{ GeV},  \\
 m_{h^\pm} & \in & [80,170] \mbox{ GeV}, \\
 m_{a_1} & \in & [4,150] \mbox{ GeV}.
\end{eqnarray*}
The limits for the various parameters has been chosen as follows: For $\tan\beta$ , $\kappa$ and $\lambda$ we impose perturbativity up to the GUT scale which effectively means that any value out side the above regions are bound to fail. (In addition some points inside these regions also fail because of this requirement.) The lower limits on $\mu$ and $m_{h^\pm}$ are dictated by experimental constraints. The upper limit on $\mu$ is not a hard one but follows from the implicit assumption that $\mu$ should be of order the electroweak scale whereas the upper limit on $m_{h^\pm}$ is given by the condition that the decay $t \to b h^+$ should be open. The reason for letting $\mu$ vary freely is mainly that this decreases the correlations between the masses of the Higgs bosons as will be discussed more below.  
Finally the lower limit on $m_{a_1}$ is chosen in order to have $a_1 \to \tau^+ \tau^-$ open whereas the upper limit follows from having $m_{h^\pm}<170$ GeV.

In order to calculate the resulting models from the inputs we use the package NMSSMTools version 3.2.0 \cite{Ellwanger:2004xm,Ellwanger:2005dv} with default settings. Among other thing this means that we impose perturbativity of the model up to the GUT scale.
Finally, in all scans we generate $\sim$ 1 M points (with flat priors in the parameters considered) which fulfill the theoretical constraints implemented in NMSSMTools.

\subsection{Current experimental constraints}

The most important constraints comes from the direct searches for Higgs bosons for which we use the package HiggsBounds version 3.7.0 \cite{Bechtle:2008jh,Bechtle:2011sb}. In addition there are also constraints from direct searches for supersymmetric particles, various flavour constraints and in principle also the anomalous magnetic moment of the muon as well as the relic density of dark matter. We have not applied the lattr two constraints for the following reasons. To investigate the amount of dark matter in the various models one would also need to vary the gaugino masses as was done in \cite{Vasquez:2012hn}. Since we keep these fixed we have not applied the dark matter constraints. On the same vain we have not applied the constraint from the anomalous magnetic moment of the muon, since this depends on the masses of the scalar partners of the muon and the neutrinos, apart from requiring $\mu$ to be positive. 

When it comes to the flavour constraints the situation is more involved in that various constraints have different level of model dependence. 
On the one hand there are constraints from tree-level mediated processes such as $B_u \to \tau^+ \nu_\tau$, which only depend on the Higgs sector and on the other hand there a constraints from loop-mediated processes such as $B_s \to \mu^+ \mu^-$ and $b \to s \gamma$ which depend on details of the supersymmetric sector of the model. In the following we will limit ourselves to applying the most severe constraints from $B_u \to \tau^+ \nu_\tau$, which limits the available parameter space in $[m_{h^\pm},\tan\beta]$  and $B_s \to \mu^+ \mu^-$ which puts limits on  $[m_{a_1},\tan\beta]$. The latter constraint is especially important since we will consider  light $a_1$ which is very constrained by the data from LHCb and CMS \cite{Aaij:2012ac,Chatrchyan:2012rg}. Finally we apply the direct constraints from searches for supersymmetric particles. For all constraints except the Higgs bosons we use NMSSMTools version 3.2.0.

\begin{figure}%[t]
\includegraphics[width=8cm]{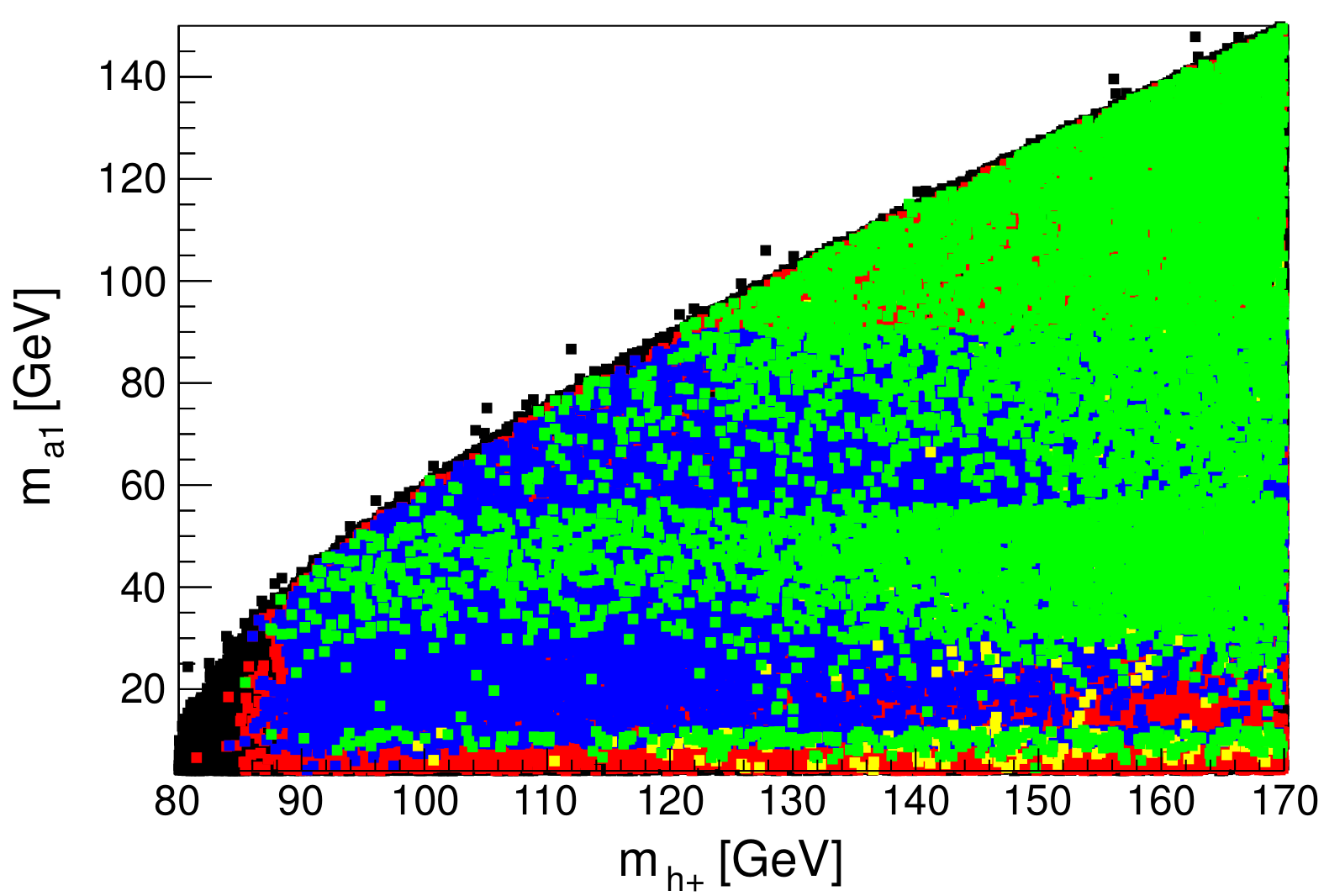}
\includegraphics[width=8cm]{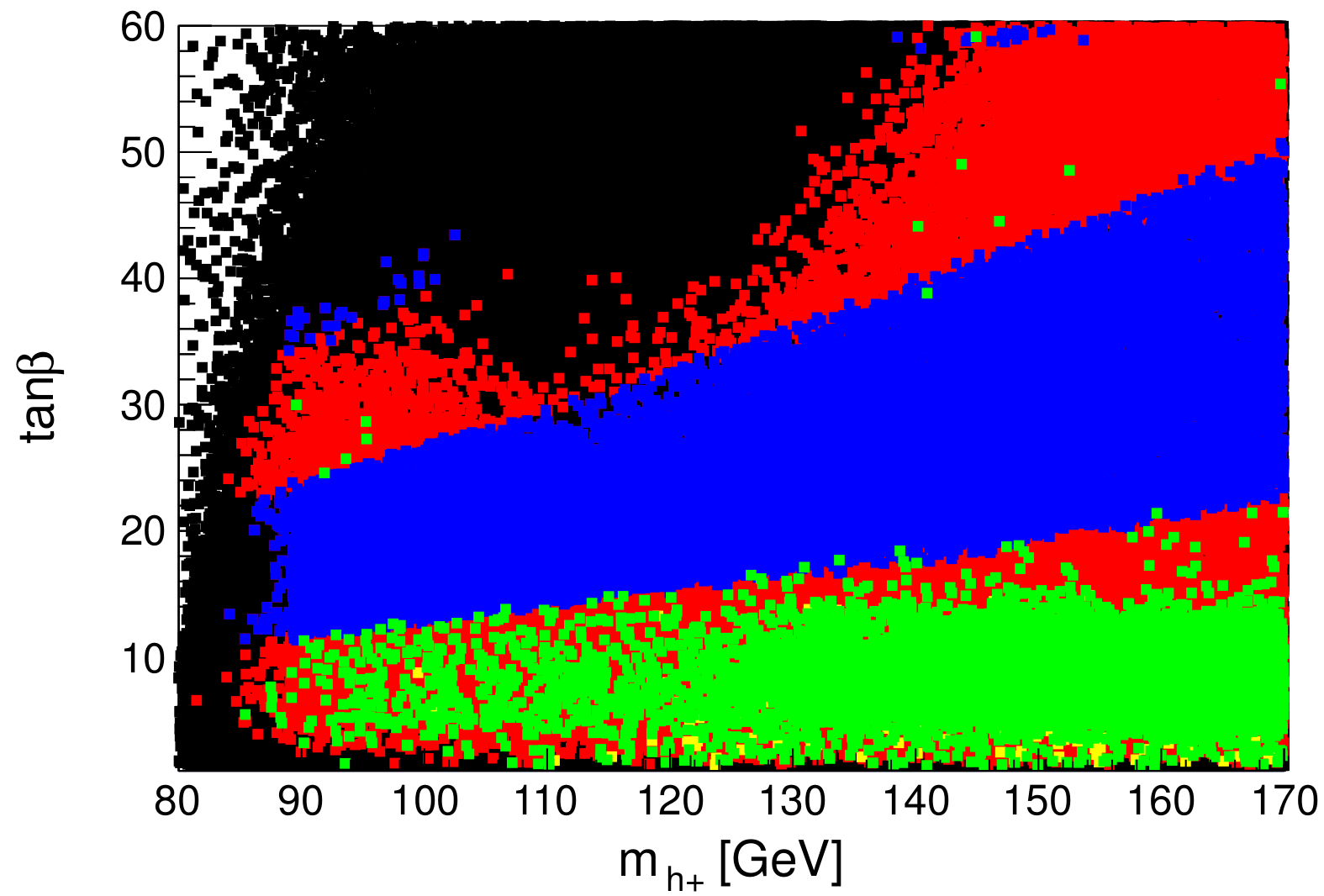}
\includegraphics[width=8cm]{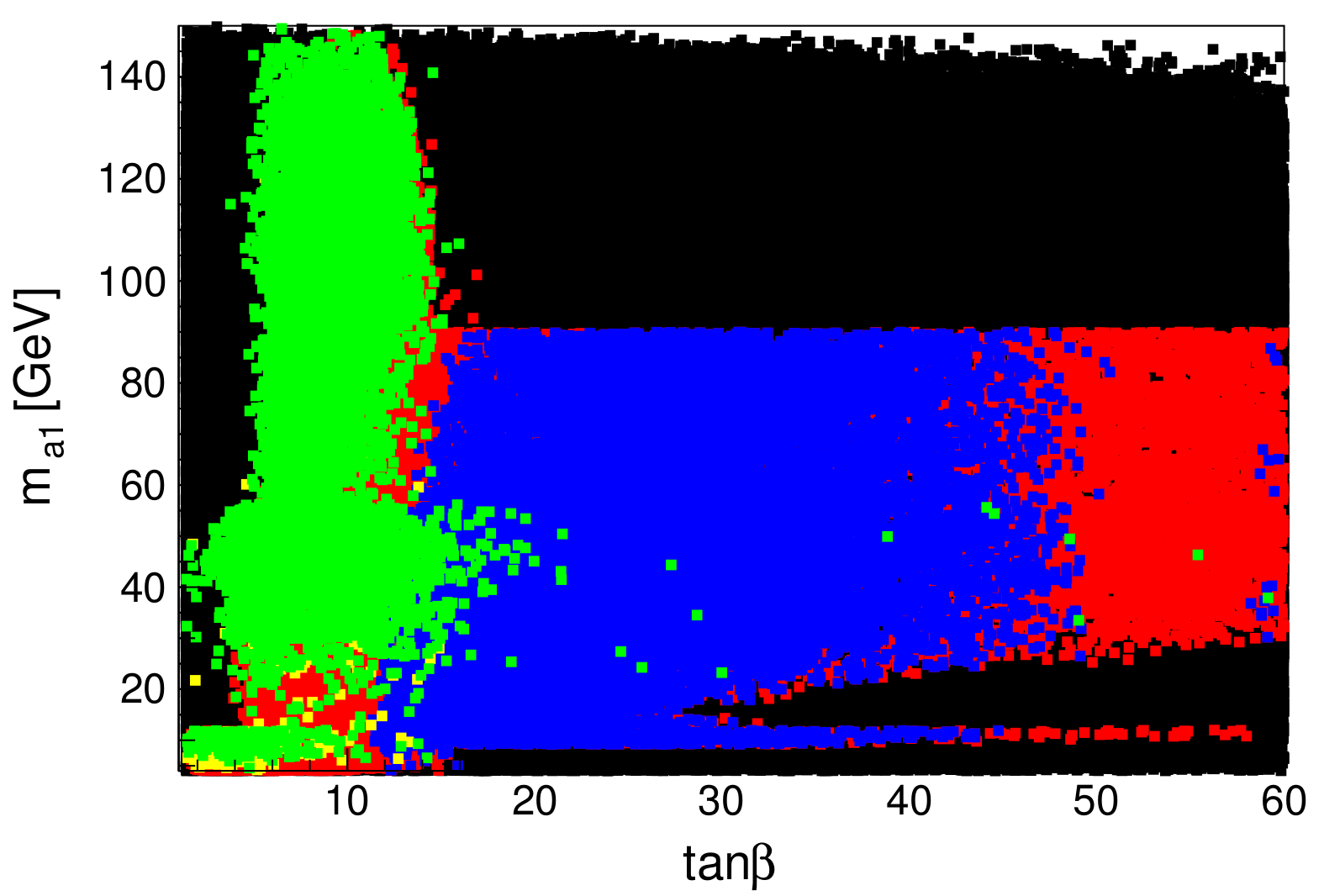}
\includegraphics[width=8cm]{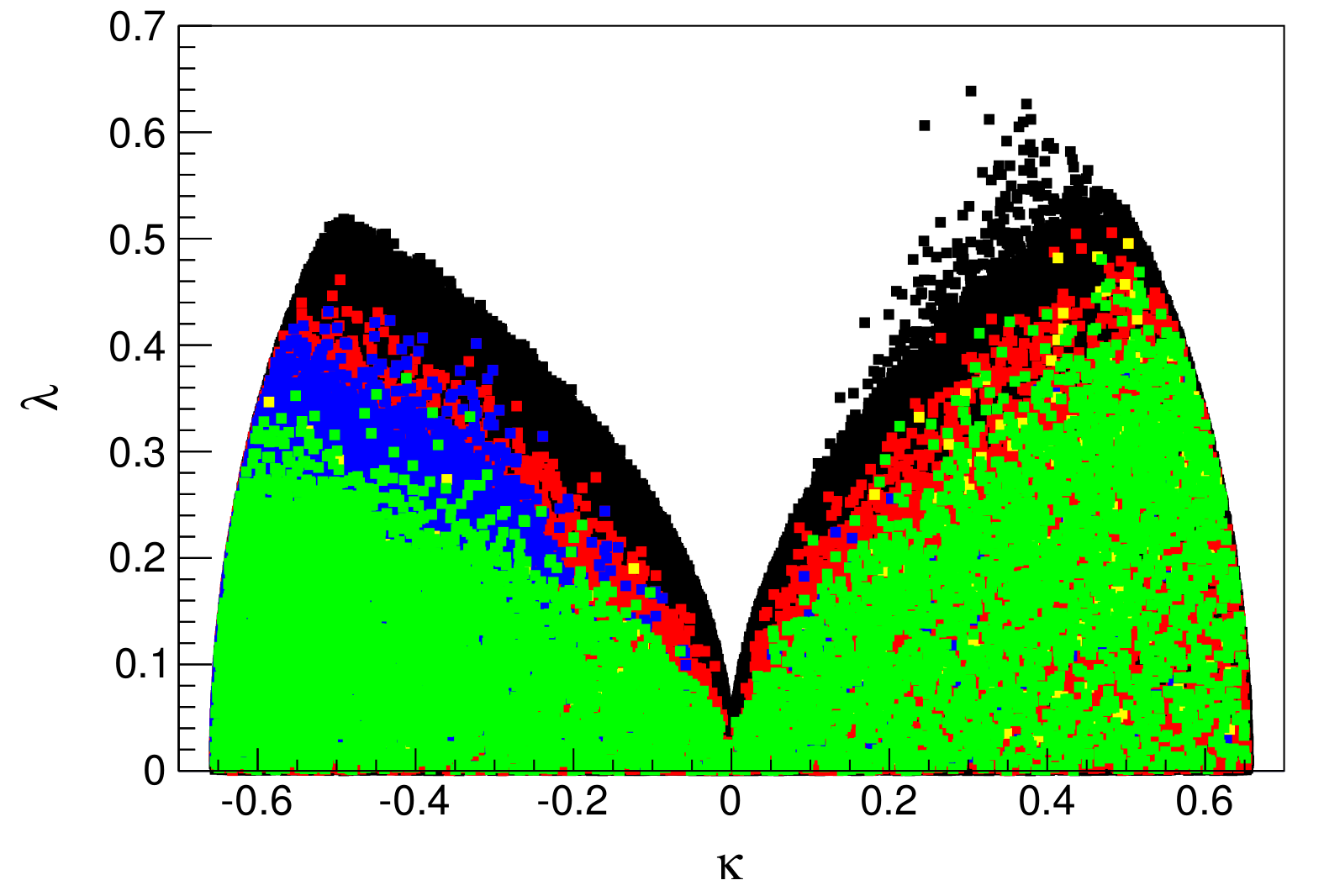}
\includegraphics[width=8cm]{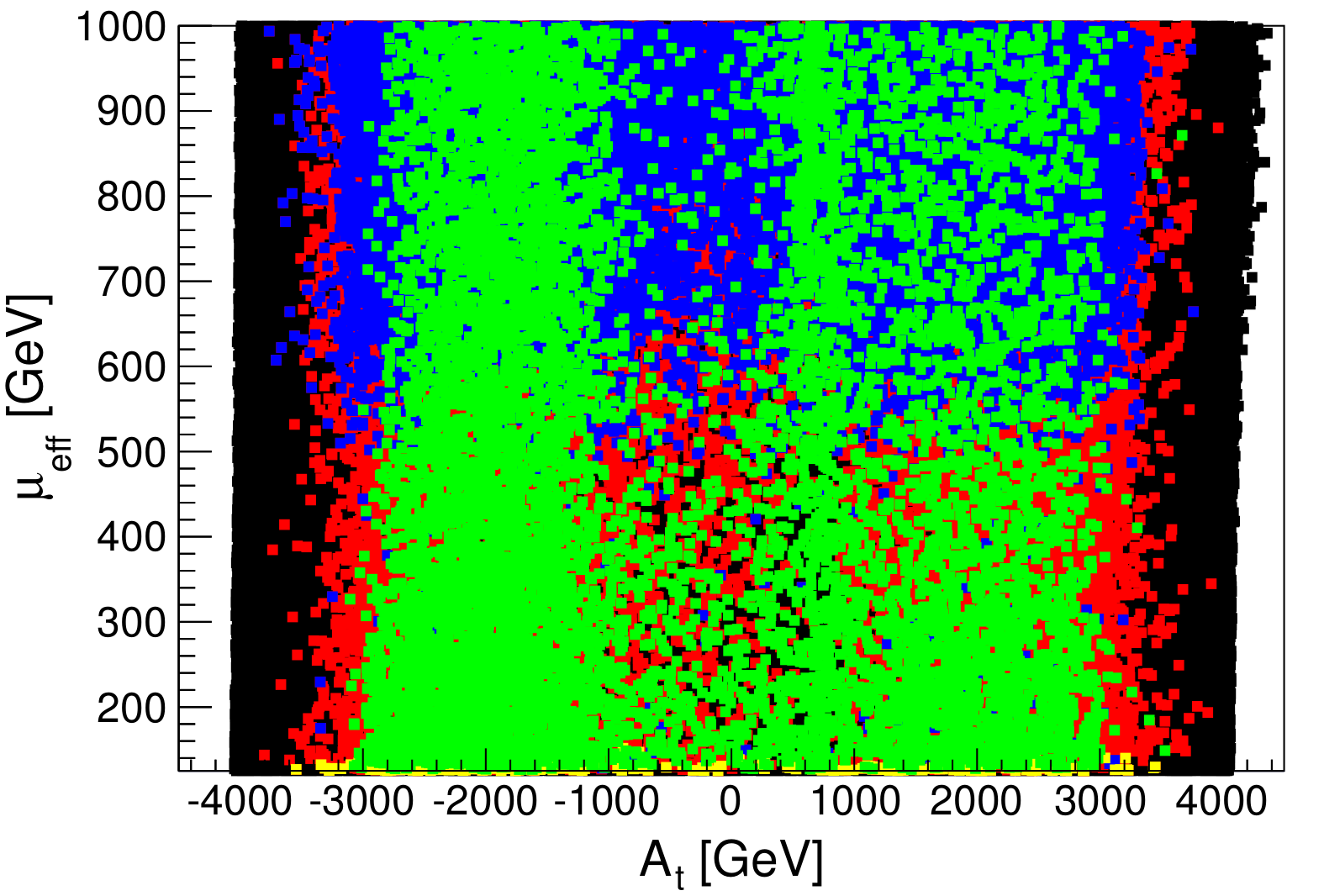}
\caption{Correlations between the parameters of the scan in the scenario under consideration (see text for details) with the various constraints applied as follows. All points in black are viable model points but are excluded by HiggsBounds. The coloured points are allowed by HiggsBounds and plotted in the following order: red points are excluded by  $B_s \to \mu^+ \mu^-$, blue points are excluded by $B_u \to \tau^+ \nu_\tau$, yellow points are excluded by direct searches for supersymmetric particles, and finally green points are allowed by all constraints considered. }
    \label{fig:JRscan}
\end{figure}

The results from the scan are displayed in Fig.~\ref{fig:JRscan} with black points being viable models from a theory point of view but excluded by the direct searches for Higgs bosons and coloured points being allowed by the same constraint.
 Of primary interest are the allowed regions in $[m_{a_1},m_{h^\pm}]$. As is clear from the figure there is a distinct region of points with $m_{a_1} \sim 2 m_B\approx10.6$ GeV allowed by all constraints (indicated by green colour) for essentially any value of  $m_{h^\pm} \gtrsim 90$ GeV. The same is also true in the constrained $m_h^{\max}$ scenario where $\mu=200$ GeV and $A_t = -\sqrt{6}M_{\rm SUSY} + \mu\cot\beta$ are fixed as was already noted in \cite{Mahmoudi:2010xp}. However, at difference to the constrained $m_h^{\max}$ scenario there are regions with larger $m_{a_1}$ that are allowed also for $m_{h^\pm} \lesssim 120$ GeV. 

Looking at the $[m_{h^\pm},\tan\beta]$-plane one clearly sees the constraint from $B_u \to \tau^+ \nu_\tau$ for intermediate $\tan\beta$ (show as blue points). For larger $\tan\beta$ there is a cancellation between the SM and $h^\pm$ contributions, which makes this region allowed by $B_u \to \tau^+ \nu_\tau$, but instead the constraints from $B_s \to \mu^+ \mu^-$ come into play (red points). It should be noted that the points excluded by $B_u \to \tau^+ \nu_\tau$ are plotted on top of the constraints from $B_s \to \mu^+ \mu^-$. Similarly the constraints from searches for supersymmetric particles are plotted (in yellow) on top of the constraints from  $B$-decays, but with the cut $\mu > 125$ GeV there are hardly any points excluded by this constraint. Finally we note that for $\tan\beta \lesssim 15$ there are allowed points in parameter space for essentially any value of $m_{h^\pm} \gtrsim 90$ GeV.
 
Turning to the  $[m_{a_1},\tan\beta]$-plane we see that the region $m_{a_1} \sim 2 m_B$ is allowed by all constraints up to $\tan\beta \lesssim 10$, whereas for larger $\tan\beta$ there are points allowed by direct Higgs boson searches but not allowed by $B_s \to \mu^+ \mu^-$ and  $B_u \to \tau^+ \nu_\tau$. Given the uncertainties related to the indirect constraints from  $B$-decays we conclude that there is a region in parameter space with $m_{a_1} \sim 2 m_B$, $m_{h^\pm}  \in [90,170]$ GeV, and $ \tan\beta \in [1,60]$ that should be searched for by the ATLAS and CMS experiments. It should also be noted that values both above and below the threshold $m_{a_1} = 2 m_B$ are allowed by the constraints.

Before turning to the signal of interest, {\it i.e.\ }$Br(h^\pm \to a_1 W)$, we also show in  Fig.~\ref{fig:JRscan} the effects of the various constraints when projected onto the $[\kappa,\lambda]$ and $[A_t,\mu]$-planes. From the first of these plots one clearly sees the constraint $\sqrt{\kappa^2+\lambda^2} \lesssim 0.7$ which arises from requiring perturbativity up to the GUT scale. From the second we see that the constraints imply $|A_t|\lesssim 3500$ GeV, which essentially follows from the radiative corrections to the lightest CP-even Higgs becoming small or even negative for large $|A_t|$ relative to the value  $M_{\rm SUSY}=1 $ TeV that we are using. We also see that
for $\mu$ there are hardly no experimental constraints in the region considered. On the other hand, if we would extend $\mu$ to values smaller than 125 GeV then all those points would be excluded by searches for supersymmetric particles.

\begin{figure}%[t]
\includegraphics[width=8cm]{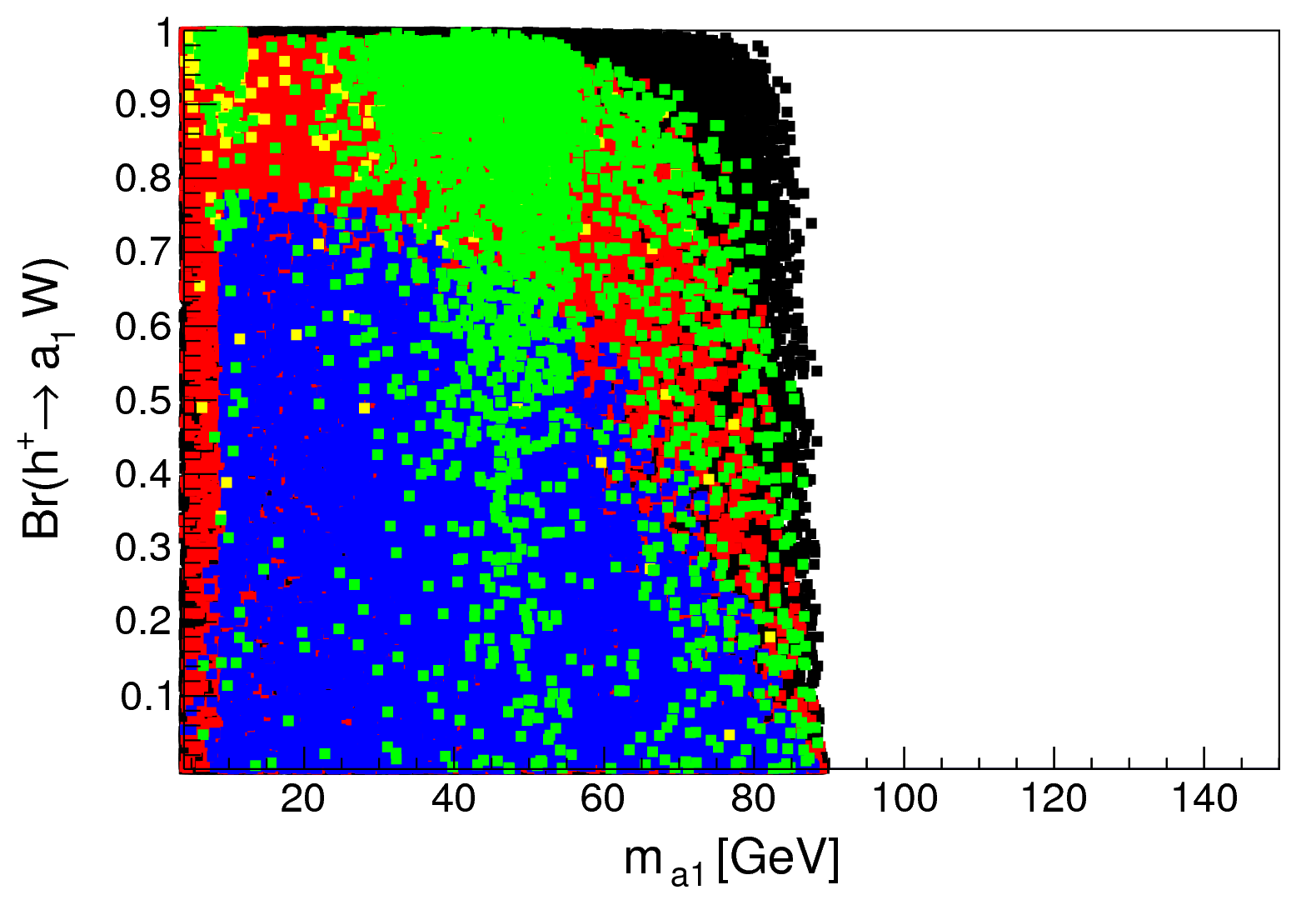}
\includegraphics[width=8cm]{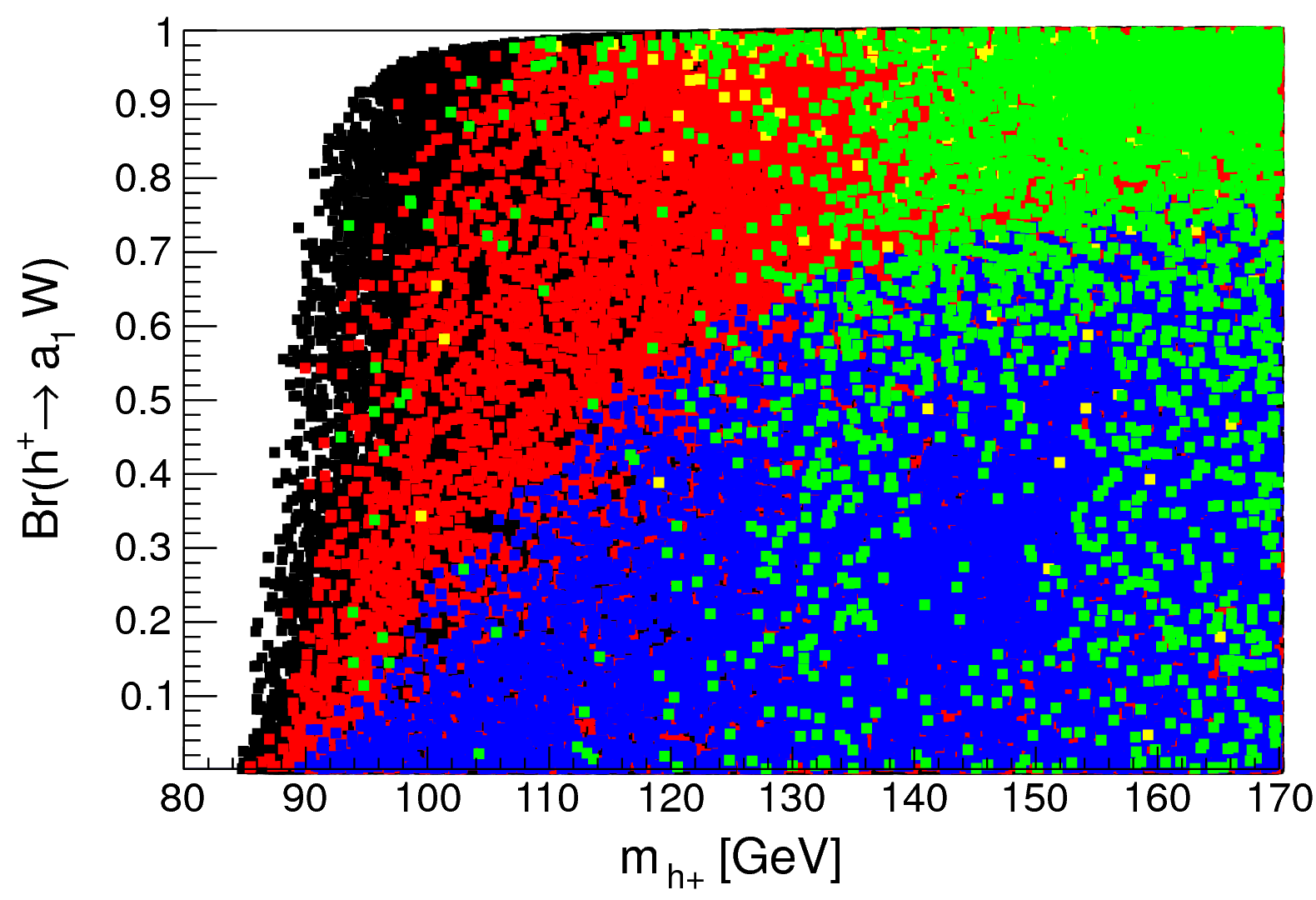}
\includegraphics[width=8cm]{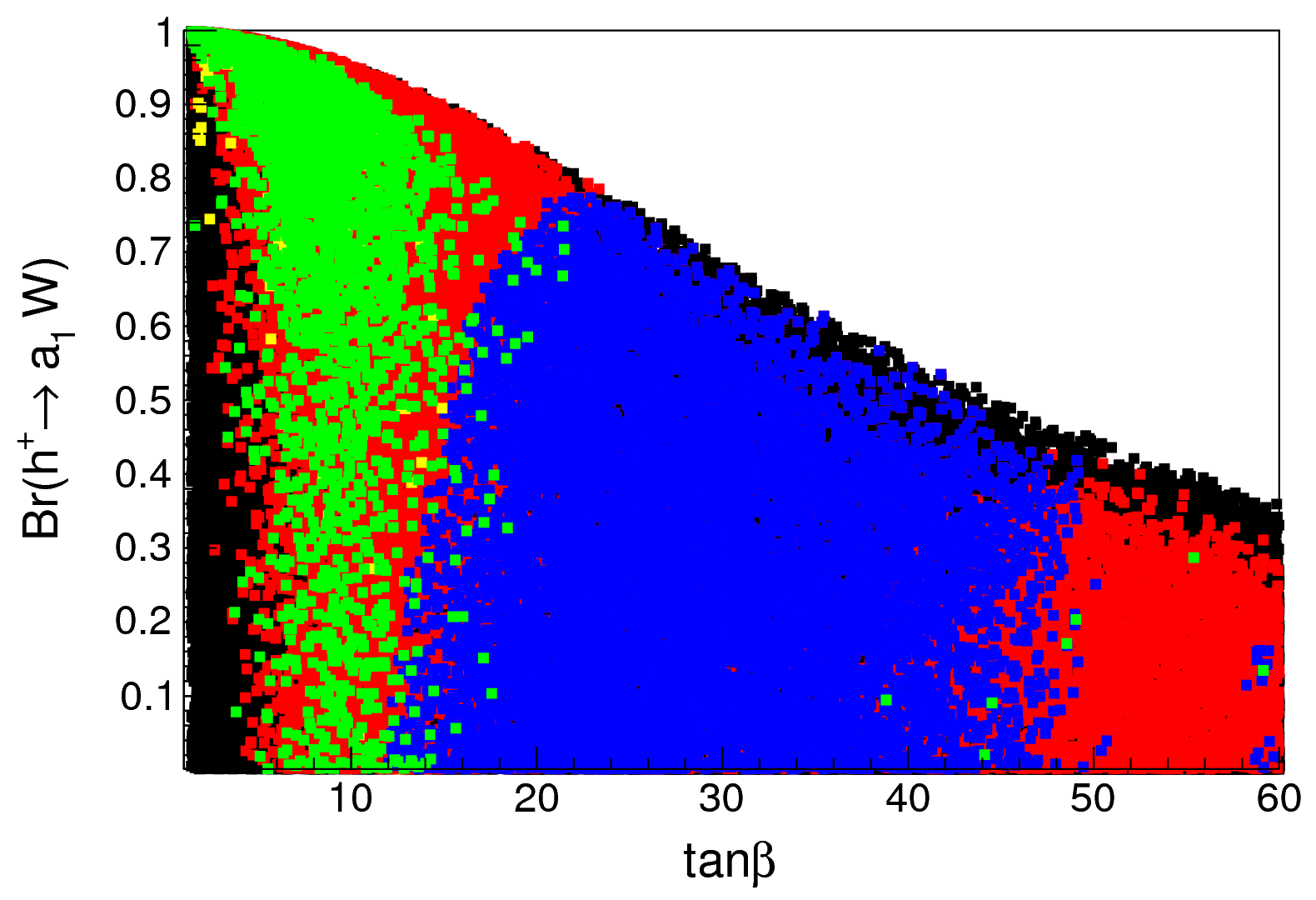}
\includegraphics[width=8cm]{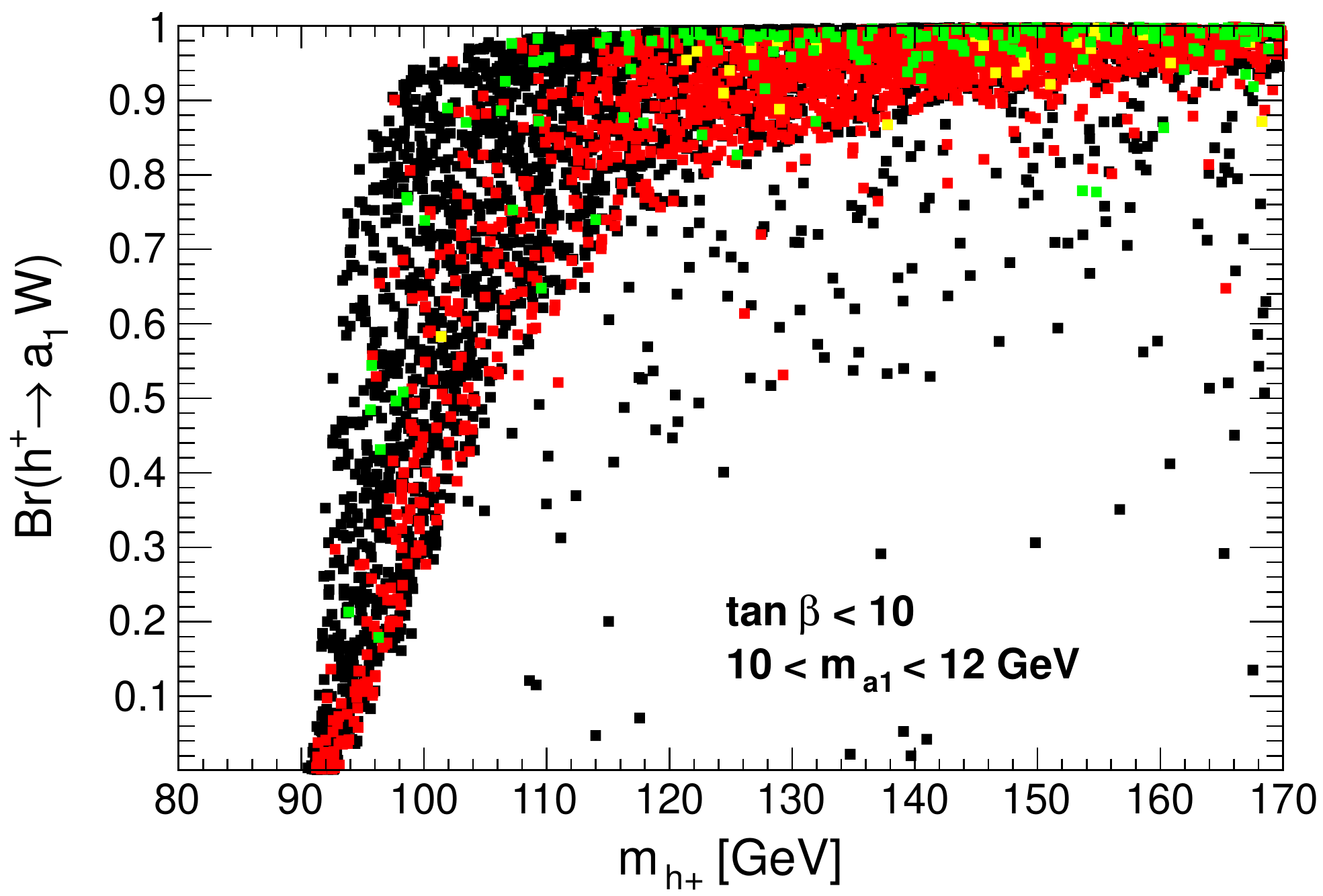}
\caption{The branching ratio for $h^\pm \to a_1 W$ when scanning over the parameters in the scenario under consideration (see text for details) with the various constraints applied. The colour coding is the same as in Fig.~\protect\ref{fig:JRscan}}
    \label{fig:BRscan}
\end{figure}

As promised we turn now to the resulting branching ratios for the decay $h^\pm \to a_1 W $. From Fig.~\ref{fig:BRscan}  we observe the following general feature, the branching ratio can be large as soon as the channel is open ($m_{a_1} < m_{h^\pm} -m_W$) except for large $\tan\beta$ where the decay  $h^\pm \to \tau \nu_\tau $ becomes dominant. Concentrating on those points that pass all the constraints considered and the region $m_{a_1} \sim 2 m_B$  we also see from the lower right plot that  $Br(h^\pm \to a_1 W ) \gtrsim 0.9$
as long as $\tan\beta \lesssim 10$ and $m_{h^\pm} \gtrsim 100$ GeV. Thus we can conclude that not only is the parameter space region $m_{a_1} \sim 2 m_B$, $m_{h^\pm}  \in [100,170]$ GeV, and $ \tan\beta \in [1,10]$ allowed - in this region the decay 
$h^\pm \to a_1 W $ is also dominant. In the next section we will exemplify how the so far unexplored region of parameter space with $m_{a_1}   \in [10,12]$ GeV can be probed by searching for $h^\pm \to a_1 W $ with $a_1 \to b\bar{b}$ in $t\bar{t}$-production at the LHC. 

\begin{figure}%[t]
\includegraphics[width=8cm]{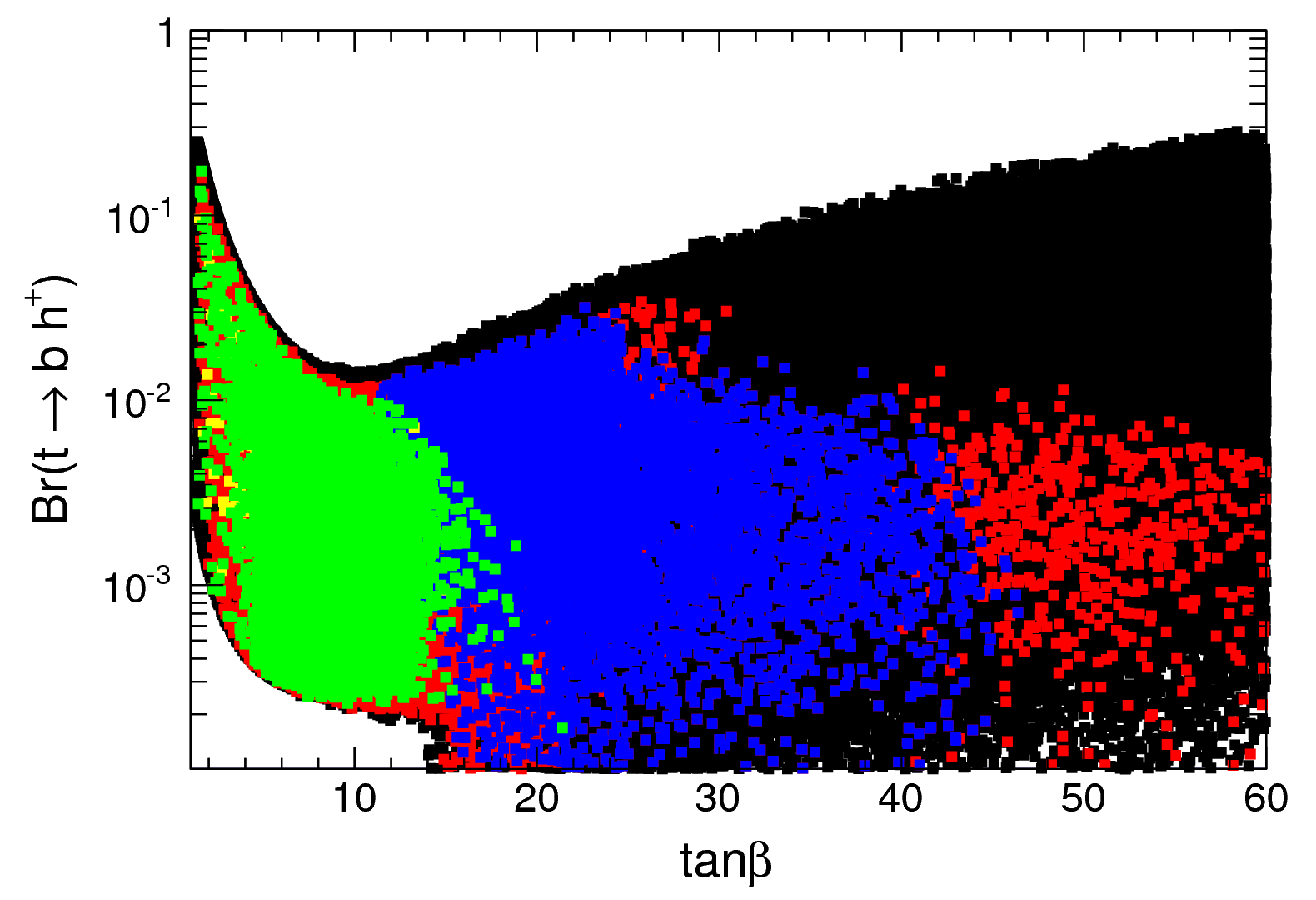}
\includegraphics[width=8cm]{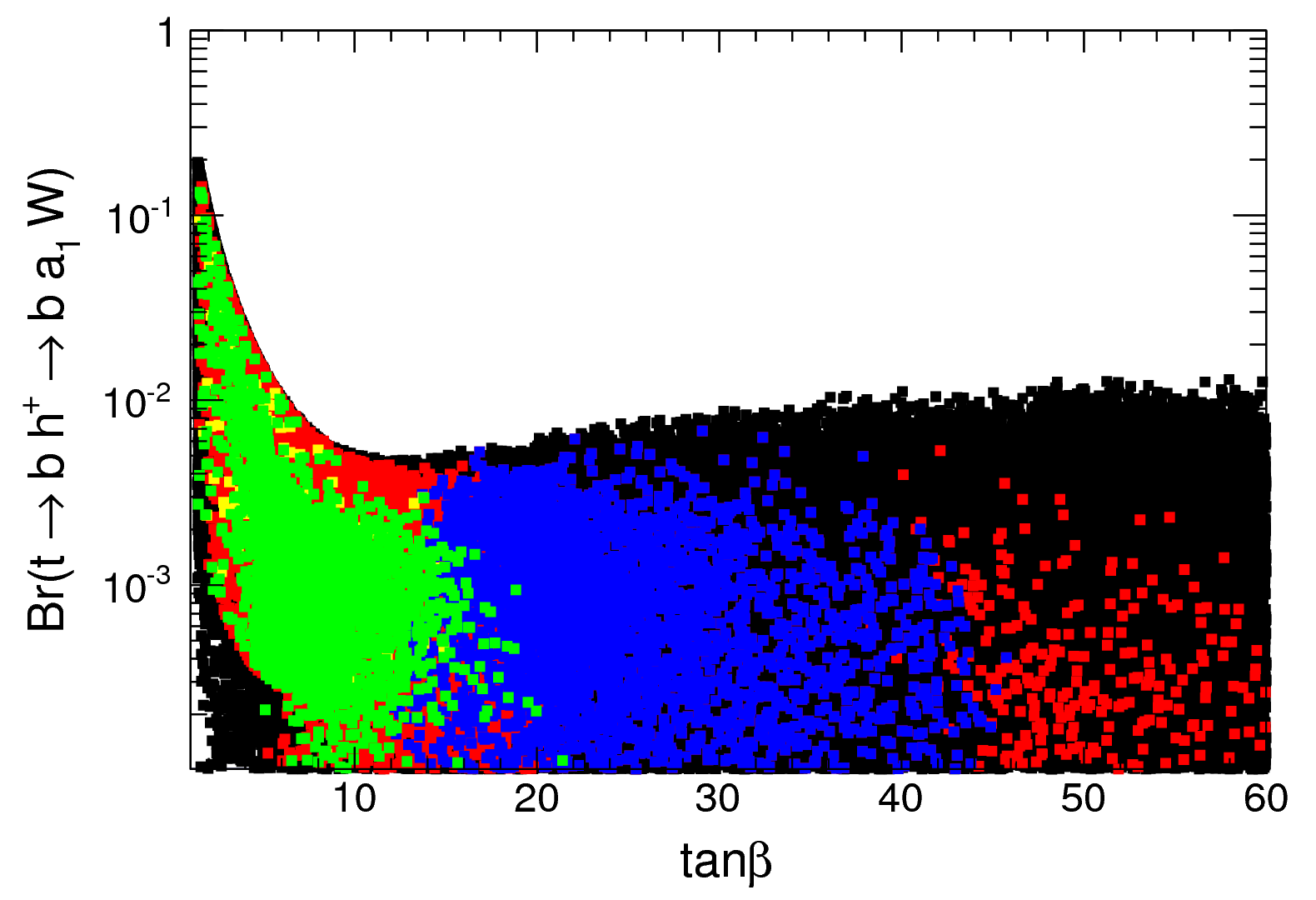}
\includegraphics[width=8cm]{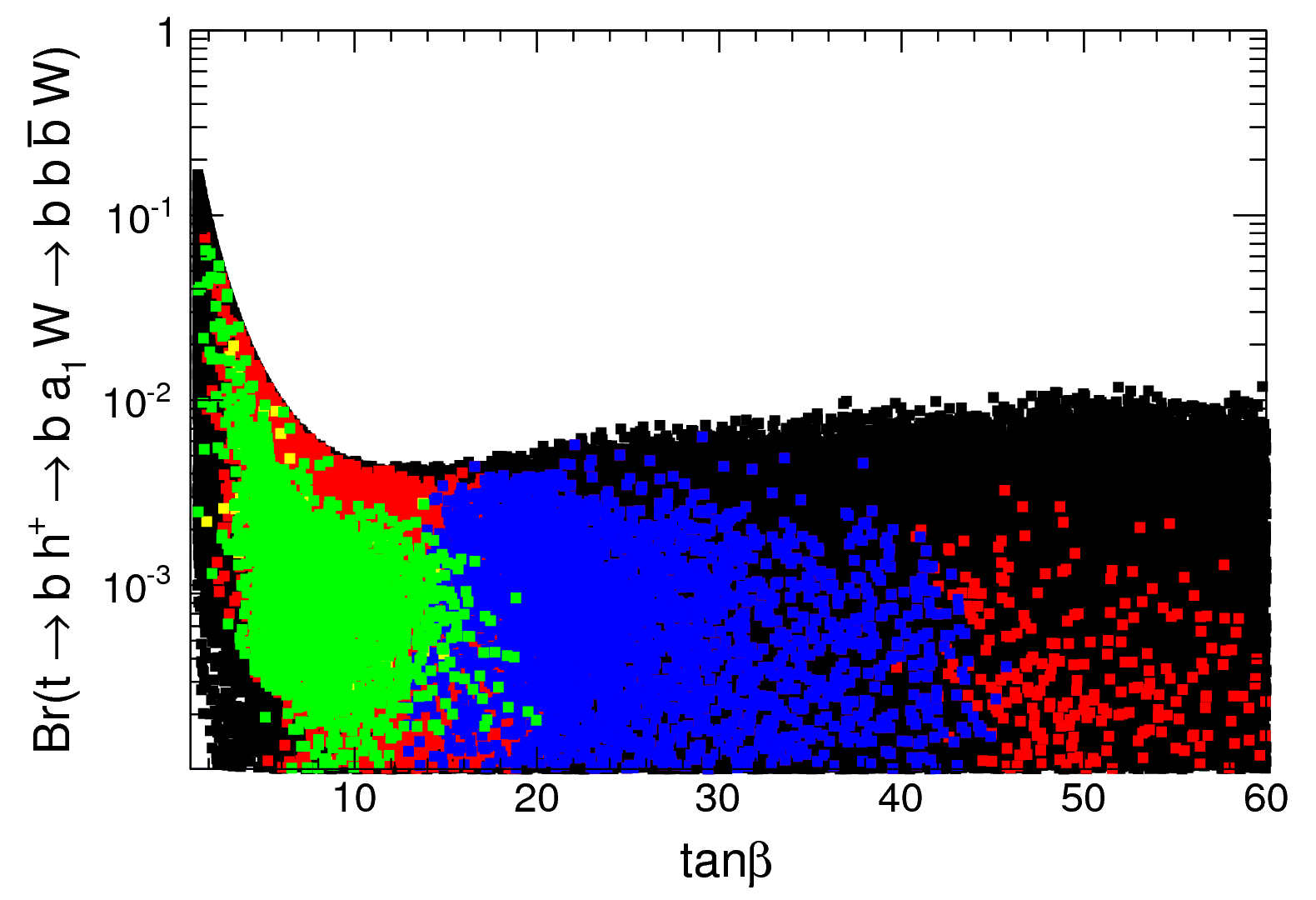}
\caption{The branching ratios for the decay chains $t \to b h^+ $, $t \to b h^+ \to b a_1 W$, and  $t \to b h^+ \to b a_1 W \to b b \bar{b} W $ respectively when scanning over the parameters in the scenario under consideration (see text for details) with the various constraints applied and only considering points with $m_{h^\pm}< 160$ GeV. The colour coding is the same as in Fig.~\protect\ref{fig:JRscan}}
    \label{fig:BRcomp}
\end{figure}

Before ending this section we also show in Fig.~\ref{fig:BRcomp} the branching rations for the decay chains of interest, {\it i.e.\ }$t \to b h^+ $, $t \to b h^+ \to b a_1 W$, and  $t \to b h^+ \to b a_1 W \to b b \bar{b} W $ as a function of $\tan\beta$ when restricting to parameter space points with $m_{h^\pm}< 160$ GeV. As can be seen from the figure, points with $Br(t \to b h^+) $ as large as $0.2$ are still allowed when the decay  $h^\pm \to a_1 W $ is included. This should be compared with the experimental constraints from ATLAS and CMS which so far have assumed that  $Br(h^\pm \to \tau \nu_{\tau}) = 1 $ giving a limit $Br(t \to b h^+) \lesssim $ few percent \cite{Aad:2012tj,CMS-light-h+}.

\section{Search strategy}
\label{sec:search}

In the following section we will perform a signal-to-background analysis for three different charged Higgs masses: $m_{h^+}=$ 100, 130 and 150 GeV respectively. For definiteness we have used $\tan\beta=50$ when simulating the signal but the end results will not depend on this value. The mass of the $a_1$ is set to 11 GeV throughout as an example of a small $a_1$ mass that is just above $b\bar{b}$-threshold. The three charged Higgs masses are chosen to illustrate different kinematic properties: at $m_{h^\pm}=150$ GeV the $b$-jet from the $t \to bh^+$ decay will be rather soft whereas the $a_1$ from the $h^+$ will be harder. For $m_{h^\pm}=100$ GeV the situation will be the opposite, and in the intermediate case $m_{h^\pm}=130$ GeV both jets can be relatively hard and in addition the available phase-space will be largest.

\begin{figure}%[t]
\includegraphics[width=8cm]{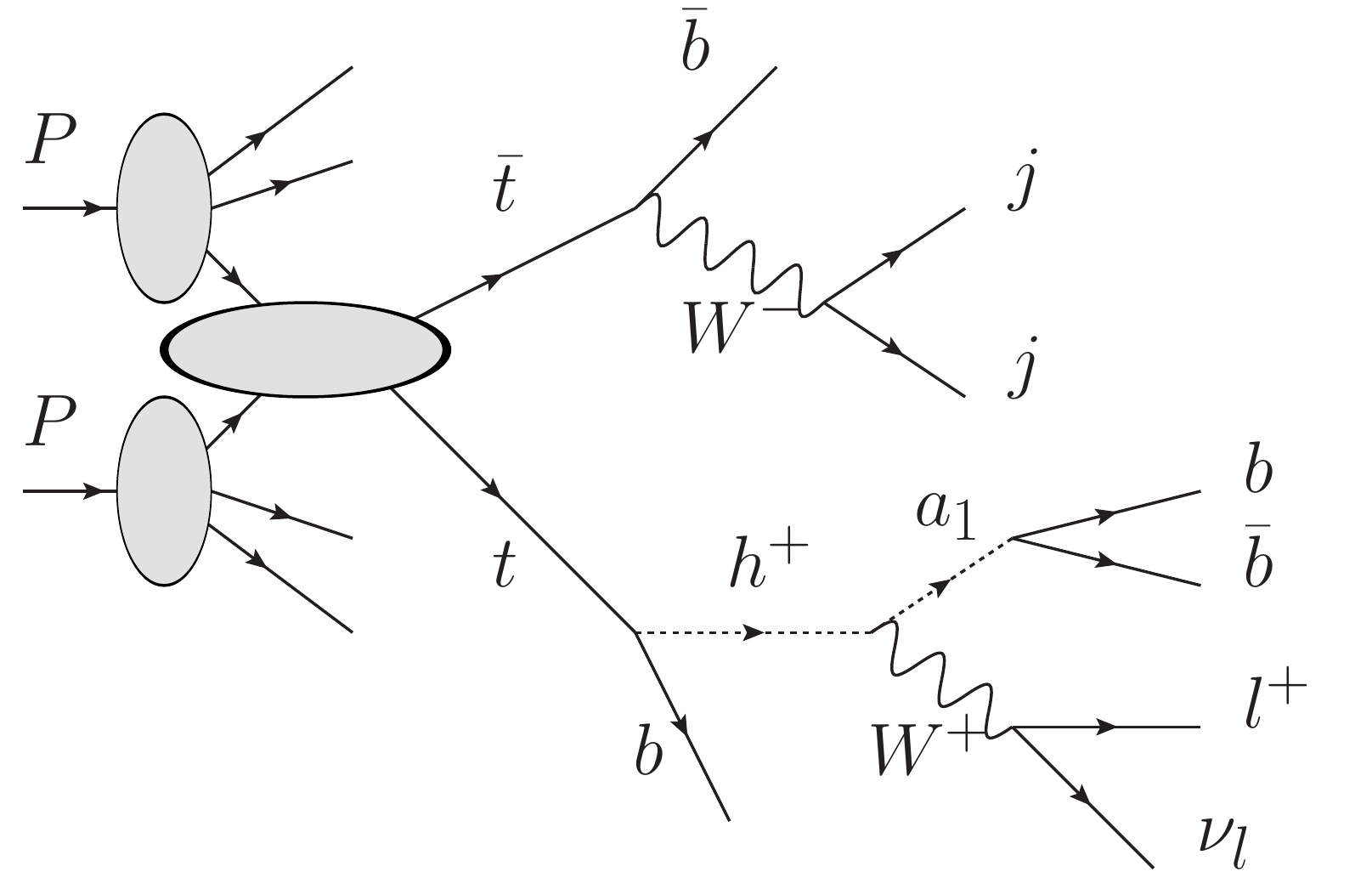}
\caption{Illustrations of $t\bar{t}$ production in proton-proton collisions with the subsequent decay chain considered in this paper.}
    \label{fig:feynman}
\end{figure}

We aim to reconstruct the signal process where one of the $t$-quarks decays leptonically via  $h^\pm \to a_1 W$ with $W \to \ell \nu_\ell$ and the other hadronically via $W\to jj$ as illustrated in Fig.~\ref{fig:feynman}. 
All cross-sections have been corrected for these enforced W decays as well as a factor of 2 for taking account of the process also with interchanged roles between the $t$ and the $\bar{t}$.  Because the $a_1$ is supposed to decay close to threshold to $b\bar{b}$, we aim for a reconstruction where the two $b$'s from the $a_1$ are clustered together to give a single $b\bar{b}$-jet.

As backgrounds to the process we consider the irreducible $t\bar{t}b\bar{b}$ as well as, because of its higher magnitude, $t\bar{t}$ with one jet being accidentally $b$-tagged (weighted with a mis-tagging probability, assumed to be 0.01~\cite{CMS-PAS-BTV-11-004,ATLAS-CONF-2012-040}). In order to include also the single top contributions to the background we have simulated the processes $\bar{t} b W^+b\bar{b}$ and $\bar{t} b W^+$ respectively, but in the following we will denote them as  $t\bar{t}b\bar{b}$ and $t\bar{t}$ for simplicity.
For other reducible backgrounds, such as $W+b \mbox{ jets}$, we assume that similar procedures can be applied as in the $t\bar{t}$ cross-section determination. For example, requiring two $b$-tagged jets reduces the $W+b \mbox{jets}$ background to $t\bar{t}$ production to about 10\%~\cite{Chatrchyan:1376068}. Several cuts are applied to strengthen the signal and to suppress the background as will be discussed in the following. 

A center-of-mass energy of 8 TeV at the Large Hadron Collider is assumed throughout the whole analysis. For the generation of the hard matrix elements, we use MadGraph 5~\cite{Alwall:2011uj} with a fixed renormalization and factorization scale (set as default to the Z mass) and the 'CTEQ6L1' parton distribution functions. To supply MadGraph with the proper parameters of the signal we have for simplicity used a simple two Higgs Doublet Model with the masses given above as implemented in the Two Higgs Doublet Model Calculator 2HDMC~\cite{Eriksson:2009ws}. The only important difference to the NMSSM then arises from the  $h^\pm \to a_1 W$ decay giving and extra factor  $\cos^2 \theta_A$ (cf. Table \ref{tab:couplings}). 
All other steps to generate complete events, such as radiation, underlying events and hadronization, are carried out using Pythia 8. We start from bare samples of 100000 events for the different signals as well as the $t\bar{t}b\bar{b}$ background whereas for the $t\bar{t}$ background we have 50 times higher statistics to start with. There is no detector simulation included in this exploratory analysis but we have simulated $b$-tagging in a simplified way as detailed below.

For all processes we use the leading order cross-sections obtained from MadGraph. On the one hand this means that there is an overall scale factor which is more or less the same for both signal and background. On the other hand the rates will be lower than what would have been result if higher order cross-sections had been used. All in all this means that the signal over background rates we find will be underestimated in this respect. For example we get a LO cross-section for 
$p p \to t \bar{t}$ process in the SM of 138 pb to be compared with the NNLL resummed result of 232 pb \cite{Beneke:2011mq}.

\subsection{Reconstruction of the leptonic $W$}

\begin{figure}%[t]
\includegraphics[width=8cm]{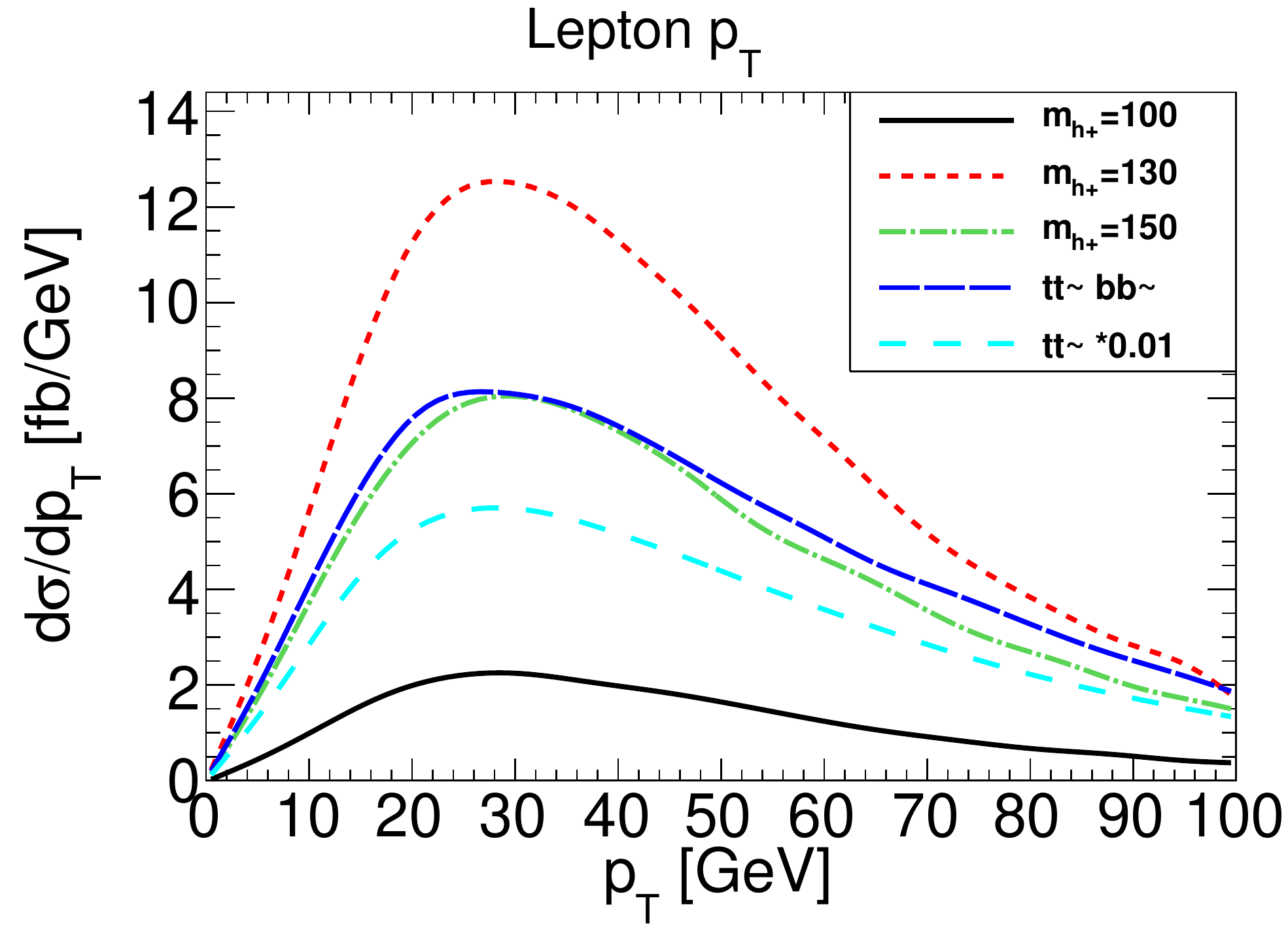}
\includegraphics[width=8cm]{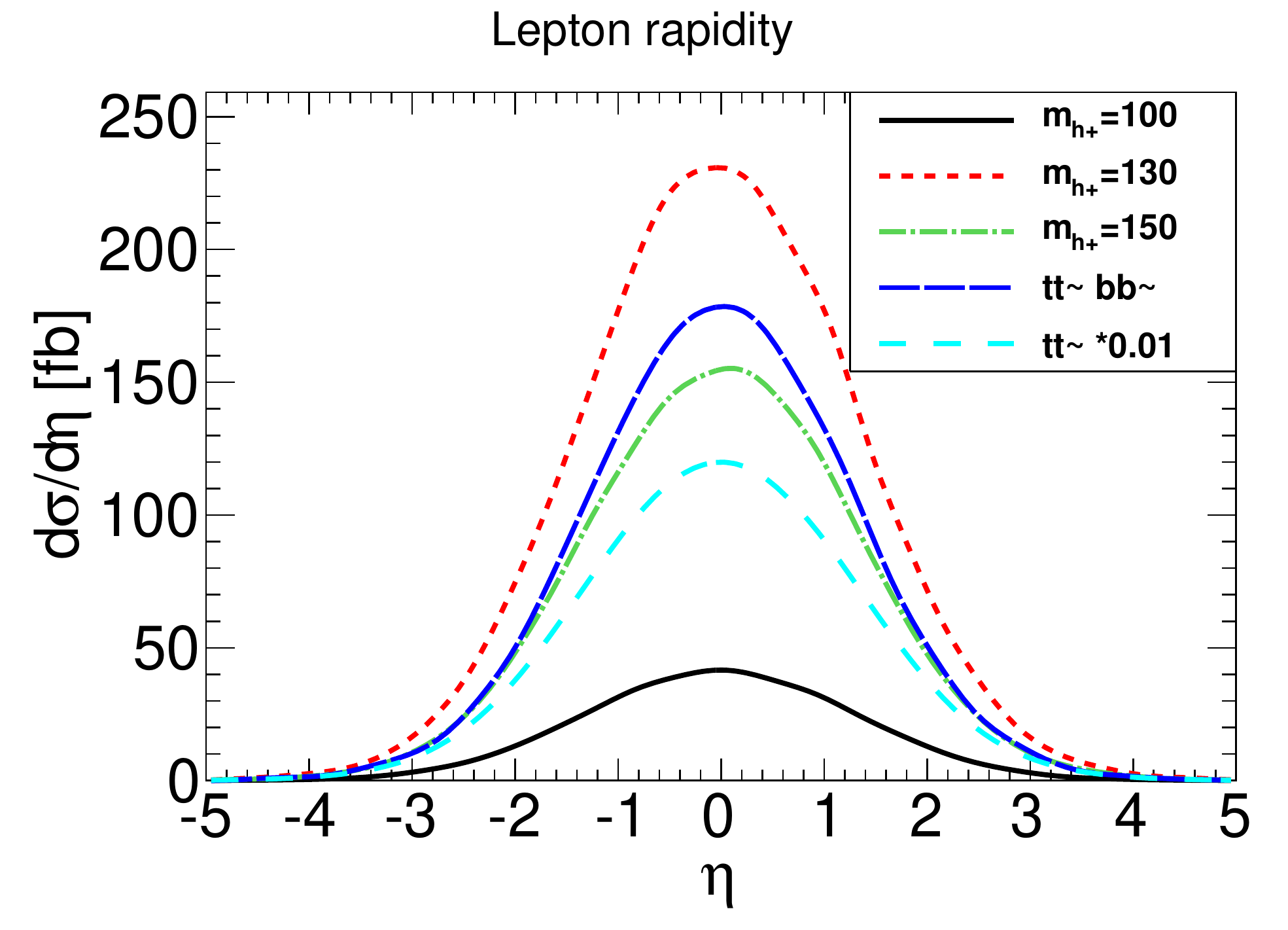}
\includegraphics[width=8cm]{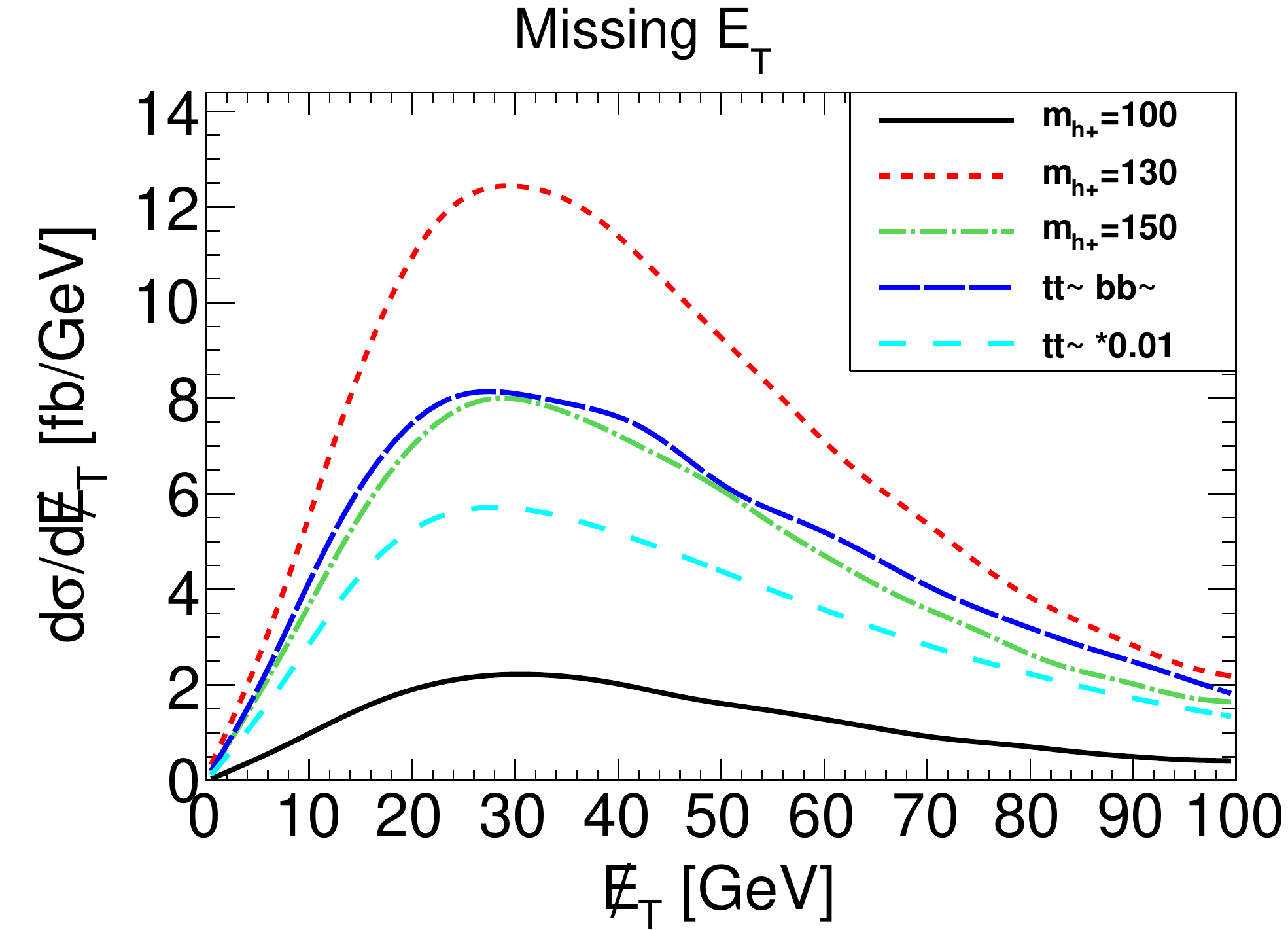}
\caption{Transverse momentum and pseudo-rapidity of the charged lepton from the $W$ decay and missing transverse energy (MET) before applying cuts.}
    \label{fig:MCTlepton}
\end{figure}

To reconstruct the leptonic $W$, we first need to identify the charged lepton ($e$ or $\mu$) associated to the hard process. 
The transverse momentum ($p_\perp$) and pseudo-rapidity ($\eta$)-distributions are shown in Fig.~\ref{fig:MCTlepton}.
After applying cuts on the lepton kinematics ($p_\perp$ $>$ 20 GeV, $|\eta|$ $<$ 2.5), we require the summed $p_\perp$ of the surrounding particles in an $(\eta,\phi)$-cone of size $\Delta R =\sqrt{\Delta \eta^2 + \Delta \phi^2} = 0.3$ around the lepton to be less than 10 GeV to call it isolated~\footnote{The numerical values for the $p_\perp$ cut and the cone size have been optimized by observation of the changes in efficiency and purity when varying the cuts}. On the whole event then, we require to have precisely one isolated lepton in the final state.

The next step in reconstructing the leptonic $W$ is to identify 
the missing energy (MET)/missing $\vec{p}_\perp$ of the event with the transverse momentum of the neutrino.
Assuming the $W$ to be on mass-shell and using a mass-less neutrino then leaves two possible solutions for the longitudinal momentum of the neutrino  $p_{\nu z}=A\pm\sqrt{D}$ with
\begin{eqnarray}
 A & = & \frac{ m_W^2 p_{lz} + 2 p_{lz} (\vec{p}_{\nu \perp}\vec{p}_{l \perp}) }{2 m_{\perp l}^2} \, , \\
 D & = & \frac{ E_l^2 }{ m_{\perp l}^4 } ( \frac{m_W^4}{4} - ({\vec{p}_{\nu \perp}})^2({\vec{p}_{l \perp}})^2 + (\vec{p}_{\nu \perp}\vec{p}_{l \perp})^2 + (\vec{p}_{\nu \perp}\vec{p}_{l \perp})m_W^2 ) \, .
\end{eqnarray}

In order to have a more accurate reconstruction of  $p_{\nu z} $ we have applied a cut $E\!\!\!\!/_\perp > 30$ GeV. In addition, 
different selection criteria for the choice of the sign have been examined,  e.g.\ the invariant mass of the reconstructed $h^\pm$. Among these, the most viable one turned out to be a simple selection of the smaller $\left| p_{\nu z} \right|$, which is correct in roughly three quarters of the signal events.

\subsection{Jet reconstruction and tagging}

For the reconstruction of the hadronic part of the event, we consider different jet clustering schemes (anti-kT~\cite{Cacciari:2008gp}, Cambridge/Aachen~\cite{Dokshitzer:1997in}, kT~\cite{Catani:1993hr,Ellis:1993tq}) as well as cone-sizes. All particles, except neutrinos and the isolated lepton, in the  rapidity range $|\eta|<4.9$ are fed into FastJet~\cite{Cacciari:2005hq,Cacciari:2011ma}. The resulting clustered jets are required to have $p_\perp > 20$ GeV. Afterwards, a simplified b-tagging is simulated for all jets in the region $|\eta|<2.5$ by comparing the $(\eta,\phi)$ of the jets to the $b$-quarks of the hard process. All jets within $\Delta R=0.4$ are then classified as $b$-jets.

\begin{figure}%[t]
\includegraphics[width=8cm]{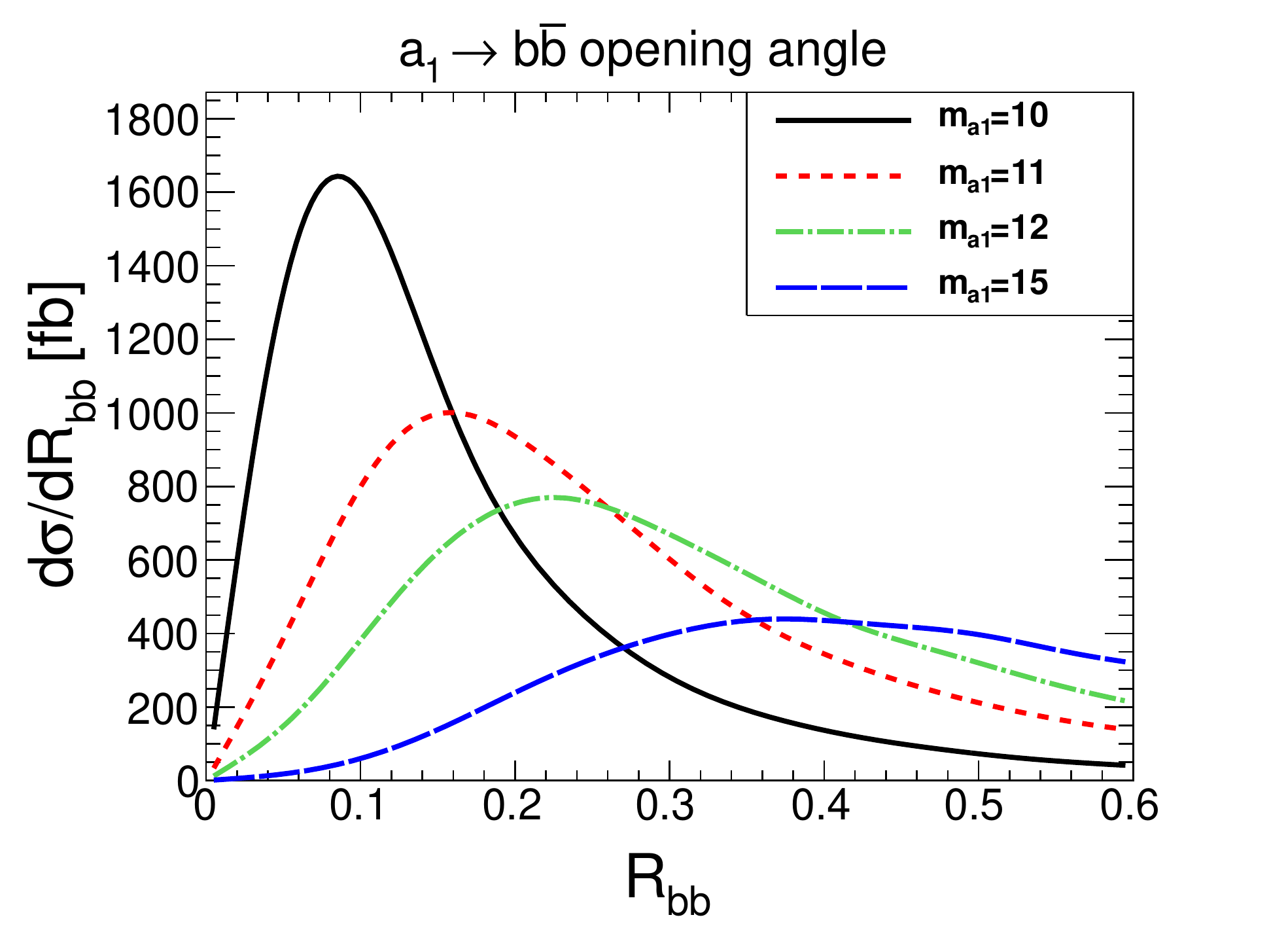}
\caption{The distance between the  two $b$-quarks coming from the $a_1$-decay on parton level for the signal with $m_{h^\pm}=130$ GeV.}
    \label{fig:rdist_mct}
\end{figure}

The jet-algorithms requires to specify the distance measure $R$ used when calculating the jet measure.
Since we want to cluster the $b\bar{b}$ pair from the $a_1$ into one jet, it is crucial that the distance between the two subjets is not too large. Fig.~\ref{fig:rdist_mct} shows the distance measure between the two $b$-quarks on parton level for different $a_1$ masses. As can be seen from the figure, for reasonable clustering cone sizes (around 0.4 to 0.6) and with $m_{a_1}$ in the region of interest for this analysis, the two $b$-quarks  will  most likely be clustered together as a single jet. 

\begin{figure}%[t]
\includegraphics[width=8cm]{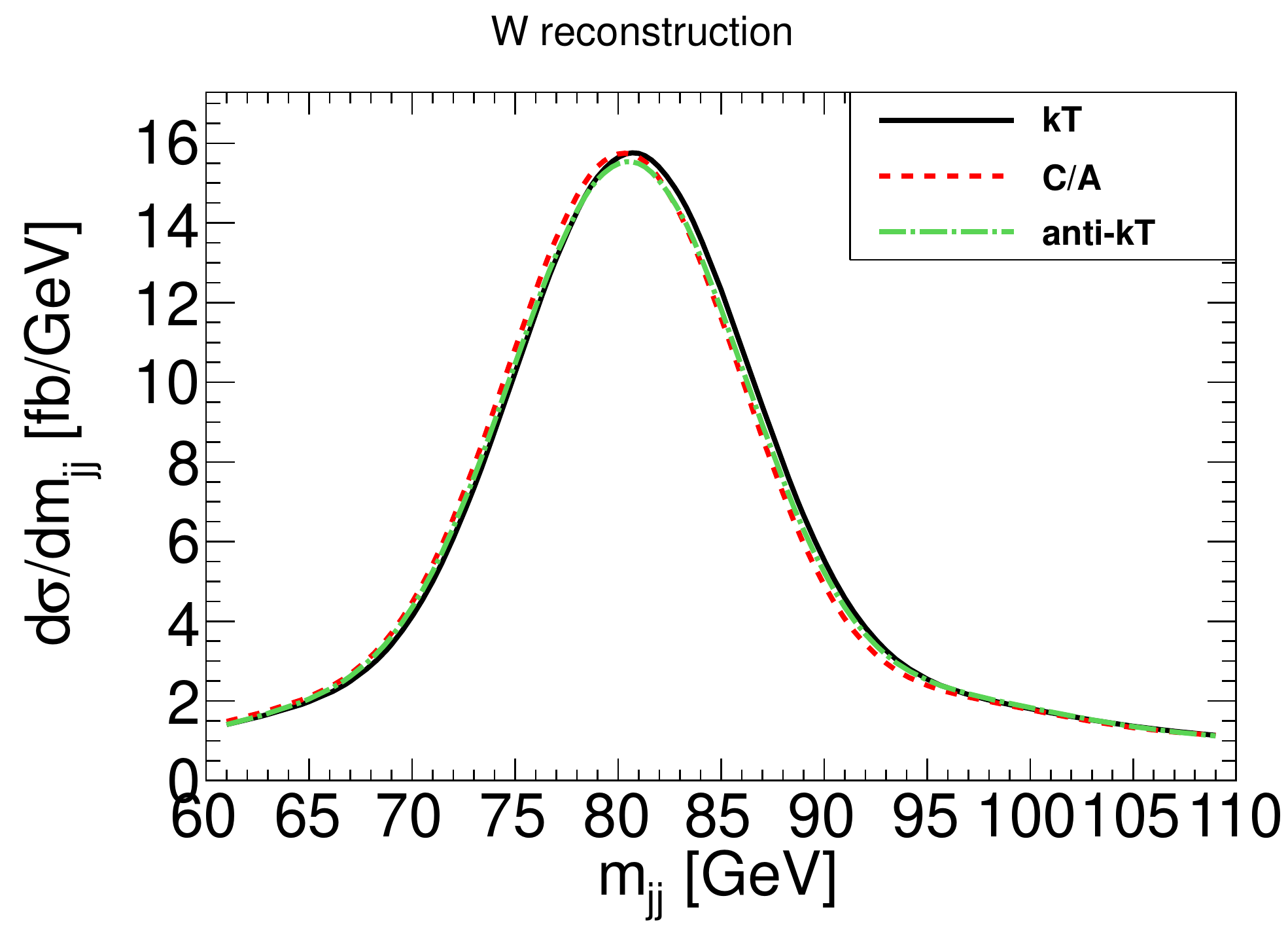}
\includegraphics[width=8cm]{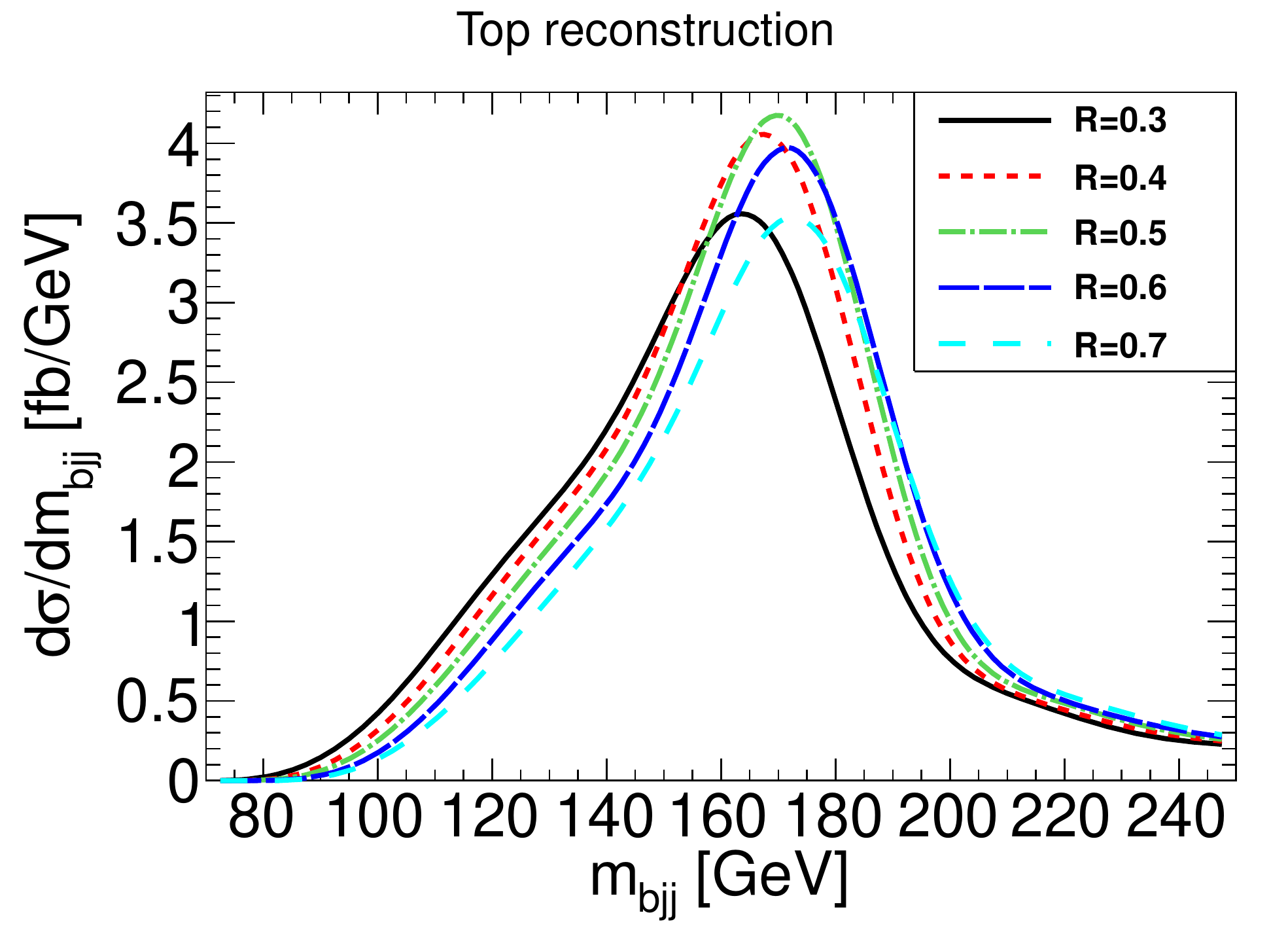}
\caption{Mass reconstruction of the hadronically decaying $W$ (left) and $t$-quark (right) for the signal with $m_{h^\pm}=130$ GeV when $R=0.5$ (left) and using the anti-kT algorithm (right).}
    \label{fig:recon}
\end{figure}

As an ideal reconstruction will now give rise to three $b$-jets, the correct reconstruction of the $h^\pm$ will be enhanced by the identification of the ``wrong'' b-jet which comes from the $\bar{t}$ together with the hadronically decaying $W^-$. The strategy to achieve this here is to first find the pair of untagged jets that is closest to the $W$ mass and then combine this pair 
with the $b$-jet that gives a mass closest to the $t$ mass. This $b$-jet will then be excluded  in the $h^\pm$ reconstruction, reducing the number of $b$-jets which have to be considered (in an event with a so far correct clustering in the desired way) to two. In addition to this, we put cuts on the quality of the reconstruction by requiring the reconstructed masses shown in Fig.~\ref{fig:recon} to be in the regions $m_W \pm 20$ GeV or $m_t \pm 30$ GeV respectively. We have checked that the reconstruction of the $W$ and $t$ masses are
quite independent of the choice of  jet clustering scheme and cone-size as is also shown in Fig.~\ref{fig:recon}.  Only the 0.3 cone size gives a slightly inferior top reconstruction, but then the $a_1$ will also not give rise to a single $b$-jet. If not mentioned differently, we thus use the anti-kT algorithm  with  $R=0.5$ in the following.

\begin{figure}%[t]
\includegraphics[width=8cm]{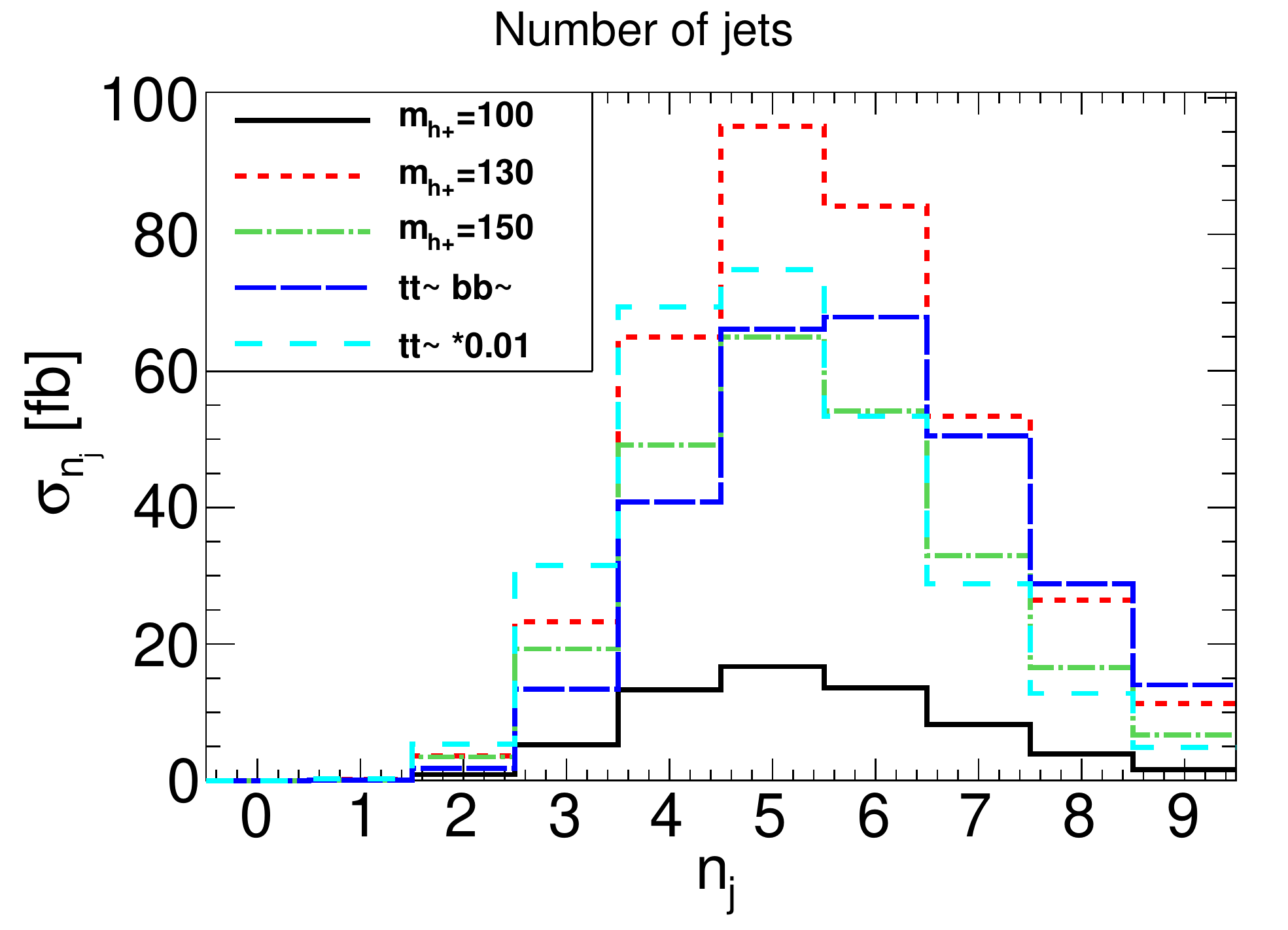}
\includegraphics[width=8cm]{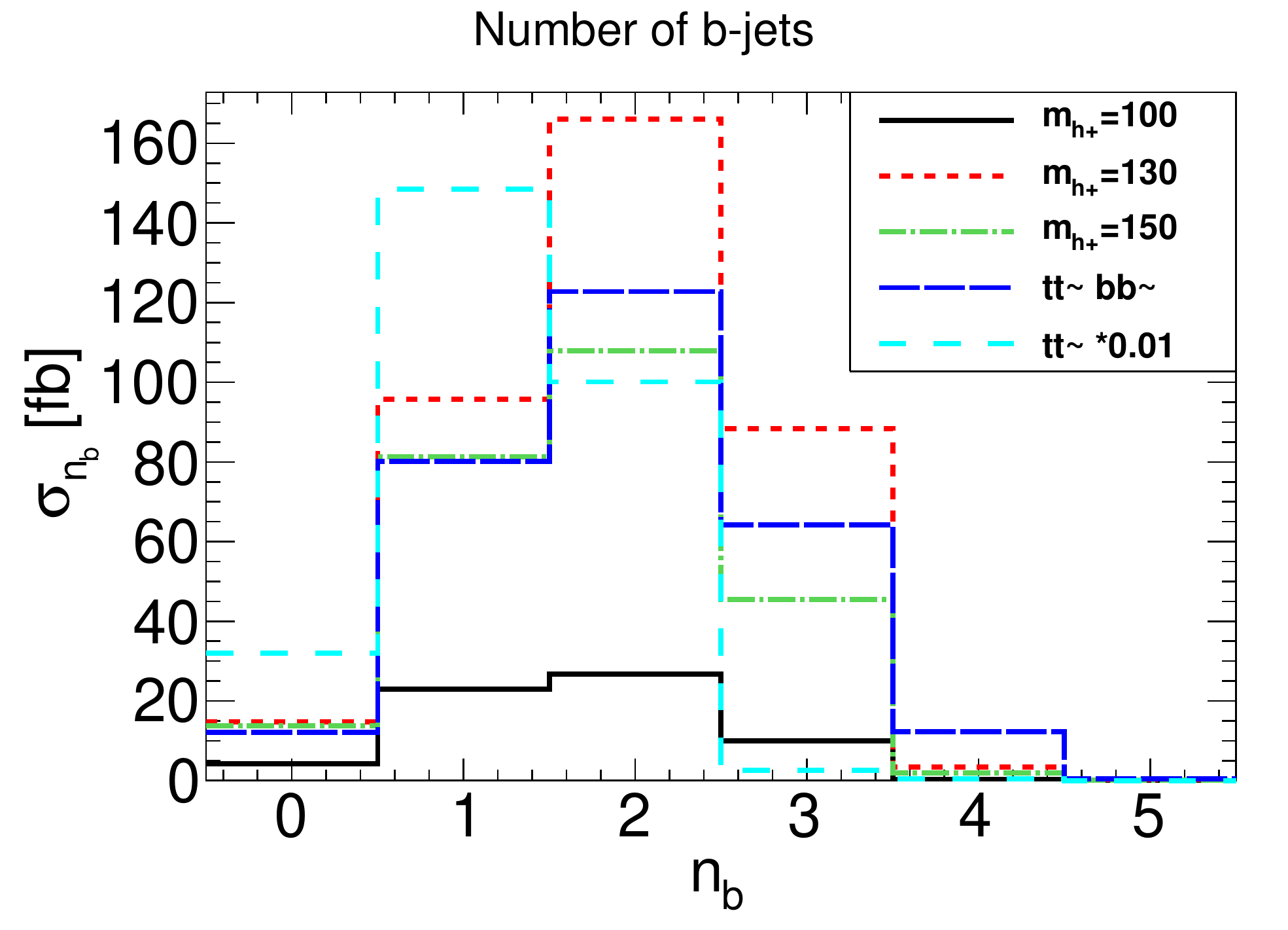}
\caption{Jet (left) and $b$-jet (right) multiplicity for the signal and backgrounds when using the anti-kT algorithm  with  $R=0.5$.}
    \label{fig:njets}
\end{figure}

Using the jet-reconstruction outlined above we obtain the jet and $b$-jet multiplicity respectively shown in Fig.~\ref{fig:njets}. As is clear from the figure, the number of jets peaks around the expected five and the number of $b$-jets is a typically smaller than ideal which is due to the limited $b$-tagging region enforced. Also note the large decrease in $b$-jet multiplicity for the $t\bar{t}$ sample from 2 to 3. Here the 3 $b$-jet sample arises from gluon splitting into $b\bar{b}$ and we do not take it into account below since that would amount to double counting with the $t\bar{t}b\bar{b}$ background. 

\begin{figure}%[t]
\includegraphics[width=8cm]{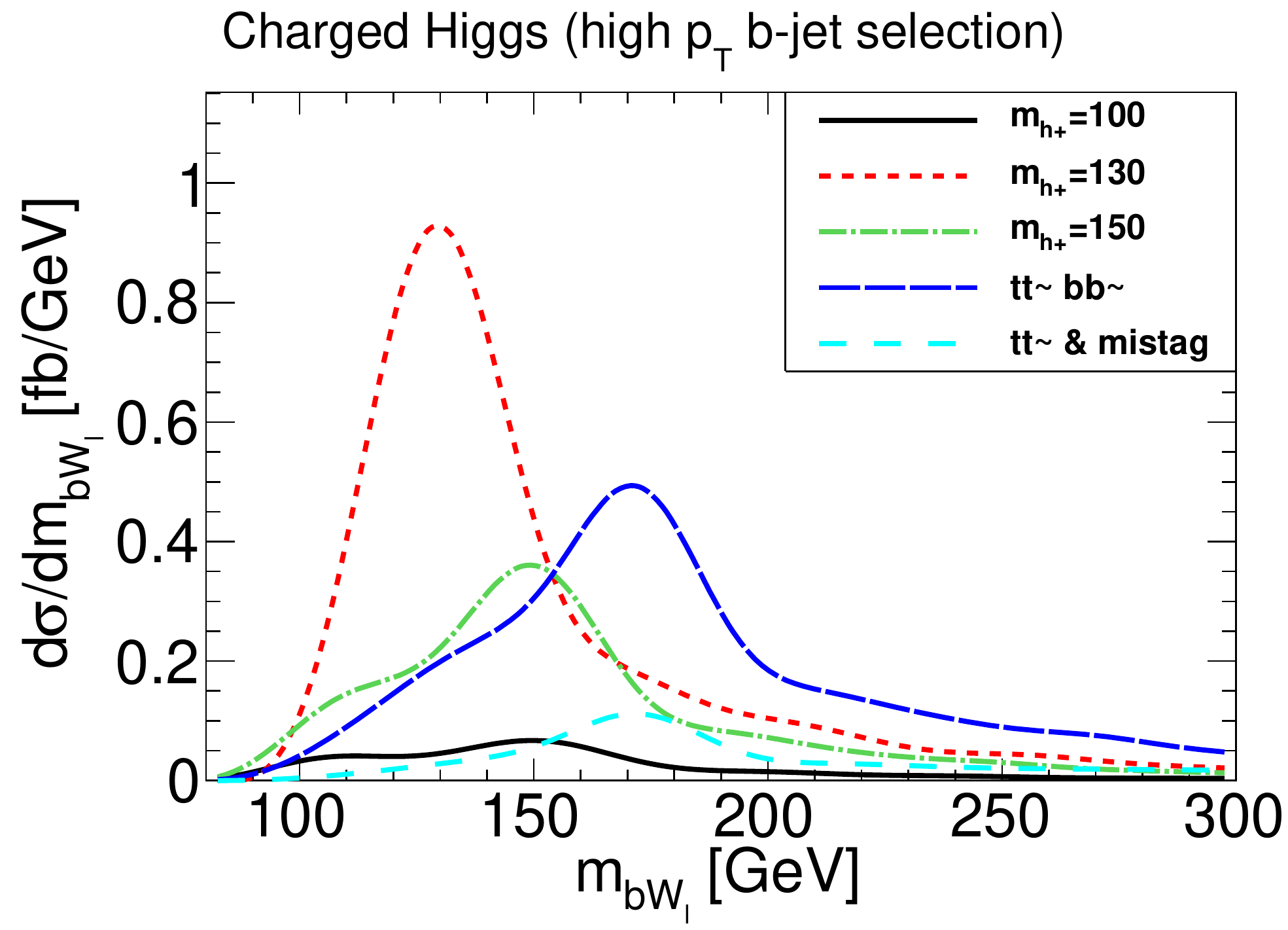}
\includegraphics[width=8cm]{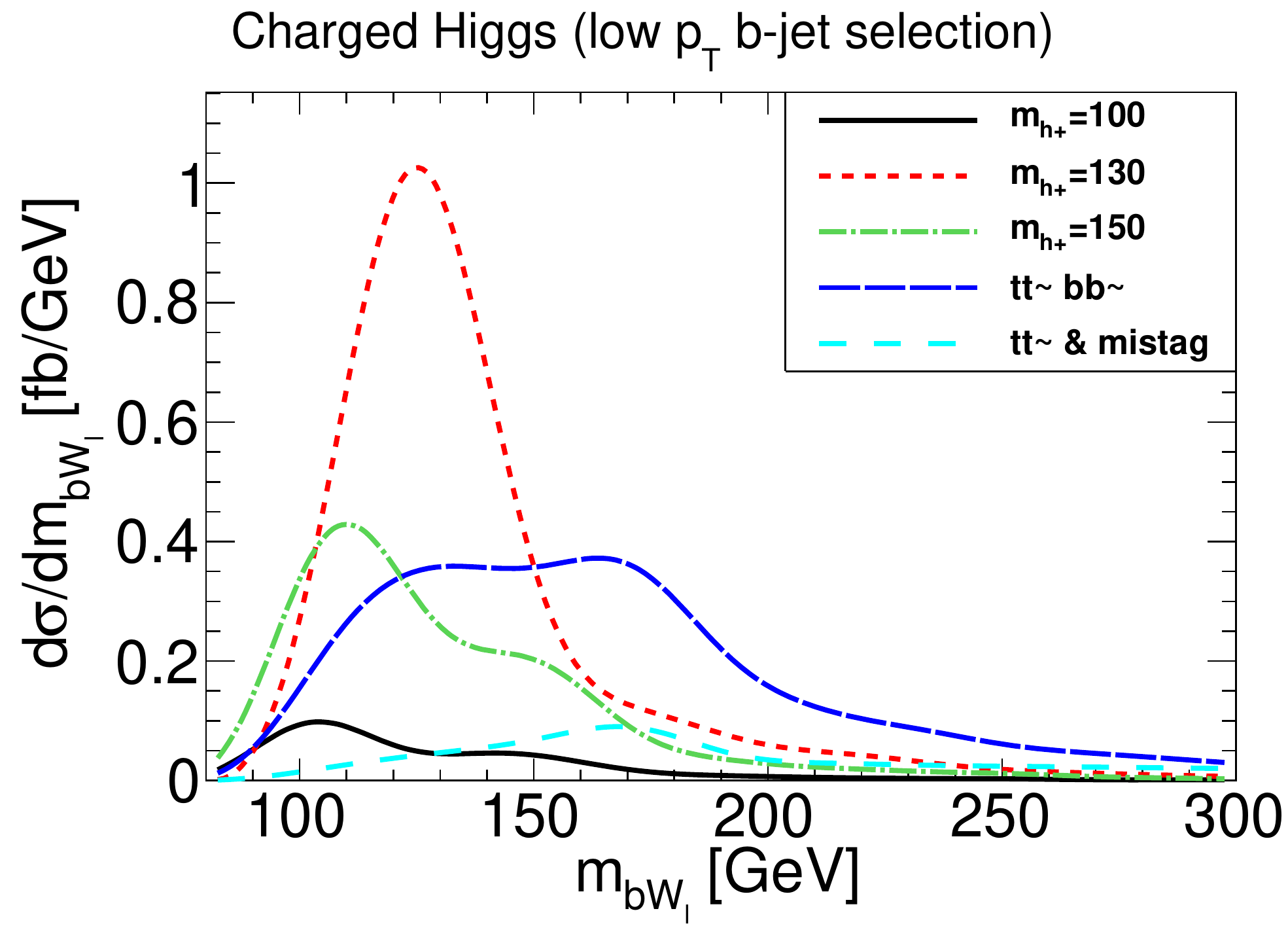}
\caption{Reconstructed charged Higgs masses, using either the higher-$p_\perp$ (left) or the lower-$p_\perp$ (right) $b$-jet combined with the leptonically decaying $W$.}
    \label{fig:signal}
\end{figure}

Given the large background from  $t\bar{t}$ for the two $b$-jet sample we   resort to requiring at least three $b$-jets. Assuming that one of the $b$-jets has been identified as coming from the $t$-quark this leaves two $b$-jets that may originate from the $h^\pm$.  
Due to the strong dependence of the hardness of the $b$-jet from the $t \to h^+ b$ decay on the $h^\pm$ mass, we consider both of these remaining solutions in general as possible candidates for the $h^\pm$ reconstruction. The resulting distributions when combining with the leptonically decaying $W$ are shown in Fig.~\ref{fig:signal}. In the $h^\pm$ reconstruction, we thus require  3 b-jets for the signals and the $t\bar{t}b\bar{b}$ background, while we require 2 $b$-jets in the $t\bar{t}$ sample. For the latter we then assume that any of the non $b$-jets inside the b-tagger region ($|\eta| < 2.5 $) can be 
mis-tagged with a probability of 0.01 per jet~\cite{CMS-PAS-BTV-11-004,ATLAS-CONF-2012-040}.

\subsection{Signal significance and reach}

To estimate the signal reach as well as its significance, we choose a common window of 90 to 160 GeV to integrate the signal as well as the backgrounds. We correct for $b$-tagging efficiencies which we assume to be 0.6 per $b$-jet. The resulting cross-sections are shown in Table~\ref{tab:xsec}.

\begin{table}[htbp]
\begin{center}
\begin{tabular}{|l|c|c|c|c|c|c|}
\cline{2-7}
\multicolumn{1}{c|}{} & \multicolumn{6}{|c|}{$\sigma_{peak}$ [fb]} \\
\cline{1-7}
\multicolumn{1}{|c|}{signal/background}  & 100 GeV & 130 GeV & 150 GeV &
 {\hspace{10pt}$t\bar{t}b\bar{b}$ \hspace{10pt}} &  {\hspace{10pt}$t\bar{t}$ \hspace{10pt}}
& \multicolumn{1}{c|}{$\Sigma$ BG} \\ \hline
High $p_\perp$ channel & 0.78 & 7.9 & 3.7 & 3.0 & 0.9 &
3.9 \\ \hline
Low $p_\perp$ channel & 0.97 & 9.2 & 4.7 & 4.8 & 1.2 &
6.0  \\ \hline
\end{tabular}
\end{center}
\caption{Integrated signal cross-sections (cf figure \protect\ref{fig:signal}) for
the three different charged Higgs masses as well as the two backgrounds.
The integration region is 90 to 160 GeV}
\label{tab:xsec}
\end{table}

S/$\sqrt{B}$ ratios for the three different mass cases are then calculated for an integrated luminosity of 20 fb$^{-1}$ and summarized in Table~\ref{tab:sbratio}. The table also shows the branching ratios at the simulated parameter points, as well as the extrapolations to  branching ratios necessary for a 5$\sigma$ discovery. From the table it is clear that for 
 $m_{h^\pm}\sim 130$ GeV, the discovery reach is maximal. For smaller or larger $m_{h^\pm} $, the limitations in phase space will reduce the branching ratio for $ h^\pm \to a_1 W$ and  $t \to h^\pm b$ respectively. In the former case this means that the standard decay channel $  h^\pm \to \tau \nu_\tau$ will also be significant.

\begin{table}[htbp]
\begin{center}
\begin{tabular}{|l|c|c|c|c|c|c|}
\cline{1-7}
\multicolumn{1}{|l|}{$m_{h^\pm}$} & \multicolumn{2}{|c|}{100 GeV} &
\multicolumn{2}{|c|}{130 GeV} & \multicolumn{2}{|c|}{150 GeV} \\
\cline{1-7}
\multicolumn{1}{|l|}{$b$-jet selection} & high $p_\perp$ & low $p_\perp$ & high $p_\perp$ &
low $p_\perp$ & high $p_\perp$ & low $p_\perp$ \\ \hline
$S/\sqrt{B}$ & 1.77 & 1.76 & 18.0 & 16.7 & 8.5 & 8.5
\\ \hline
BR($t \to b h^+ \to b a_1 W \to b b \bar{b} W $) & \multicolumn{2}{c|}{0.0051} & \multicolumn{2}{c|}{0.022} &
\multicolumn{2}{c|}{0.015} \\ \hline
BR$_{crit}$ & 0.014 & 0.014 & 0.0060 & 0.0065 & 0.0085 & 0.0085 \\
\hline
\end{tabular}
\end{center}
\caption{The S/$\sqrt{B}$ ratios obtained for an
integrated luminosity of 20 fb$^{-1}$ for the different signals considered together with the branching ratios for the
total decay chain $t \to b h^+ \to b a_1 W \to b b \bar{b} W $ of the respective parameter points  and  a
 linear extrapolation to the critical branching ratio necessary for a 5$\sigma$ discovery.}
\label{tab:sbratio}
\end{table}

Before ending this section we note that similarly to the standard decay modes of the charged Higgs boson it should be possible to use the spin-correlations between the decay products of the two top quarks as a way of enhancing the signal \cite{Eriksson:2007fx,Bernreuther:2008ju,Godbole:2011vw}.

\section{Compatibility with possible Higgs signal}
\label{sec:comp}

Recently the ATLAS and CMS experiments announced the combined results of the SM Higgs ($\phi$) searches using $\approx 5$  fb$^{-1}$ of data from 2011 \cite{ATLAS:2012ae,Chatrchyan:2012tx}.
In short they can be summarized as follows: the CMS experiment has ruled out the region  $m_{\phi}  \in [129,600]$ GeV  at the 95~\% confidence level as expected whereas in the region [118,129] GeV they have not been able to make any exclusion at all even though they expected to be able to do so at the 95~\% confidence level, the ATLAS experiment has similarly   ruled out the regions 
 $m_{\phi}  \in [112.9,115.5]$ GeV, $m_{\phi}  \in [131,238]$ GeV, and  $m_{\phi}  \in [251,466]$ GeV  at the 95~\% confidence level whereas they had expected to rule out the region $[125,519]$ GeV.

Instead, both experiments have found an excess of events in the regions $m_{\phi} \sim 126$ GeV and $m_{\phi} \sim 124$ GeV for ATLAS and CMS respectively, which when corrected for the so called look-elsewhere-effect and combining the different channels has a statistical significance of about $2\sigma$ for each of the two experiments. Although not statistically significant these results have stirred a lot of excitement in the high-energy physics community \cite{Hall:2011aa,Heinemeyer:2011aa,Arbey:2011ab,Carena:2011aa,Kadastik:2011aa,Ellwanger:2011aa,Cao:2011sn,Cao:2012fz,Vasquez:2012hn,Ellwanger:2012ke}. It will most likely not be possible to draw any final conclusions about whether this is a true signal or not before the end of the run in 2012 which will give another $\sim 20$ fb$^{-1}$ of data.

One of the most important properties of the possible signal at the LHC is that the $\phi \to \gamma \gamma$ channel is similar to what is expected from the SM. Therefore we start by considering the would be signal from the $h_1$ as well as the $h_2$ compared to what is expected in the SM for a Higgs boson ($\phi$) with the same mass. 
Assuming that gluon-gluon fusion dominates the production, which we have verified is always the case in the scenarios we consider, this ratio is given by 
\begin{equation}
R_{gg\gamma\gamma}^{h_i} = 
 \frac{\sigma(gg\to h_i)_{\rm NMSSM}}{\sigma(gg\to \phi)_{\rm SM}}
 \frac{Br(h_i\to\gamma\gamma)_{\rm NMSSM}}{Br(\phi\to\gamma\gamma)_{\rm SM}} 
 \simeq 
 \frac{\Gamma(h_i \to gg)_{\rm NMSSM}}{\Gamma(\phi \to gg )_{\rm SM}}
 \frac{Br(h_i\to\gamma\gamma)_{\rm NMSSM}}{Br(\phi\to\gamma\gamma)_{\rm SM}} 
\end{equation}
where in the second equality, following for example~\cite{Ellwanger:2011aa,Cao:2012fz}, we have made the implicit assumption that the difference in radiative corrections for the production and decay processes are canceled in the ratio.

\begin{figure}%[t]
\includegraphics[width=8cm]{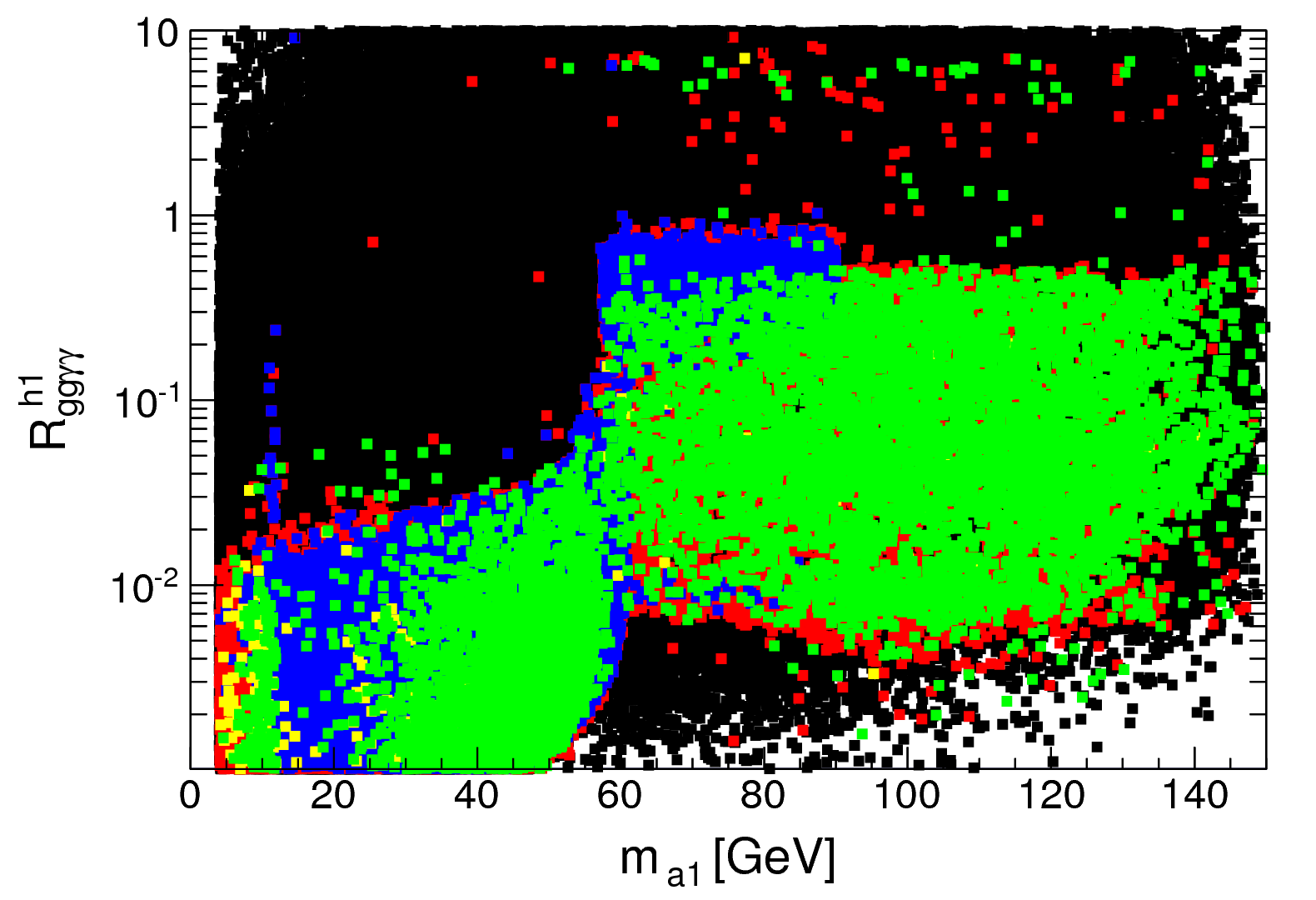}
\includegraphics[width=8cm]{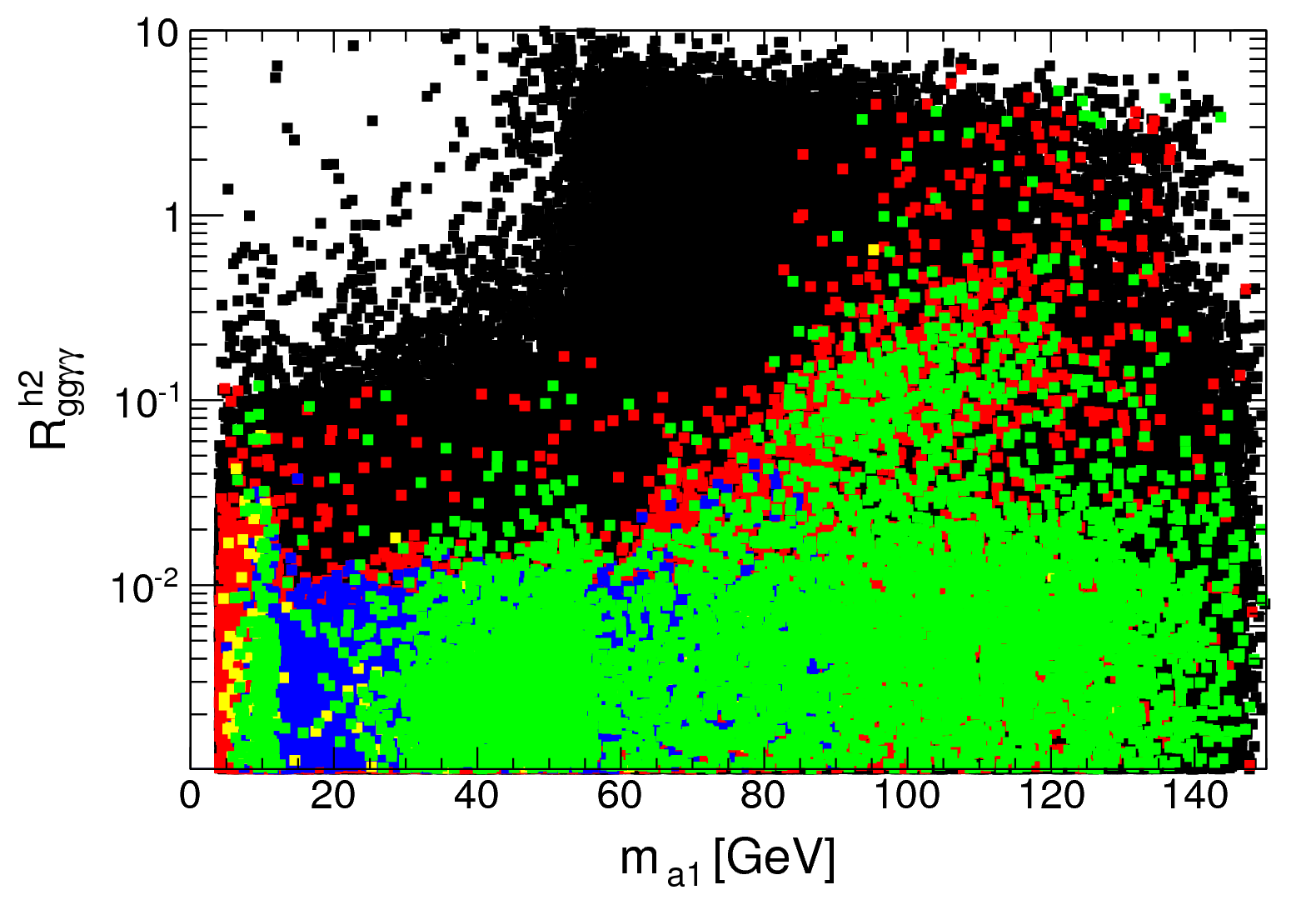}
\includegraphics[width=8cm]{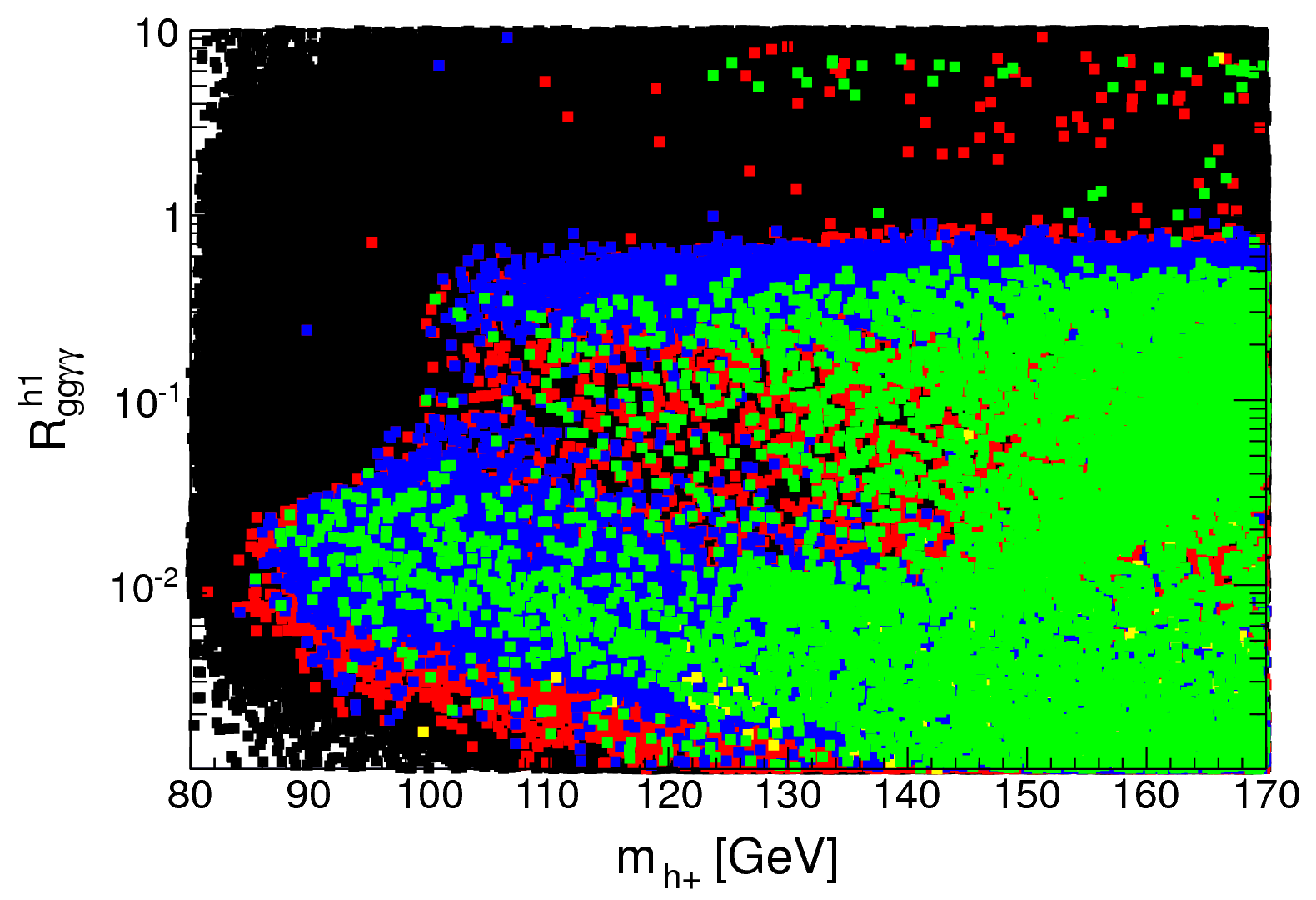}
\includegraphics[width=8cm]{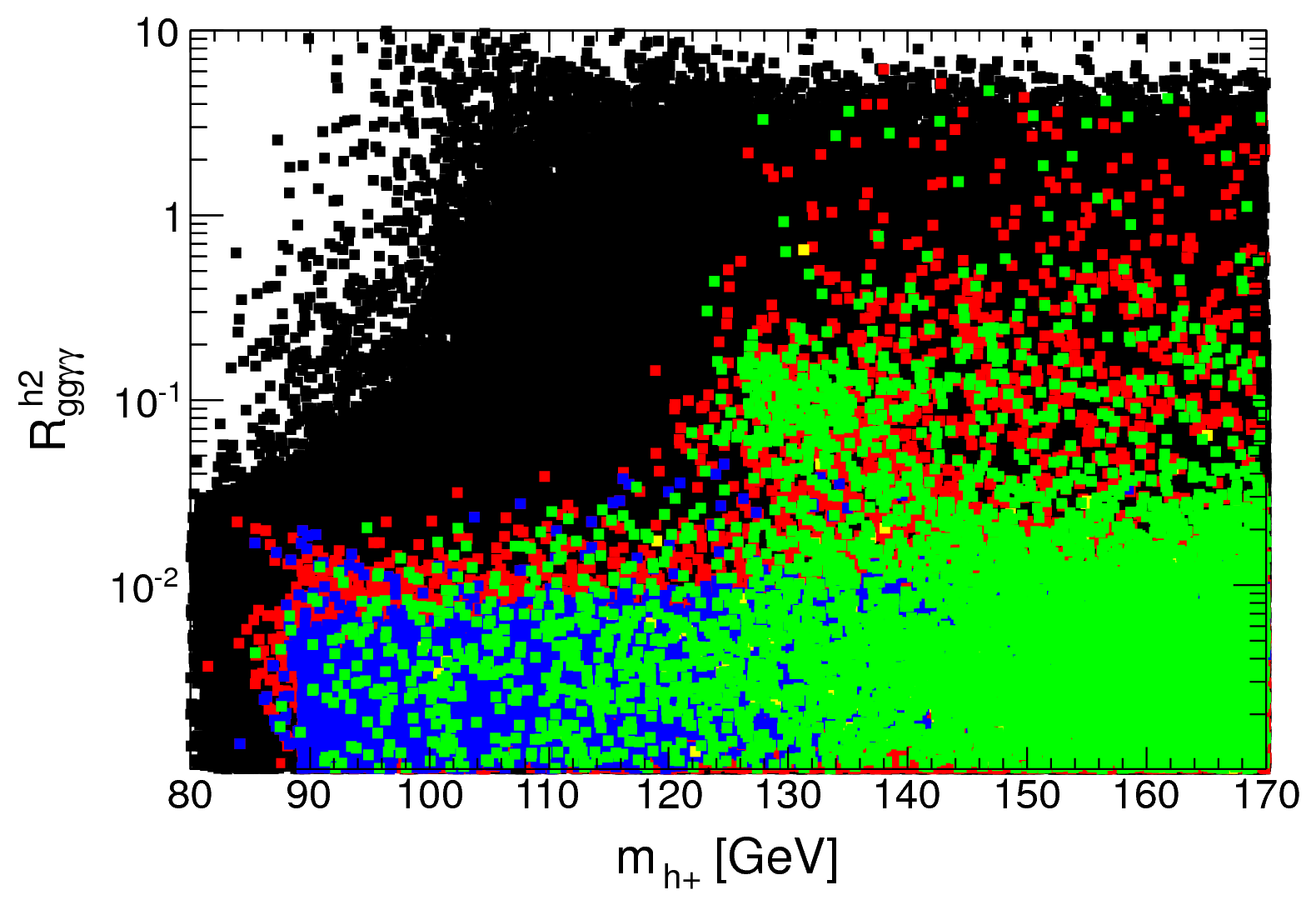}
\includegraphics[width=8cm]{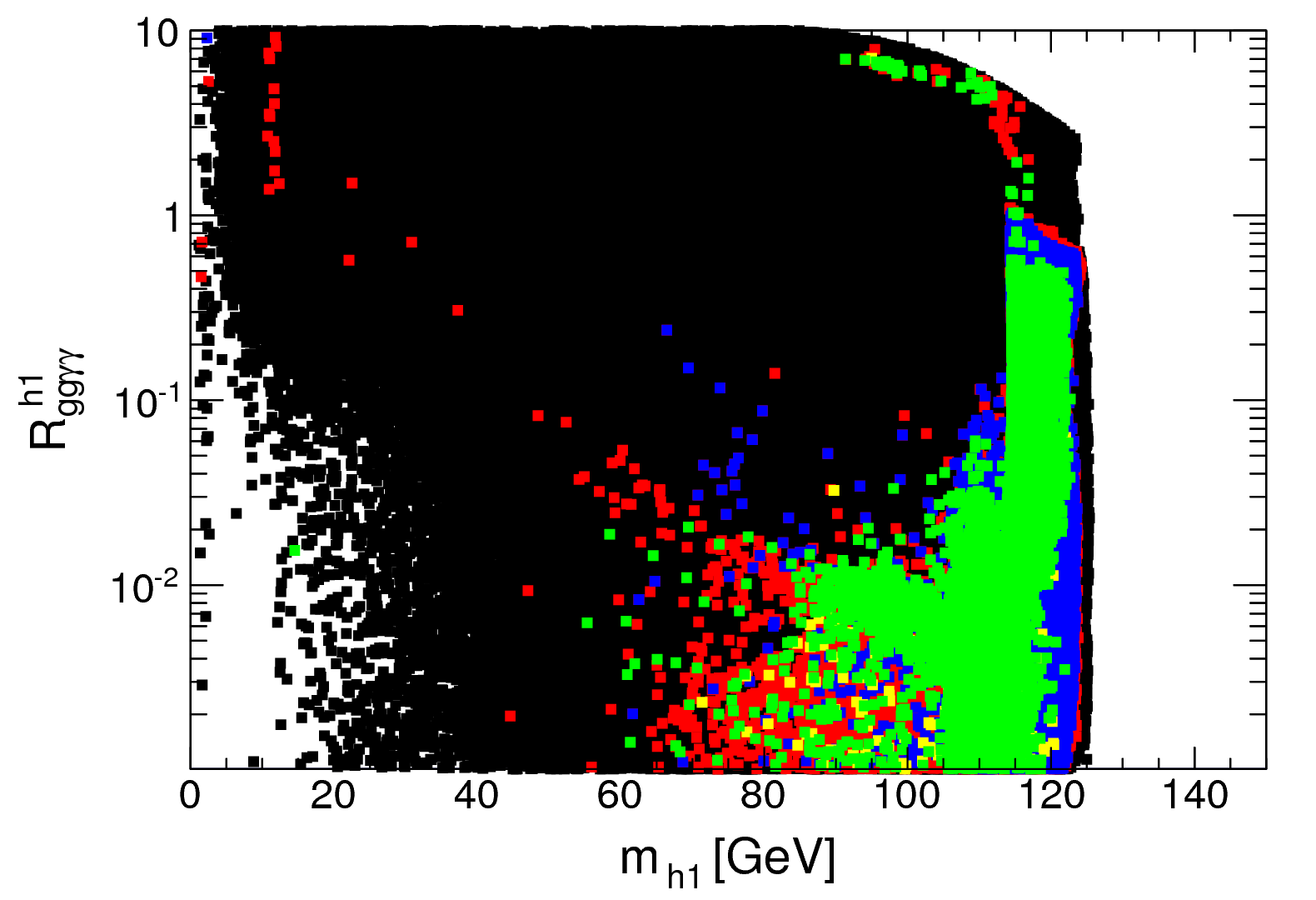}
\includegraphics[width=8cm]{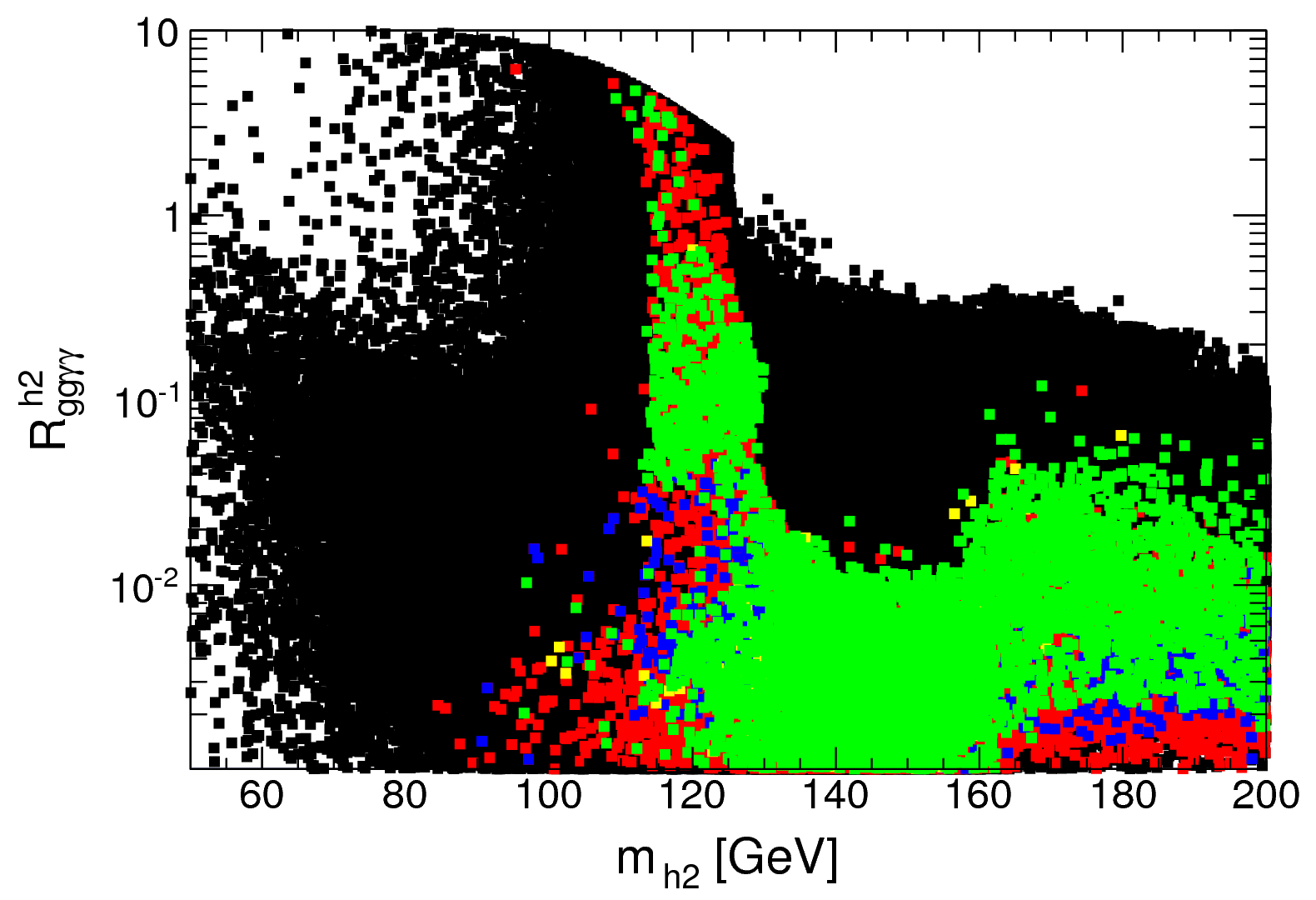}
\caption{The signal for $h_1 \to \gamma \gamma$ (left) and $h_2 \to \gamma \gamma$ (right) relative to the standard model when scanning over the parameters in the scenario under consideration (see text for details) with the various constraints applied. The colour coding is the same as in 
Fig.~\protect\ref{fig:JRscan}}
    \label{fig:JRcomp}
\end{figure}

The results obtained for this quantity when using the same scan as in section \ref{sec:expcon} are displayed in 
Fig.~\ref{fig:JRcomp}. From the figure it is clear that if the possible signal seen at the LHC is the $h_1$ 
then we would have to have $m_{a_1} \gtrsim 60$ GeV. Even so $R_{gg\gamma\gamma}^{h_1}$ does not become larger than $\sim 0.5$ for $m_{h_1}\gtrsim 120$ GeV in our scan and there appears to be an upper bound 
$m_{h_1}\lesssim 122$ GeV. However, the difference in mass compared to the possible observation is so small that it should be considered to be within the theoretical uncertainty~\footnote{For example, including the full one-loop corrections and the two-loop ones from the $t$ and $b$ Yukawa couplings could push the $h_1$ mass bit higher in compliance with the possible signal. Similarly increasing $M_{\rm SUSY}$ would also increase $m_{h_1}$.}. In any case it is clear that it is hardly possible to have a light $a_1$ if it is the $h_1$ that has been seen at the LHC. 

Next we turn to the possibility that it is the $h_2$ that has been seen by the LHC experiments. The results of the scan are also shown in Fig.~\ref{fig:JRcomp}. As can be seen from the figure the results are more promising in this case. There are points with $R_{gg\gamma\gamma}^{h_2} \gtrsim 0.1$ also for small $m_{a_1}$ and intermediate $m_{h^\pm}$ that have $m_{h_2}$ in the region indicated by the LHC experiments.

\begin{figure}%[t]
\includegraphics[width=8cm]{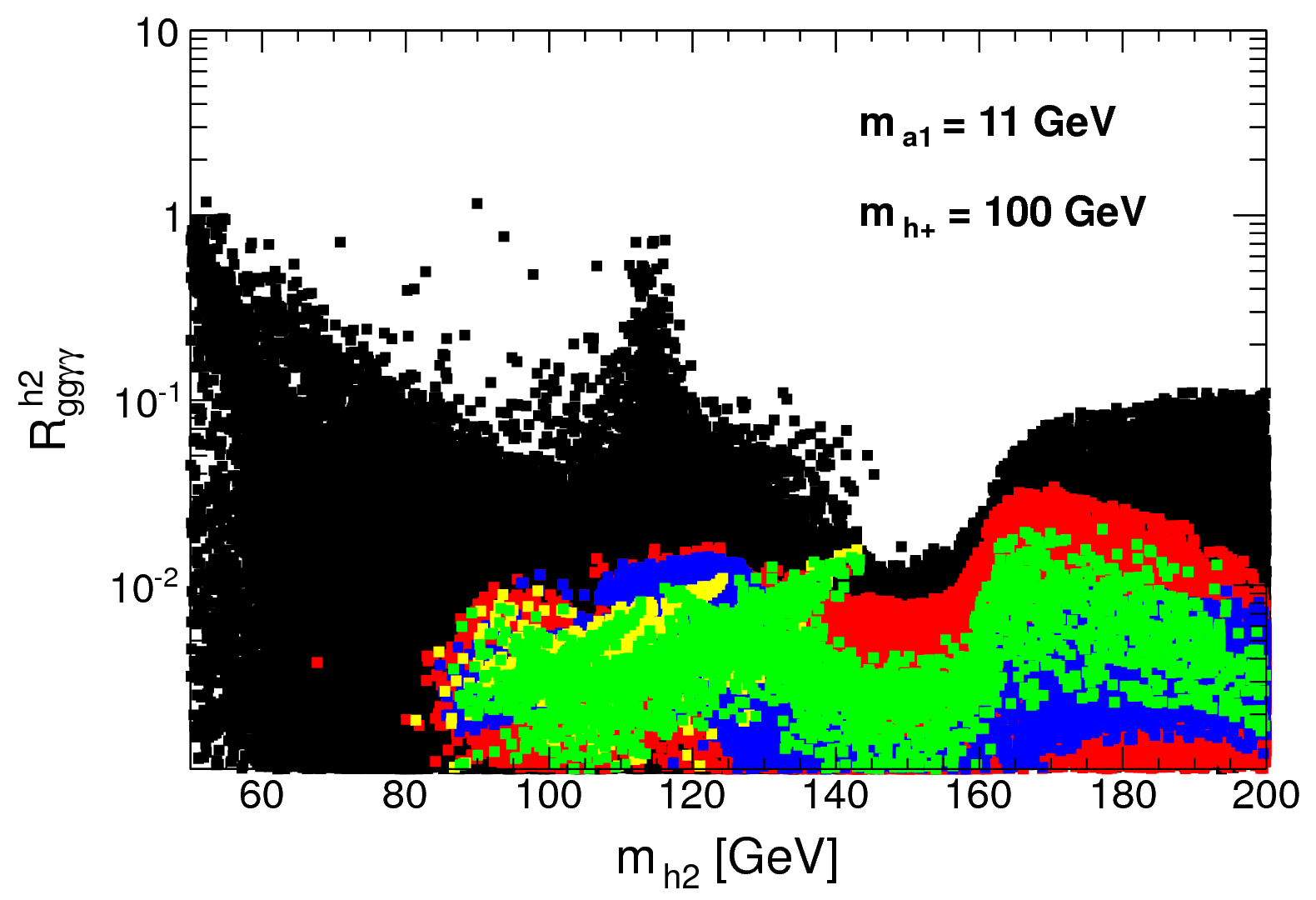}
\includegraphics[width=8cm]{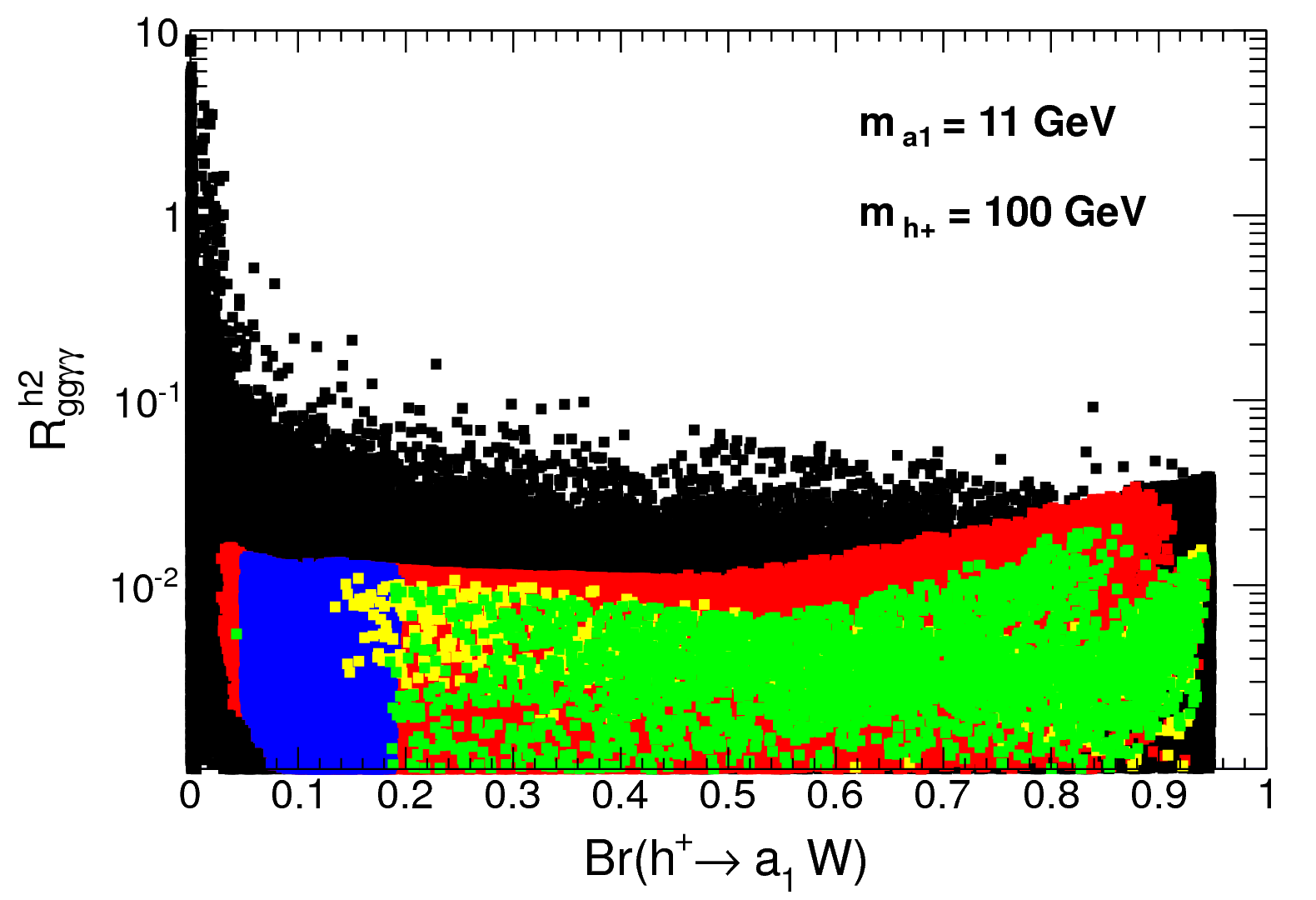}
\includegraphics[width=8cm]{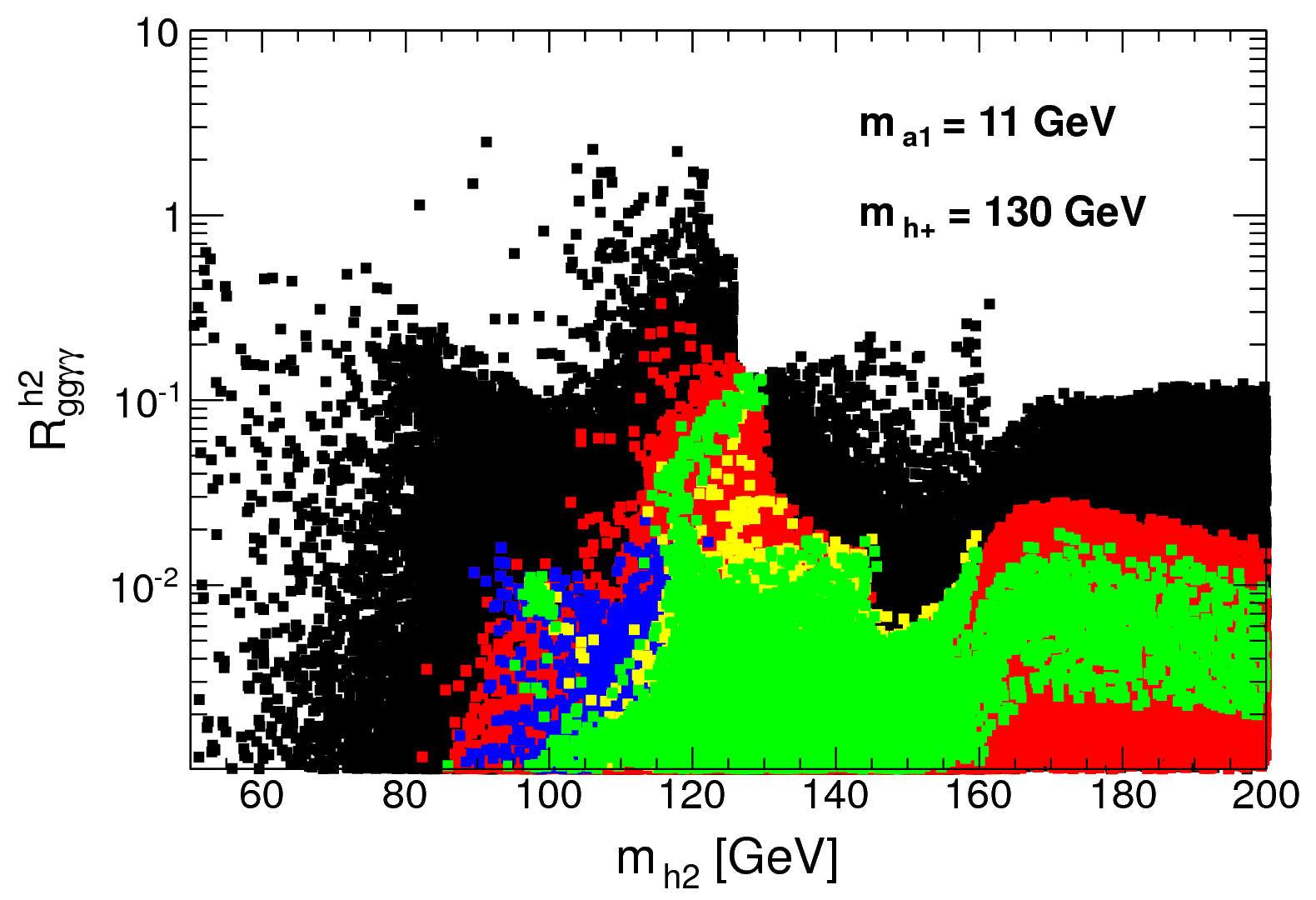}
\includegraphics[width=8cm]{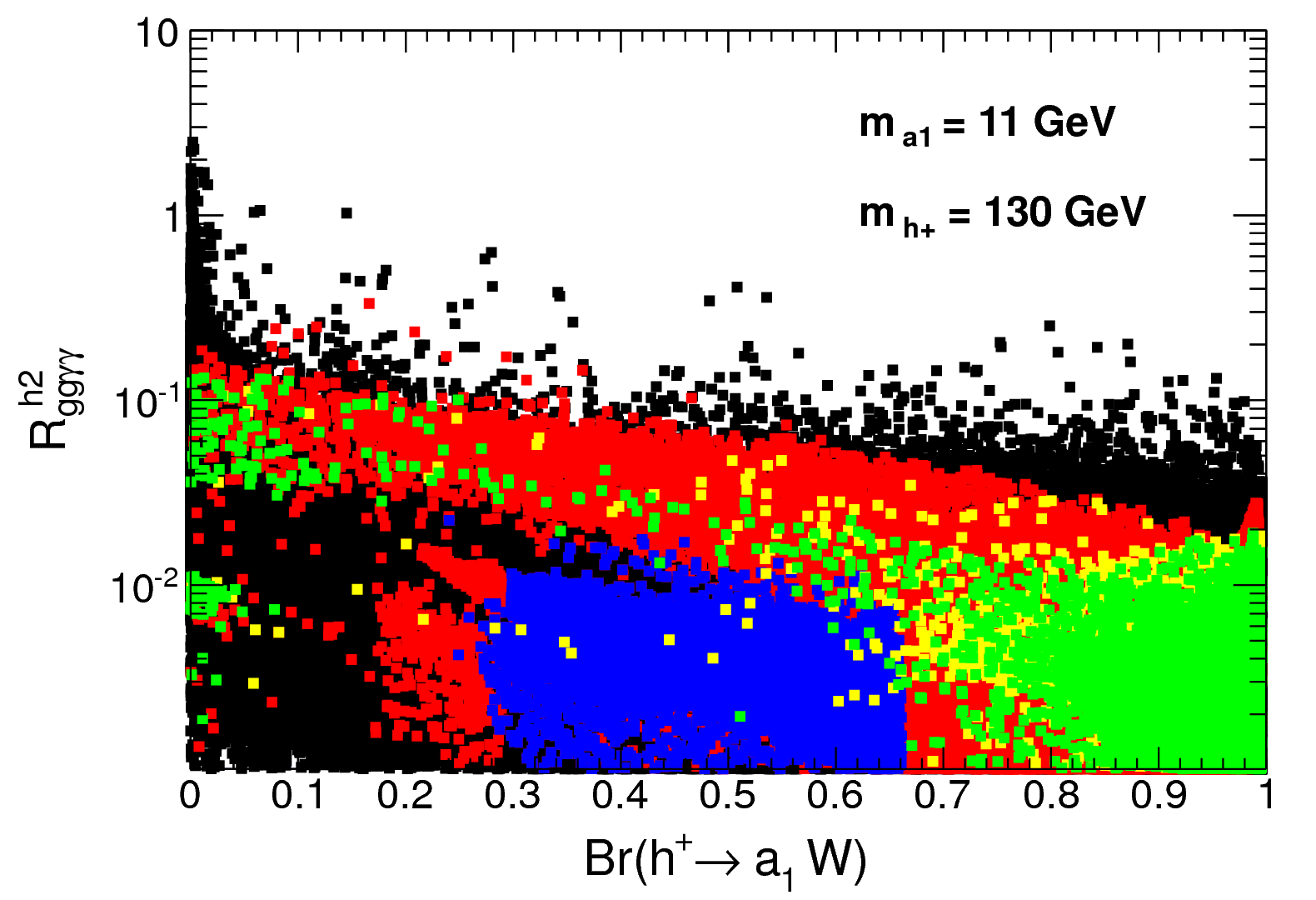}
\includegraphics[width=8cm]{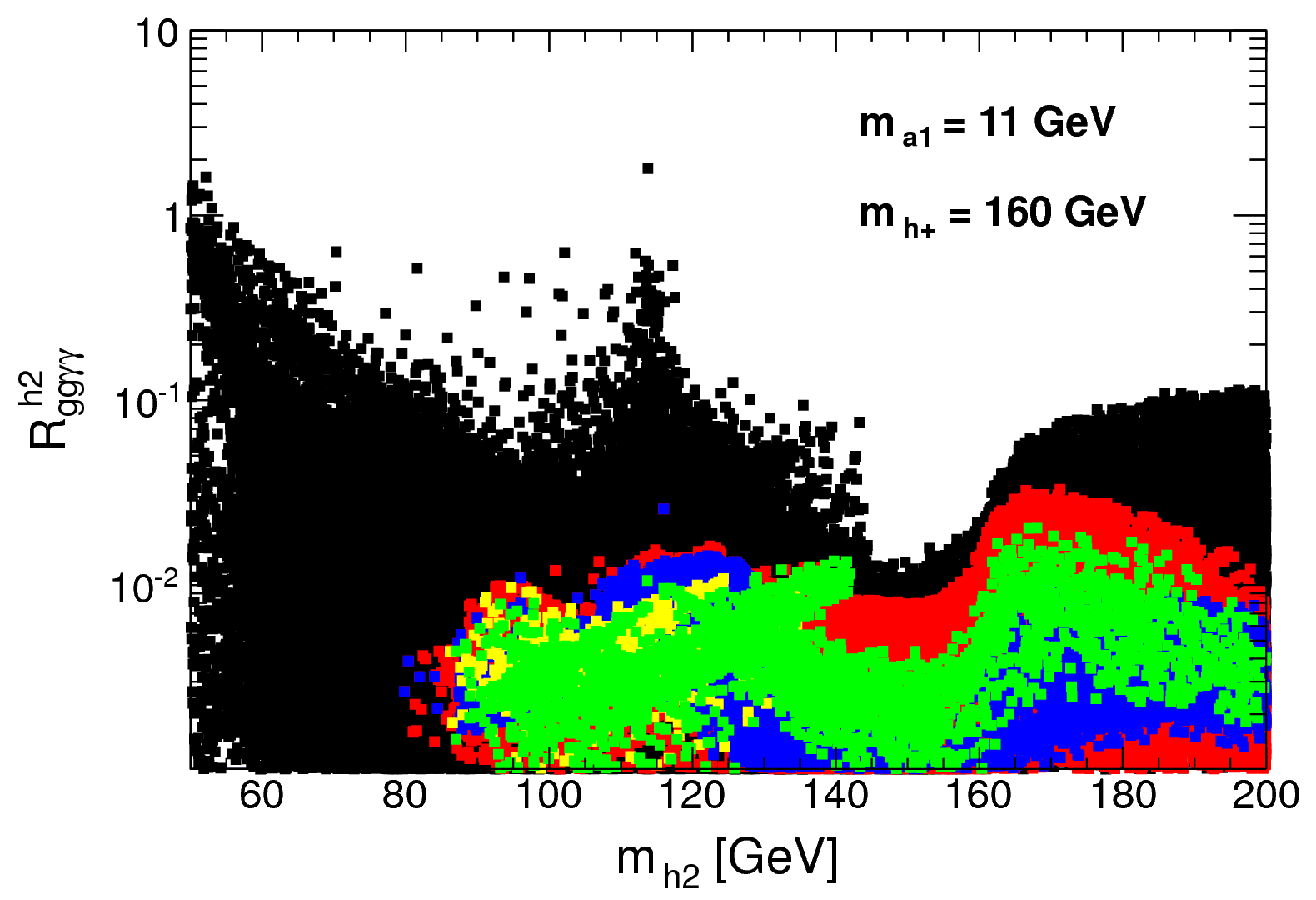}
\includegraphics[width=8cm]{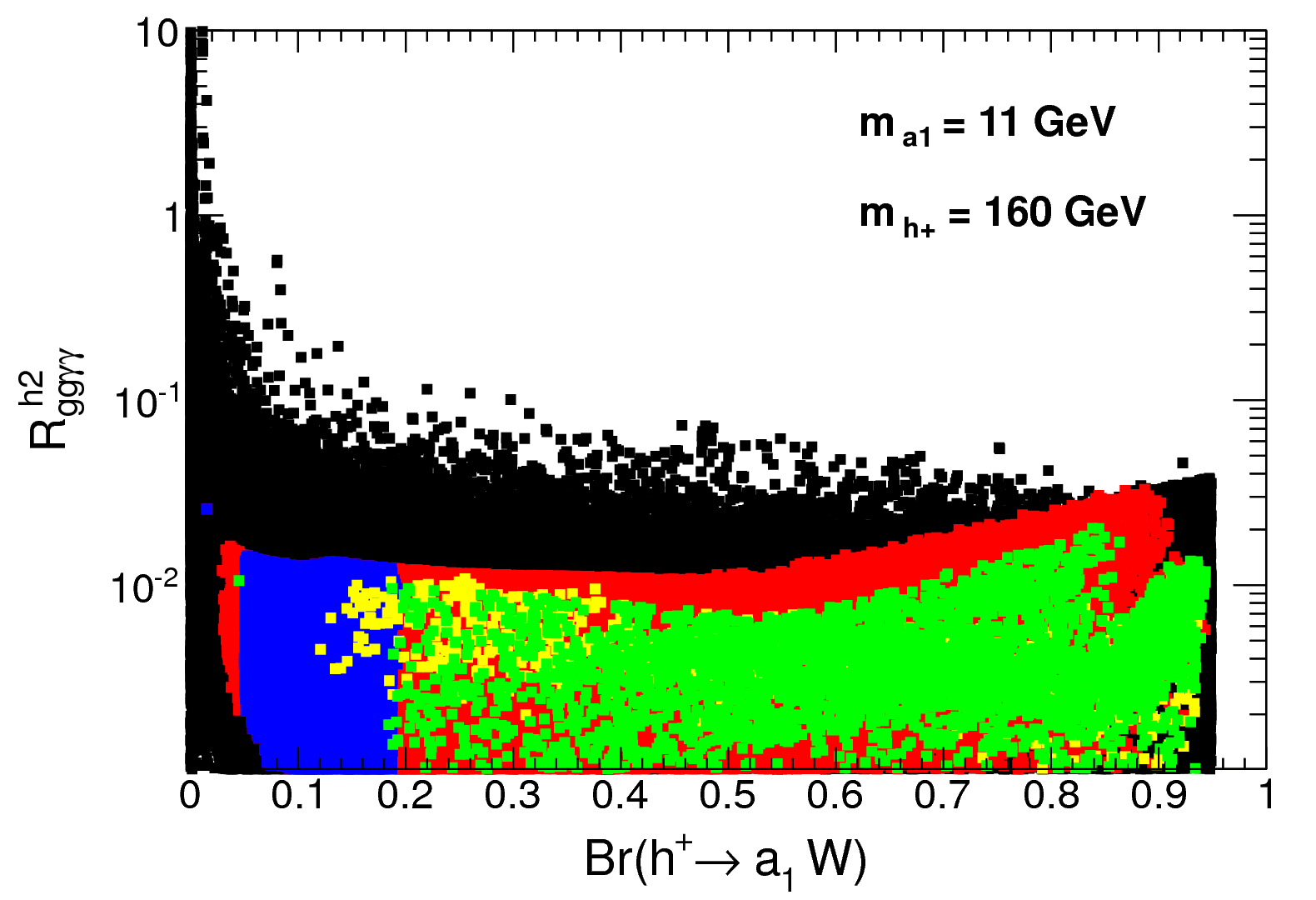}
\caption{The signal for $h_2 \to \gamma \gamma$ relative to the standard model (left) and the branching fraction for $h^\pm \to a_1 W $ (right) in a scan with $m_{a_1}=11$ GeV and $m_{h^\pm}$ fixed at 100 (top), 130 (middle), and 160 (bottom) GeV respectively. The colour coding for the various constraints is the same as in Fig.~\protect\ref{fig:JRscan}}
    \label{fig:JRscan11}
\end{figure}

In order to explore this possibility more closely we show in Fig.~\ref{fig:JRscan11} the results when fixing the $a_1$ mass to $11$ GeV as was done in section \ref{sec:search} for the three cases $m_{h^\pm}=100, 130, 160$ GeV. 
%In addition we limit the scan to small $\lambda$ and $\kappa$ ($\lambda \in [0,0.2]$, $\kappa \in [-0.1,0.1]$) as will be explained below. 
As can be seen from the figure, for the intermediate charged Higgs mass it is possible to reach $R_{gg\gamma\gamma}^{h_2} \sim 0.15$. However it should be noted that in this case, as also shown in the figure, the branching fraction for $h^\pm \to a_1 W $ is quite small meaning that the standard decay channel $h^\pm \to \tau \nu_\tau$ can be used even though with a slightly reduced branching fraction.

The last logical possibility would be that it is the $h_3$ that has been observed by the CERN experiments. However, in the scenarios we consider it turns out that $R_{gg\gamma\gamma}^{h_3}$ is always small.

The trends seen in Fig.~\ref{fig:JRscan11} can be understood on more general grounds from the difficulties of having a light $a_1$ with mass $m_{a_1}< m_{h_i} /2 $ and at the same time have a large $R_{gg\gamma\gamma}^{h_i}$. The first problem is that unless the triple Higgs coupling $g_{h_i a_1 a_1}$ is small the decay  $ h_i \to a_1 a_1$ will become dominant. Looking at the structures of $g_{h_i a_1 a_1}$, which for example can be found in \cite{Ellwanger:2009dp}, this means that both $\lambda$ and $\kappa$ typically have to be small. Secondly the decay $ h_i \to a_1 Z$ will also dominate if  $m_{a_1}< m_{h_i}-m_Z $ unless the corresponding reduced coupling given in table \ref{tab:couplings} is small. In other words we need $(\cos\beta \mathbf{S}_{i1} - \sin\beta \mathbf{S}_{i2})\cos\theta_A$ to be small. There are essentially three ways to achieve this. If $\cos\theta_A$ is small then $a_1$ will be mainly singlet like and decouple. However, then $ h^\pm \to a_1 W$ will also be small. The second possibility is that 
$\mathbf{S}_{i1}$ and $\mathbf{S}_{i2}$ are small, but then the would-be-signal would be mainly singlet-like and not produced in the first place, so $\mathbf{S}_{i3}$ has to be small. Finally then, the combination $(\cos\beta \mathbf{S}_{i1} - \sin\beta \mathbf{S}_{i2})$ could be small but then the complementary combination $(\sin\beta \mathbf{S}_{i1} + \cos\beta \mathbf{S}_{i2})$ would have to be large giving an increased coupling $h_i \to VV$. All in all this means that is difficult to have a  light ${a_1}$ that is not decoupled and still have a large $R_{gg\gamma\gamma}^{h_i}$, although it cannot be completely ruled out.

\section{Summary and conclusions}
\label{sec:sum}

We have considered a well motivated class of supersymmetric extensions of the standard model with non-minimal Higgs sector -- namely the CP-conserving NMSSM. In these types of models the additional Higgs singlet can modify the phenomenology of the Higgs sector in many different ways. In this paper we have specifically addressed the possibility of having a light CP-odd Higgs boson close to, but still above the $b\bar{b}$ threshold. This in turn means that the light charged Higgs boson, with mass $m_{h^\pm}< m_t$, can decay into $a_1 W$ in addition to the standard decays $ {h^\pm} \to \tau \nu_\tau$, thus invalidating the interpretations made of charged Higgs bosons searches assuming that $ Br({h^\pm}\to  \tau \nu_\tau)=1$.

When investigating the viability of these types of scenarios we have found that the experimental constraints from direct searches are quite weak even when taking the latest constraints from LHC into account. The constraints from indirect searches in $B$-decays are more constraining but also more model dependent. Even when including the results from the most important ones, namely $B_u \to \tau \nu_\tau$ and $B_s \to \mu^+ \mu^-$, we still find a region of parameter space that is allowed with $\tan\beta \in [1,10]$. This is precisely the same region as the one where the decay $ {h^\pm} \to a_1 W$ can be dominant.

The phenomenology of these types of scenarios is special in that, due to its low mass, the $a_1$ will decay into a single $b\bar{b}$ jet. Even so we have shown that it is possible to reconstruct the decay $ {h^\pm} \to a_1 W$ using standard jet finding algorithms when the $W$ decays leptonically. This requires to use the missing transverse momentum to calculate the four momentum of the $W$, which can then be combined with the $a_1$-jet to give a mass peak at  $m_{h^\pm}$. The other $t$ quark is assumed to decay hadronically according to $t \to b jj$ giving an additional handle to identify the events of interest.
The most important background is thus the irreducible one from $t\bar{t} b\bar{b}$ production but we have also taken into account
 the $t\bar{t}$ background by considering the possibility of mis-tagging ordinary jets as $b$-jets.

Based on our study we find that with an integrated luminosity of 20 fb$^{-1}$ it should be possible to discover a charged Higgs bosons in these types of scenarios as long as the combined branching fraction for the decay chain  $t \to b h^+ \to b a_1 W \to b b \bar{b} W $ is larger than $\approx 0.01$. 

Finally we have also investigated the phenomenological consequences of the possible Higgs signal seen at the LHC on the types of scenarios we consider. We find that it is difficult to have a light $a_1$ that is not decoupled and at the same time have a combined production and decay into $\gamma\gamma$ for one of the CP-even Higgs bosons with mass $\sim 125$ GeV. As a consequence we have not been able to find regions of parameter space where the  $ {h^\pm} \to a_1 W$ decay dominates and at the same time are compatible with the possible Higgs signal. This means that irrespectively of whether the possible Higgs signal is substantiated or not, the LHC experiments should be able to either discover or put very tight constraints on a light charged Higgs bosons also in the NMSSM. 

{\bf Note added:}\\
On July 4, 2012 the ATLAS and CMS experiments announced the discovery of a new Higgs-like particle in the same mass region as the previous observations already cited in the text. 

\acknowledgments
We thank Oscar St{\aa}l for helpful communications. Furthermore, 
we would like to thank Stefan Prestel for his kind advice
and help, especially regarding the technical realization.
This work is supported in part by the Swedish Research Council grant 621-2011-5333.

\bibliography{paper}

%merlin.mbs 2010-03-15 4.21a (PWD, AO, DPC)
\begin{thebibliography}{54}%
\makeatletter
\providecommand \@ifxundefined [1]{%
 \@ifx{#1\undefined}
}%
\providecommand \@ifnum [1]{%
 \ifnum #1\expandafter \@firstoftwo
 \else \expandafter \@secondoftwo
 \fi
}%
\providecommand \@ifx [1]{%
 \ifx #1\expandafter \@firstoftwo
 \else \expandafter \@secondoftwo
 \fi
}%
\providecommand \natexlab [1]{#1}%
\providecommand \enquote  [1]{``#1''}%
\providecommand \bibnamefont  [1]{#1}%
\providecommand \bibfnamefont [1]{#1}%
\providecommand \citenamefont [1]{#1}%
\providecommand \href@noop [0]{\@secondoftwo}%
\providecommand \href [0]{\begingroup \@sanitize@url \@href}%
\providecommand \@href[1]{\@@startlink{#1}\@@href}%
\providecommand \@@href[1]{\endgroup#1\@@endlink}%
\providecommand \@sanitize@url [0]{\catcode `\\12\catcode `\$12\catcode
  `\&12\catcode `\#12\catcode `\^12\catcode `\_12\catcode `\%12\relax}%
\providecommand \@@startlink[1]{}%
\providecommand \@@endlink[0]{}%
\providecommand \url  [0]{\begingroup\@sanitize@url \@url }%
\providecommand \@url [1]{\endgroup\@href {#1}{\urlprefix }}%
\providecommand \urlprefix  [0]{URL }%
\providecommand \Eprint [0]{\href }%
\@ifxundefined \urlstyle {%
  \providecommand \doi  [0]{\begingroup \@sanitize@url \@doi}%
  \providecommand \@doi [1]{\endgroup \@@startlink {\doibase
  #1}doi:\discretionary {}{}{}#1\@@endlink }%
}{%
  \providecommand \doi  [0]{doi:\discretionary{}{}{}\begingroup
  \urlstyle{rm}\Url }%
}%
\providecommand \doibase [0]{http://dx.doi.org/}%
\providecommand \Doi [0]{\begingroup \@sanitize@url \@Doi }%
\providecommand \@Doi  [1]{\endgroup\@@startlink{\doibase#1}\@@Doi}%
\providecommand \@@Doi [1]{#1\@@endlink}%
\providecommand \selectlanguage [0]{\@gobble}%
\providecommand \bibinfo  [0]{\@secondoftwo}%
\providecommand \bibfield  [0]{\@secondoftwo}%
\providecommand \translation [1]{[#1]}%
\providecommand \BibitemOpen [0]{}%
\providecommand \bibitemStop [0]{}%
\providecommand \bibitemNoStop [0]{.\EOS\space}%
\providecommand \EOS [0]{\spacefactor3000\relax}%
\providecommand \BibitemShut  [1]{\csname bibitem#1\endcsname}%
%</preamble>
\bibitem [{\citenamefont {Nilles}(1984)}]{Nilles:1983ge}%
  \BibitemOpen
  \bibfield  {author} {\bibinfo {author} {\bibfnamefont {H.~P.}\ \bibnamefont
  {Nilles}},\ }\Doi {10.1016/0370-1573(84)90008-5} {\bibfield  {journal}
  {\bibinfo  {journal} {Phys.Rept.},\ }\textbf {\bibinfo {volume} {110}},\
  \bibinfo {pages} {1} (\bibinfo {year} {1984})}\BibitemShut {NoStop}%
%%CITATION = PRPLC,110,1;%%
\bibitem [{\citenamefont {Haber}\ and\ \citenamefont
  {Kane}(1985)}]{Haber:1984rc}%
  \BibitemOpen
  \bibfield  {author} {\bibinfo {author} {\bibfnamefont {H.~E.}\ \bibnamefont
  {Haber}}\ and\ \bibinfo {author} {\bibfnamefont {G.~L.}\ \bibnamefont
  {Kane}},\ }\Doi {10.1016/0370-1573(85)90051-1} {\bibfield  {journal}
  {\bibinfo  {journal} {Phys.Rept.},\ }\textbf {\bibinfo {volume} {117}},\
  \bibinfo {pages} {75} (\bibinfo {year} {1985})}\BibitemShut {NoStop}%
%%CITATION = PRPLC,117,75;%%
\bibitem [{\citenamefont {Djouadi}(2008)}]{Djouadi:2005gj}%
  \BibitemOpen
  \bibfield  {author} {\bibinfo {author} {\bibfnamefont {A.}~\bibnamefont
  {Djouadi}},\ }\Doi {10.1016/j.physrep.2007.10.005} {\bibfield  {journal}
  {\bibinfo  {journal} {Phys.Rept.},\ }\textbf {\bibinfo {volume} {459}},\
  \bibinfo {pages} {1} (\bibinfo {year} {2008})},\ \Eprint
  {http://arxiv.org/abs/hep-ph/0503173} {arXiv:hep-ph/0503173 [hep-ph]}
  \BibitemShut {NoStop}%
%%CITATION = HEP-PH/0503173;%%
\bibitem [{\citenamefont {Martin}(1997)}]{Martin:1997ns}%
  \BibitemOpen
  \bibfield  {author} {\bibinfo {author} {\bibfnamefont {S.~P.}\ \bibnamefont
  {Martin}},\ }\href@noop {} { (\bibinfo {year} {1997})},\ \Eprint
  {http://arxiv.org/abs/hep-ph/9709356} {arXiv:hep-ph/9709356 [hep-ph]}
  \BibitemShut {NoStop}%
%%CITATION = HEP-PH/9709356;%%
\bibitem [{\citenamefont {Goldberg}(1983)}]{Goldberg:1983nd}%
  \BibitemOpen
  \bibfield  {author} {\bibinfo {author} {\bibfnamefont {H.}~\bibnamefont
  {Goldberg}},\ }\Doi {10.1103/PhysRevLett.50.1419,
  10.1103/PhysRevLett.50.1419} {\bibfield  {journal} {\bibinfo  {journal}
  {Phys.Rev.Lett.},\ }\textbf {\bibinfo {volume} {50}},\ \bibinfo {pages}
  {1419} (\bibinfo {year} {1983})}\BibitemShut {NoStop}%
%%CITATION = PRLTA,50,1419;%%
\bibitem [{\citenamefont {Ellis}\ \emph {et~al.}(1984)\citenamefont {Ellis},
  \citenamefont {Hagelin}, \citenamefont {Nanopoulos}, \citenamefont {Olive},\
  and\ \citenamefont {Srednicki}}]{Ellis:1983ew}%
  \BibitemOpen
  \bibfield  {author} {\bibinfo {author} {\bibfnamefont {J.~R.}\ \bibnamefont
  {Ellis}}, \bibinfo {author} {\bibfnamefont {J.}~\bibnamefont {Hagelin}},
  \bibinfo {author} {\bibfnamefont {D.~V.}\ \bibnamefont {Nanopoulos}},
  \bibinfo {author} {\bibfnamefont {K.~A.}\ \bibnamefont {Olive}}, \ and\
  \bibinfo {author} {\bibfnamefont {M.}~\bibnamefont {Srednicki}},\ }\Doi
  {10.1016/0550-3213(84)90461-9} {\bibfield  {journal} {\bibinfo  {journal}
  {Nucl.Phys.},\ }\textbf {\bibinfo {volume} {B238}},\ \bibinfo {pages} {453}
  (\bibinfo {year} {1984})}\BibitemShut {NoStop}%
%%CITATION = NUPHA,B238,453;%%
\bibitem [{\citenamefont {Maniatis}(2010)}]{Maniatis:2009re}%
  \BibitemOpen
  \bibfield  {author} {\bibinfo {author} {\bibfnamefont {M.}~\bibnamefont
  {Maniatis}},\ }\Doi {10.1142/S0217751X10049827} {\bibfield  {journal}
  {\bibinfo  {journal} {Int.J.Mod.Phys.},\ }\textbf {\bibinfo {volume} {A25}},\
  \bibinfo {pages} {3505} (\bibinfo {year} {2010})},\ \Eprint
  {http://arxiv.org/abs/0906.0777} {arXiv:0906.0777 [hep-ph]} \BibitemShut
  {NoStop}%
%%CITATION = ARXIV:0906.0777;%%
\bibitem [{\citenamefont {Ellwanger}\ \emph {et~al.}(2010)\citenamefont
  {Ellwanger}, \citenamefont {Hugonie},\ and\ \citenamefont
  {Teixeira}}]{Ellwanger:2009dp}%
  \BibitemOpen
  \bibfield  {author} {\bibinfo {author} {\bibfnamefont {U.}~\bibnamefont
  {Ellwanger}}, \bibinfo {author} {\bibfnamefont {C.}~\bibnamefont {Hugonie}},
  \ and\ \bibinfo {author} {\bibfnamefont {A.~M.}\ \bibnamefont {Teixeira}},\
  }\Doi {10.1016/j.physrep.2010.07.001} {\bibfield  {journal} {\bibinfo
  {journal} {Phys.Rept.},\ }\textbf {\bibinfo {volume} {496}},\ \bibinfo
  {pages} {1} (\bibinfo {year} {2010})},\ \Eprint
  {http://arxiv.org/abs/0910.1785} {arXiv:0910.1785 [hep-ph]} \BibitemShut
  {NoStop}%
%%CITATION = ARXIV:0910.1785;%%
\bibitem [{\citenamefont {Drees}\ \emph {et~al.}(1998)\citenamefont {Drees},
  \citenamefont {Ma}, \citenamefont {Pandita}, \citenamefont {Roy},\ and\
  \citenamefont {Vempati}}]{Drees:1998pw}%
  \BibitemOpen
  \bibfield  {author} {\bibinfo {author} {\bibfnamefont {M.}~\bibnamefont
  {Drees}}, \bibinfo {author} {\bibfnamefont {E.}~\bibnamefont {Ma}}, \bibinfo
  {author} {\bibfnamefont {P.}~\bibnamefont {Pandita}}, \bibinfo {author}
  {\bibfnamefont {D.}~\bibnamefont {Roy}}, \ and\ \bibinfo {author}
  {\bibfnamefont {S.~K.}\ \bibnamefont {Vempati}},\ }\Doi
  {10.1016/S0370-2693(98)00712-6} {\bibfield  {journal} {\bibinfo  {journal}
  {Phys.Lett.},\ }\textbf {\bibinfo {volume} {B433}},\ \bibinfo {pages} {346}
  (\bibinfo {year} {1998})},\ \Eprint {http://arxiv.org/abs/hep-ph/9805242}
  {arXiv:hep-ph/9805242 [hep-ph]} \BibitemShut {NoStop}%
%%CITATION = HEP-PH/9805242;%%
\bibitem [{\citenamefont {Drees}\ \emph {et~al.}(1999)\citenamefont {Drees},
  \citenamefont {Guchait},\ and\ \citenamefont {Roy}}]{Drees:1999sb}%
  \BibitemOpen
  \bibfield  {author} {\bibinfo {author} {\bibfnamefont {M.}~\bibnamefont
  {Drees}}, \bibinfo {author} {\bibfnamefont {M.}~\bibnamefont {Guchait}}, \
  and\ \bibinfo {author} {\bibfnamefont {D.}~\bibnamefont {Roy}},\ }\Doi
  {10.1016/S0370-2693(99)01329-5} {\bibfield  {journal} {\bibinfo  {journal}
  {Phys.Lett.},\ }\textbf {\bibinfo {volume} {B471}},\ \bibinfo {pages} {39}
  (\bibinfo {year} {1999})},\ \Eprint {http://arxiv.org/abs/hep-ph/9909266}
  {arXiv:hep-ph/9909266 [hep-ph]} \BibitemShut {NoStop}%
%%CITATION = HEP-PH/9909266;%%
\bibitem [{\citenamefont {Akeroyd}\ \emph {et~al.}(2008)\citenamefont
  {Akeroyd}, \citenamefont {Arhrib},\ and\ \citenamefont
  {Yan}}]{Akeroyd:2007yj}%
  \BibitemOpen
  \bibfield  {author} {\bibinfo {author} {\bibfnamefont {A.}~\bibnamefont
  {Akeroyd}}, \bibinfo {author} {\bibfnamefont {A.}~\bibnamefont {Arhrib}}, \
  and\ \bibinfo {author} {\bibfnamefont {Q.-S.}\ \bibnamefont {Yan}},\ }\Doi
  {10.1140/epjc/s10052-008-0617-3} {\bibfield  {journal} {\bibinfo  {journal}
  {Eur.Phys.J.},\ }\textbf {\bibinfo {volume} {C55}},\ \bibinfo {pages} {653}
  (\bibinfo {year} {2008})},\ \Eprint {http://arxiv.org/abs/0712.3933}
  {arXiv:0712.3933 [hep-ph]} \BibitemShut {NoStop}%
%%CITATION = ARXIV:0712.3933;%%
\bibitem [{\citenamefont {Abdallah}\ \emph {et~al.}(2004)\citenamefont
  {Abdallah} \emph {et~al.}}]{Abdallah:2003wd}%
  \BibitemOpen
  \bibfield  {author} {\bibinfo {author} {\bibfnamefont {J.}~\bibnamefont
  {Abdallah}} \emph {et~al.} (\bibinfo {collaboration} {DELPHI
  Collaboration}),\ }\Doi {10.1140/epjc/s2004-01732-6} {\bibfield  {journal}
  {\bibinfo  {journal} {Eur.Phys.J.},\ }\textbf {\bibinfo {volume} {C34}},\
  \bibinfo {pages} {399} (\bibinfo {year} {2004})},\ \Eprint
  {http://arxiv.org/abs/hep-ex/0404012} {arXiv:hep-ex/0404012 [hep-ex]}
  \BibitemShut {NoStop}%
%%CITATION = HEP-EX/0404012;%%
\bibitem [{\citenamefont {Erbacher}\ \emph {et~al.}(2010)\citenamefont
  {Erbacher}, \citenamefont {Ivanov},\ and\ \citenamefont
  {Johnson}}]{CDF-10104}%
  \BibitemOpen
  \bibfield  {author} {\bibinfo {author} {\bibfnamefont {R.}~\bibnamefont
  {Erbacher}}, \bibinfo {author} {\bibfnamefont {A.}~\bibnamefont {Ivanov}}, \
  and\ \bibinfo {author} {\bibfnamefont {W.}~\bibnamefont {Johnson}} (\bibinfo
  {collaboration} {CDF collaboration}),\ }\href
  {http://www-cdf.fnal.gov/physics/new/top/2009/tprop/nMSSMhiggs/} {\bibfield
  {journal} {\bibinfo  {journal} {CDF note 10104}} (\bibinfo {year}
  {2010})}\BibitemShut {NoStop}%
\bibitem [{\citenamefont {Dermisek}\ and\ \citenamefont
  {Gunion}(2009)}]{Dermisek:2008uu}%
  \BibitemOpen
  \bibfield  {author} {\bibinfo {author} {\bibfnamefont {R.}~\bibnamefont
  {Dermisek}}\ and\ \bibinfo {author} {\bibfnamefont {J.~F.}\ \bibnamefont
  {Gunion}},\ }\Doi {10.1103/PhysRevD.79.055014} {\bibfield  {journal}
  {\bibinfo  {journal} {Phys.Rev.},\ }\textbf {\bibinfo {volume} {D79}},\
  \bibinfo {pages} {055014} (\bibinfo {year} {2009})},\ \Eprint
  {http://arxiv.org/abs/0811.3537} {arXiv:0811.3537 [hep-ph]} \BibitemShut
  {NoStop}%
%%CITATION = ARXIV:0811.3537;%%
\bibitem [{\citenamefont {Dermisek}\ and\ \citenamefont
  {Gunion}(2010){\natexlab{a}}}]{Dermisek:2009fd}%
  \BibitemOpen
  \bibfield  {author} {\bibinfo {author} {\bibfnamefont {R.}~\bibnamefont
  {Dermisek}}\ and\ \bibinfo {author} {\bibfnamefont {J.~F.}\ \bibnamefont
  {Gunion}},\ }\Doi {10.1103/PhysRevD.81.055001} {\bibfield  {journal}
  {\bibinfo  {journal} {Phys.Rev.},\ }\textbf {\bibinfo {volume} {D81}},\
  \bibinfo {pages} {055001} (\bibinfo {year} {2010}{\natexlab{a}})},\ \Eprint
  {http://arxiv.org/abs/0911.2460} {arXiv:0911.2460 [hep-ph]} \BibitemShut
  {NoStop}%
%%CITATION = ARXIV:0911.2460;%%
\bibitem [{\citenamefont {Dermisek}\ and\ \citenamefont
  {Gunion}(2010){\natexlab{b}}}]{Dermisek:2010mg}%
  \BibitemOpen
  \bibfield  {author} {\bibinfo {author} {\bibfnamefont {R.}~\bibnamefont
  {Dermisek}}\ and\ \bibinfo {author} {\bibfnamefont {J.~F.}\ \bibnamefont
  {Gunion}},\ }\Doi {10.1103/PhysRevD.81.075003} {\bibfield  {journal}
  {\bibinfo  {journal} {Phys.Rev.},\ }\textbf {\bibinfo {volume} {D81}},\
  \bibinfo {pages} {075003} (\bibinfo {year} {2010}{\natexlab{b}})},\ \Eprint
  {http://arxiv.org/abs/1002.1971} {arXiv:1002.1971 [hep-ph]} \BibitemShut
  {NoStop}%
%%CITATION = ARXIV:1002.1971;%%
\bibitem [{\citenamefont {Mahmoudi}\ \emph {et~al.}(2011)\citenamefont
  {Mahmoudi}, \citenamefont {Rathsman}, \citenamefont {Stal},\ and\
  \citenamefont {Zeune}}]{Mahmoudi:2010xp}%
  \BibitemOpen
  \bibfield  {author} {\bibinfo {author} {\bibfnamefont {F.}~\bibnamefont
  {Mahmoudi}}, \bibinfo {author} {\bibfnamefont {J.}~\bibnamefont {Rathsman}},
  \bibinfo {author} {\bibfnamefont {O.}~\bibnamefont {Stal}}, \ and\ \bibinfo
  {author} {\bibfnamefont {L.}~\bibnamefont {Zeune}},\ }\Doi
  {10.1140/epjc/s10052-011-1608-3} {\bibfield  {journal} {\bibinfo  {journal}
  {Eur.Phys.J.},\ }\textbf {\bibinfo {volume} {C71}},\ \bibinfo {pages} {1608}
  (\bibinfo {year} {2011})},\ \Eprint {http://arxiv.org/abs/1012.4490}
  {arXiv:1012.4490 [hep-ph]} \BibitemShut {NoStop}%
%%CITATION = ARXIV:1012.4490;%%
\bibitem [{\citenamefont {Stal}\ and\ \citenamefont
  {Weiglein}(2012)}]{Stal:2011cz}%
  \BibitemOpen
  \bibfield  {author} {\bibinfo {author} {\bibfnamefont {O.}~\bibnamefont
  {Stal}}\ and\ \bibinfo {author} {\bibfnamefont {G.}~\bibnamefont
  {Weiglein}},\ }\Doi {10.1007/JHEP01(2012)071} {\bibfield  {journal} {\bibinfo
   {journal} {JHEP},\ }\textbf {\bibinfo {volume} {1201}},\ \bibinfo {pages}
  {071} (\bibinfo {year} {2012})},\ \Eprint {http://arxiv.org/abs/1108.0595}
  {arXiv:1108.0595 [hep-ph]} \BibitemShut {NoStop}%
%%CITATION = ARXIV:1108.0595;%%
\bibitem [{\citenamefont {Carena}\ \emph {et~al.}(2003)\citenamefont {Carena},
  \citenamefont {Heinemeyer}, \citenamefont {Wagner},\ and\ \citenamefont
  {Weiglein}}]{Carena:2002qg}%
  \BibitemOpen
  \bibfield  {author} {\bibinfo {author} {\bibfnamefont {M.~S.}\ \bibnamefont
  {Carena}}, \bibinfo {author} {\bibfnamefont {S.}~\bibnamefont {Heinemeyer}},
  \bibinfo {author} {\bibfnamefont {C.~E.~M.}\ \bibnamefont {Wagner}}, \ and\
  \bibinfo {author} {\bibfnamefont {G.}~\bibnamefont {Weiglein}},\ }\Doi
  {10.1140/epjc/s2002-01084-3} {\bibfield  {journal} {\bibinfo  {journal} {Eur.
  Phys. J.},\ }\textbf {\bibinfo {volume} {C26}},\ \bibinfo {pages} {601}
  (\bibinfo {year} {2003})},\ \Eprint {http://arxiv.org/abs/hep-ph/0202167}
  {arXiv:hep-ph/0202167} \BibitemShut {NoStop}%
%%CITATION = HEP-PH/0202167;%%
\bibitem [{\citenamefont {Ellwanger}\ \emph {et~al.}(2005)\citenamefont
  {Ellwanger}, \citenamefont {Gunion},\ and\ \citenamefont
  {Hugonie}}]{Ellwanger:2004xm}%
  \BibitemOpen
  \bibfield  {author} {\bibinfo {author} {\bibfnamefont {U.}~\bibnamefont
  {Ellwanger}}, \bibinfo {author} {\bibfnamefont {J.~F.}\ \bibnamefont
  {Gunion}}, \ and\ \bibinfo {author} {\bibfnamefont {C.}~\bibnamefont
  {Hugonie}},\ }\Doi {10.1088/1126-6708/2005/02/066} {\bibfield  {journal}
  {\bibinfo  {journal} {JHEP},\ }\textbf {\bibinfo {volume} {02}},\ \bibinfo
  {pages} {066} (\bibinfo {year} {2005})},\ \Eprint
  {http://arxiv.org/abs/hep-ph/0406215} {arXiv:hep-ph/0406215} \BibitemShut
  {NoStop}%
%%CITATION = HEP-PH/0406215;%%
\bibitem [{\citenamefont {Ellwanger}\ and\ \citenamefont
  {Hugonie}(2006)}]{Ellwanger:2005dv}%
  \BibitemOpen
  \bibfield  {author} {\bibinfo {author} {\bibfnamefont {U.}~\bibnamefont
  {Ellwanger}}\ and\ \bibinfo {author} {\bibfnamefont {C.}~\bibnamefont
  {Hugonie}},\ }\Doi {10.1016/j.cpc.2006.04.004} {\bibfield  {journal}
  {\bibinfo  {journal} {Comput. Phys. Commun.},\ }\textbf {\bibinfo {volume}
  {175}},\ \bibinfo {pages} {290} (\bibinfo {year} {2006})},\ \Eprint
  {http://arxiv.org/abs/hep-ph/0508022} {arXiv:hep-ph/0508022} \BibitemShut
  {NoStop}%
%%CITATION = HEP-PH/0508022;%%
\bibitem [{\citenamefont {Bechtle}\ \emph {et~al.}(2010)\citenamefont
  {Bechtle}, \citenamefont {Brein}, \citenamefont {Heinemeyer}, \citenamefont
  {Weiglein},\ and\ \citenamefont {Williams}}]{Bechtle:2008jh}%
  \BibitemOpen
  \bibfield  {author} {\bibinfo {author} {\bibfnamefont {P.}~\bibnamefont
  {Bechtle}}, \bibinfo {author} {\bibfnamefont {O.}~\bibnamefont {Brein}},
  \bibinfo {author} {\bibfnamefont {S.}~\bibnamefont {Heinemeyer}}, \bibinfo
  {author} {\bibfnamefont {G.}~\bibnamefont {Weiglein}}, \ and\ \bibinfo
  {author} {\bibfnamefont {K.~E.}\ \bibnamefont {Williams}},\ }\Doi
  {10.1016/j.cpc.2009.09.003} {\bibfield  {journal} {\bibinfo  {journal}
  {Comput.Phys.Commun.},\ }\textbf {\bibinfo {volume} {181}},\ \bibinfo {pages}
  {138} (\bibinfo {year} {2010})},\ \Eprint {http://arxiv.org/abs/0811.4169}
  {arXiv:0811.4169 [hep-ph]} \BibitemShut {NoStop}%
%%CITATION = ARXIV:0811.4169;%%
\bibitem [{\citenamefont {Bechtle}\ \emph {et~al.}(2011)\citenamefont
  {Bechtle}, \citenamefont {Brein}, \citenamefont {Heinemeyer}, \citenamefont
  {Weiglein},\ and\ \citenamefont {Williams}}]{Bechtle:2011sb}%
  \BibitemOpen
  \bibfield  {author} {\bibinfo {author} {\bibfnamefont {P.}~\bibnamefont
  {Bechtle}}, \bibinfo {author} {\bibfnamefont {O.}~\bibnamefont {Brein}},
  \bibinfo {author} {\bibfnamefont {S.}~\bibnamefont {Heinemeyer}}, \bibinfo
  {author} {\bibfnamefont {G.}~\bibnamefont {Weiglein}}, \ and\ \bibinfo
  {author} {\bibfnamefont {K.~E.}\ \bibnamefont {Williams}},\ }\Doi
  {10.1016/j.cpc.2011.07.015} {\bibfield  {journal} {\bibinfo  {journal}
  {Comput.Phys.Commun.},\ }\textbf {\bibinfo {volume} {182}},\ \bibinfo {pages}
  {2605} (\bibinfo {year} {2011})},\ \Eprint {http://arxiv.org/abs/1102.1898}
  {arXiv:1102.1898 [hep-ph]} \BibitemShut {NoStop}%
%%CITATION = ARXIV:1102.1898;%%
\bibitem [{\citenamefont {Vasquez}\ \emph {et~al.}(2012)\citenamefont
  {Vasquez}, \citenamefont {Belanger}, \citenamefont {Boehm}, \citenamefont
  {Da~Silva}, \citenamefont {Richardson} \emph {et~al.}}]{Vasquez:2012hn}%
  \BibitemOpen
  \bibfield  {author} {\bibinfo {author} {\bibfnamefont {D.~A.}\ \bibnamefont
  {Vasquez}}, \bibinfo {author} {\bibfnamefont {G.}~\bibnamefont {Belanger}},
  \bibinfo {author} {\bibfnamefont {C.}~\bibnamefont {Boehm}}, \bibinfo
  {author} {\bibfnamefont {J.}~\bibnamefont {Da~Silva}}, \bibinfo {author}
  {\bibfnamefont {P.}~\bibnamefont {Richardson}},  \emph {et~al.},\ }\href@noop
  {} { (\bibinfo {year} {2012})},\ \Eprint {http://arxiv.org/abs/1203.3446}
  {arXiv:1203.3446 [hep-ph]} \BibitemShut {NoStop}%
%%CITATION = ARXIV:1203.3446;%%
\bibitem [{\citenamefont {Aaij}\ \emph {et~al.}(2012)\citenamefont {Aaij} \emph
  {et~al.}}]{Aaij:2012ac}%
  \BibitemOpen
  \bibfield  {author} {\bibinfo {author} {\bibfnamefont {R.}~\bibnamefont
  {Aaij}} \emph {et~al.} (\bibinfo {collaboration} {LHCb collaboration}),\
  }\href@noop {} { (\bibinfo {year} {2012})},\ \Eprint
  {http://arxiv.org/abs/1203.4493} {arXiv:1203.4493 [hep-ex]} \BibitemShut
  {NoStop}%
%%CITATION = ARXIV:1203.4493;%%
\bibitem [{\citenamefont {Chatrchyan}\ \emph
  {et~al.}(2012){\natexlab{a}}\citenamefont {Chatrchyan} \emph
  {et~al.}}]{Chatrchyan:2012rg}%
  \BibitemOpen
  \bibfield  {author} {\bibinfo {author} {\bibfnamefont {S.}~\bibnamefont
  {Chatrchyan}} \emph {et~al.} (\bibinfo {collaboration} {CMS Collaboration}),\
  }\href@noop {} {\bibfield  {journal} {\bibinfo  {journal} {JHEP},\ }\textbf
  {\bibinfo {volume} {1204}},\ \bibinfo {pages} {033} (\bibinfo {year}
  {2012}{\natexlab{a}})},\ \Eprint {http://arxiv.org/abs/1203.3976}
  {arXiv:1203.3976 [hep-ex]} \BibitemShut {NoStop}%
%%CITATION = ARXIV:1203.3976;%%
\bibitem [{\citenamefont {Aad}\ \emph {et~al.}(2012){\natexlab{a}}\citenamefont
  {Aad} \emph {et~al.}}]{Aad:2012tj}%
  \BibitemOpen
  \bibfield  {author} {\bibinfo {author} {\bibfnamefont {G.}~\bibnamefont
  {Aad}} \emph {et~al.} (\bibinfo {collaboration} {ATLAS Collaboration}),\
  }\href@noop {} { (\bibinfo {year} {2012}{\natexlab{a}})},\ \Eprint
  {http://arxiv.org/abs/1204.2760} {arXiv:1204.2760 [hep-ex]} \BibitemShut
  {NoStop}%
%%CITATION = ARXIV:1204.2760;%%
\bibitem [{\citenamefont {Chatrchyan}\ \emph
  {et~al.}(2012){\natexlab{b}}\citenamefont {Chatrchyan} \emph
  {et~al.}}]{CMS-light-h+}%
  \BibitemOpen
  \bibfield  {author} {\bibinfo {author} {\bibfnamefont {S.}~\bibnamefont
  {Chatrchyan}} \emph {et~al.} (\bibinfo {collaboration} {CMS Collaboration}),\
  }\href@noop {} { (\bibinfo {year} {2012}{\natexlab{b}})},\ \Eprint
  {http://arxiv.org/abs/1205.5736} {arXiv:1205.5736 [hep-ex]} \BibitemShut
  {NoStop}%
%%CITATION = ARXIV:1205.5736;%%
\bibitem [{CMS(2012)}]{CMS-PAS-BTV-11-004}%
  \BibitemOpen
  \href@noop {} {\emph {\bibinfo {title} {b-Jet Identification in the CMS
  Experiment}}},\ \bibinfo {type} {Tech. Rep.}\ \bibinfo {number}
  {CMS-PAS-BTV-11-004}\ (\bibinfo {year} {2012})\BibitemShut {NoStop}%
\bibitem [{ATL(2012)}]{ATLAS-CONF-2012-040}%
  \BibitemOpen
  \href@noop {} {\emph {\bibinfo {title} {Measurement of the Mistag Rate with 5
  fb$^{−1}$ of Data Collected by the ATLAS Detector}}},\ \bibinfo {type}
  {Tech. Rep.}\ \bibinfo {number} {ATLAS-CONF-2012-040}\ (\bibinfo
  {institution} {CERN},\ \bibinfo {address} {Geneva},\ \bibinfo {year}
  {2012})\BibitemShut {NoStop}%
\bibitem [{\citenamefont {Chatrchyan}\ \emph {et~al.}(2011)\citenamefont
  {Chatrchyan} \emph {et~al.}}]{Chatrchyan:1376068}%
  \BibitemOpen
  \bibfield  {author} {\bibinfo {author} {\bibfnamefont {S.}~\bibnamefont
  {Chatrchyan}} \emph {et~al.} (\bibinfo {collaboration} {CMS Collaboration}),\
  }\href@noop {} {\bibfield  {journal} {\bibinfo  {journal} {Phys. Rev. D},\
  }\textbf {\bibinfo {volume} {84}},\ \bibinfo {pages} {092004. 39 p} (\bibinfo
  {year} {2011})}\BibitemShut {NoStop}%
\bibitem [{\citenamefont {Alwall}\ \emph {et~al.}(2011)\citenamefont {Alwall},
  \citenamefont {Herquet}, \citenamefont {Maltoni}, \citenamefont {Mattelaer},\
  and\ \citenamefont {Stelzer}}]{Alwall:2011uj}%
  \BibitemOpen
  \bibfield  {author} {\bibinfo {author} {\bibfnamefont {J.}~\bibnamefont
  {Alwall}}, \bibinfo {author} {\bibfnamefont {M.}~\bibnamefont {Herquet}},
  \bibinfo {author} {\bibfnamefont {F.}~\bibnamefont {Maltoni}}, \bibinfo
  {author} {\bibfnamefont {O.}~\bibnamefont {Mattelaer}}, \ and\ \bibinfo
  {author} {\bibfnamefont {T.}~\bibnamefont {Stelzer}},\ }\Doi
  {10.1007/JHEP06(2011)128} {\bibfield  {journal} {\bibinfo  {journal} {JHEP},\
  }\textbf {\bibinfo {volume} {1106}},\ \bibinfo {pages} {128} (\bibinfo {year}
  {2011})},\ \Eprint {http://arxiv.org/abs/1106.0522} {arXiv:1106.0522
  [hep-ph]} \BibitemShut {NoStop}%
%%CITATION = ARXIV:1106.0522;%%
\bibitem [{\citenamefont {Eriksson}\ \emph {et~al.}(2010)\citenamefont
  {Eriksson}, \citenamefont {Rathsman},\ and\ \citenamefont
  {Stal}}]{Eriksson:2009ws}%
  \BibitemOpen
  \bibfield  {author} {\bibinfo {author} {\bibfnamefont {D.}~\bibnamefont
  {Eriksson}}, \bibinfo {author} {\bibfnamefont {J.}~\bibnamefont {Rathsman}},
  \ and\ \bibinfo {author} {\bibfnamefont {O.}~\bibnamefont {Stal}},\ }\Doi
  {10.1016/j.cpc.2009.09.011} {\bibfield  {journal} {\bibinfo  {journal}
  {Comput.Phys.Commun.},\ }\textbf {\bibinfo {volume} {181}},\ \bibinfo {pages}
  {189} (\bibinfo {year} {2010})},\ \Eprint {http://arxiv.org/abs/0902.0851}
  {arXiv:0902.0851 [hep-ph]} \BibitemShut {NoStop}%
%%CITATION = ARXIV:0902.0851;%%
\bibitem [{\citenamefont {Beneke}\ \emph {et~al.}(2012)\citenamefont {Beneke},
  \citenamefont {Falgari}, \citenamefont {Klein},\ and\ \citenamefont
  {Schwinn}}]{Beneke:2011mq}%
  \BibitemOpen
  \bibfield  {author} {\bibinfo {author} {\bibfnamefont {M.}~\bibnamefont
  {Beneke}}, \bibinfo {author} {\bibfnamefont {P.}~\bibnamefont {Falgari}},
  \bibinfo {author} {\bibfnamefont {S.}~\bibnamefont {Klein}}, \ and\ \bibinfo
  {author} {\bibfnamefont {C.}~\bibnamefont {Schwinn}},\ }\Doi
  {10.1016/j.nuclphysb.2011.10.021} {\bibfield  {journal} {\bibinfo  {journal}
  {Nucl.Phys.},\ }\textbf {\bibinfo {volume} {B855}},\ \bibinfo {pages} {695}
  (\bibinfo {year} {2012})},\ \Eprint {http://arxiv.org/abs/1109.1536}
  {arXiv:1109.1536 [hep-ph]} \BibitemShut {NoStop}%
%%CITATION = ARXIV:1109.1536;%%
\bibitem [{\citenamefont {Cacciari}\ \emph {et~al.}(2008)\citenamefont
  {Cacciari}, \citenamefont {Salam},\ and\ \citenamefont
  {Soyez}}]{Cacciari:2008gp}%
  \BibitemOpen
  \bibfield  {author} {\bibinfo {author} {\bibfnamefont {M.}~\bibnamefont
  {Cacciari}}, \bibinfo {author} {\bibfnamefont {G.~P.}\ \bibnamefont {Salam}},
  \ and\ \bibinfo {author} {\bibfnamefont {G.}~\bibnamefont {Soyez}},\ }\Doi
  {10.1088/1126-6708/2008/04/063} {\bibfield  {journal} {\bibinfo  {journal}
  {JHEP},\ }\textbf {\bibinfo {volume} {0804}},\ \bibinfo {pages} {063}
  (\bibinfo {year} {2008})},\ \Eprint {http://arxiv.org/abs/0802.1189}
  {arXiv:0802.1189 [hep-ph]} \BibitemShut {NoStop}%
%%CITATION = ARXIV:0802.1189;%%
\bibitem [{\citenamefont {Dokshitzer}\ \emph {et~al.}(1997)\citenamefont
  {Dokshitzer}, \citenamefont {Leder}, \citenamefont {Moretti},\ and\
  \citenamefont {Webber}}]{Dokshitzer:1997in}%
  \BibitemOpen
  \bibfield  {author} {\bibinfo {author} {\bibfnamefont {Y.~L.}\ \bibnamefont
  {Dokshitzer}}, \bibinfo {author} {\bibfnamefont {G.}~\bibnamefont {Leder}},
  \bibinfo {author} {\bibfnamefont {S.}~\bibnamefont {Moretti}}, \ and\
  \bibinfo {author} {\bibfnamefont {B.}~\bibnamefont {Webber}},\ }\href@noop {}
  {\bibfield  {journal} {\bibinfo  {journal} {JHEP},\ }\textbf {\bibinfo
  {volume} {9708}},\ \bibinfo {pages} {001} (\bibinfo {year} {1997})},\ \Eprint
  {http://arxiv.org/abs/hep-ph/9707323} {arXiv:hep-ph/9707323 [hep-ph]}
  \BibitemShut {NoStop}%
%%CITATION = HEP-PH/9707323;%%
\bibitem [{\citenamefont {Catani}\ \emph {et~al.}(1993)\citenamefont {Catani},
  \citenamefont {Dokshitzer}, \citenamefont {Seymour},\ and\ \citenamefont
  {Webber}}]{Catani:1993hr}%
  \BibitemOpen
  \bibfield  {author} {\bibinfo {author} {\bibfnamefont {S.}~\bibnamefont
  {Catani}}, \bibinfo {author} {\bibfnamefont {Y.~L.}\ \bibnamefont
  {Dokshitzer}}, \bibinfo {author} {\bibfnamefont {M.}~\bibnamefont {Seymour}},
  \ and\ \bibinfo {author} {\bibfnamefont {B.}~\bibnamefont {Webber}},\ }\Doi
  {10.1016/0550-3213(93)90166-M} {\bibfield  {journal} {\bibinfo  {journal}
  {Nucl.Phys.},\ }\textbf {\bibinfo {volume} {B406}},\ \bibinfo {pages} {187}
  (\bibinfo {year} {1993})}\BibitemShut {NoStop}%
%%CITATION = NUPHA,B406,187;%%
\bibitem [{\citenamefont {Ellis}\ and\ \citenamefont
  {Soper}(1993)}]{Ellis:1993tq}%
  \BibitemOpen
  \bibfield  {author} {\bibinfo {author} {\bibfnamefont {S.~D.}\ \bibnamefont
  {Ellis}}\ and\ \bibinfo {author} {\bibfnamefont {D.~E.}\ \bibnamefont
  {Soper}},\ }\Doi {10.1103/PhysRevD.48.3160} {\bibfield  {journal} {\bibinfo
  {journal} {Phys.Rev.},\ }\textbf {\bibinfo {volume} {D48}},\ \bibinfo {pages}
  {3160} (\bibinfo {year} {1993})},\ \Eprint
  {http://arxiv.org/abs/hep-ph/9305266} {arXiv:hep-ph/9305266 [hep-ph]}
  \BibitemShut {NoStop}%
%%CITATION = HEP-PH/9305266;%%
\bibitem [{\citenamefont {Cacciari}\ and\ \citenamefont
  {Salam}(2006)}]{Cacciari:2005hq}%
  \BibitemOpen
  \bibfield  {author} {\bibinfo {author} {\bibfnamefont {M.}~\bibnamefont
  {Cacciari}}\ and\ \bibinfo {author} {\bibfnamefont {G.~P.}\ \bibnamefont
  {Salam}},\ }\Doi {10.1016/j.physletb.2006.08.037} {\bibfield  {journal}
  {\bibinfo  {journal} {Phys.Lett.},\ }\textbf {\bibinfo {volume} {B641}},\
  \bibinfo {pages} {57} (\bibinfo {year} {2006})},\ \Eprint
  {http://arxiv.org/abs/hep-ph/0512210} {arXiv:hep-ph/0512210 [hep-ph]}
  \BibitemShut {NoStop}%
%%CITATION = HEP-PH/0512210;%%
\bibitem [{\citenamefont {Cacciari}\ \emph {et~al.}(2012)\citenamefont
  {Cacciari}, \citenamefont {Salam},\ and\ \citenamefont
  {Soyez}}]{Cacciari:2011ma}%
  \BibitemOpen
  \bibfield  {author} {\bibinfo {author} {\bibfnamefont {M.}~\bibnamefont
  {Cacciari}}, \bibinfo {author} {\bibfnamefont {G.~P.}\ \bibnamefont {Salam}},
  \ and\ \bibinfo {author} {\bibfnamefont {G.}~\bibnamefont {Soyez}},\
  }\href@noop {} {\bibfield  {journal} {\bibinfo  {journal} {Eur.Phys.J.},\
  }\textbf {\bibinfo {volume} {C72}},\ \bibinfo {pages} {1896} (\bibinfo {year}
  {2012})},\ \Eprint {http://arxiv.org/abs/1111.6097} {arXiv:1111.6097
  [hep-ph]} \BibitemShut {NoStop}%
%%CITATION = ARXIV:1111.6097;%%
\bibitem [{\citenamefont {Eriksson}\ \emph {et~al.}(2008)\citenamefont
  {Eriksson}, \citenamefont {Ingelman}, \citenamefont {Rathsman},\ and\
  \citenamefont {Stal}}]{Eriksson:2007fx}%
  \BibitemOpen
  \bibfield  {author} {\bibinfo {author} {\bibfnamefont {D.}~\bibnamefont
  {Eriksson}}, \bibinfo {author} {\bibfnamefont {G.}~\bibnamefont {Ingelman}},
  \bibinfo {author} {\bibfnamefont {J.}~\bibnamefont {Rathsman}}, \ and\
  \bibinfo {author} {\bibfnamefont {O.}~\bibnamefont {Stal}},\ }\Doi
  {10.1088/1126-6708/2008/01/024} {\bibfield  {journal} {\bibinfo  {journal}
  {JHEP},\ }\textbf {\bibinfo {volume} {0801}},\ \bibinfo {pages} {024}
  (\bibinfo {year} {2008})},\ \Eprint {http://arxiv.org/abs/0710.5906}
  {arXiv:0710.5906 [hep-ph]} \BibitemShut {NoStop}%
%%CITATION = ARXIV:0710.5906;%%
\bibitem [{\citenamefont {Bernreuther}(2008)}]{Bernreuther:2008ju}%
  \BibitemOpen
  \bibfield  {author} {\bibinfo {author} {\bibfnamefont {W.}~\bibnamefont
  {Bernreuther}},\ }\Doi {10.1088/0954-3899/35/8/083001} {\bibfield  {journal}
  {\bibinfo  {journal} {J.Phys.G},\ }\textbf {\bibinfo {volume} {G35}},\
  \bibinfo {pages} {083001} (\bibinfo {year} {2008})},\ \Eprint
  {http://arxiv.org/abs/0805.1333} {arXiv:0805.1333 [hep-ph]} \BibitemShut
  {NoStop}%
%%CITATION = ARXIV:0805.1333;%%
\bibitem [{\citenamefont {Godbole}\ \emph {et~al.}(2012)\citenamefont
  {Godbole}, \citenamefont {Hartgring}, \citenamefont {Niessen},\ and\
  \citenamefont {White}}]{Godbole:2011vw}%
  \BibitemOpen
  \bibfield  {author} {\bibinfo {author} {\bibfnamefont {R.~M.}\ \bibnamefont
  {Godbole}}, \bibinfo {author} {\bibfnamefont {L.}~\bibnamefont {Hartgring}},
  \bibinfo {author} {\bibfnamefont {I.}~\bibnamefont {Niessen}}, \ and\
  \bibinfo {author} {\bibfnamefont {C.~D.}\ \bibnamefont {White}},\ }\Doi
  {10.1007/JHEP01(2012)011} {\bibfield  {journal} {\bibinfo  {journal} {JHEP},\
  }\textbf {\bibinfo {volume} {1201}},\ \bibinfo {pages} {011} (\bibinfo {year}
  {2012})},\ \Eprint {http://arxiv.org/abs/1111.0759} {arXiv:1111.0759
  [hep-ph]} \BibitemShut {NoStop}%
%%CITATION = ARXIV:1111.0759;%%
\bibitem [{\citenamefont {Aad}\ \emph {et~al.}(2012){\natexlab{b}}\citenamefont
  {Aad} \emph {et~al.}}]{ATLAS:2012ae}%
  \BibitemOpen
  \bibfield  {author} {\bibinfo {author} {\bibfnamefont {G.}~\bibnamefont
  {Aad}} \emph {et~al.} (\bibinfo {collaboration} {ATLAS Collaboration}),\
  }\href@noop {} {\bibfield  {journal} {\bibinfo  {journal} {Phys.Lett.},\
  }\textbf {\bibinfo {volume} {B710}},\ \bibinfo {pages} {49} (\bibinfo {year}
  {2012}{\natexlab{b}})},\ \Eprint {http://arxiv.org/abs/1202.1408}
  {arXiv:1202.1408 [hep-ex]} \BibitemShut {NoStop}%
%%CITATION = ARXIV:1202.1408;%%
\bibitem [{\citenamefont {Chatrchyan}\ \emph
  {et~al.}(2012){\natexlab{c}}\citenamefont {Chatrchyan} \emph
  {et~al.}}]{Chatrchyan:2012tx}%
  \BibitemOpen
  \bibfield  {author} {\bibinfo {author} {\bibfnamefont {S.}~\bibnamefont
  {Chatrchyan}} \emph {et~al.} (\bibinfo {collaboration} {CMS Collaboration}),\
  }\href@noop {} {\bibfield  {journal} {\bibinfo  {journal} {Phys.Lett.},\
  }\textbf {\bibinfo {volume} {B710}},\ \bibinfo {pages} {26} (\bibinfo {year}
  {2012}{\natexlab{c}})},\ \Eprint {http://arxiv.org/abs/1202.1488}
  {arXiv:1202.1488 [hep-ex]} \BibitemShut {NoStop}%
%%CITATION = ARXIV:1202.1488;%%
\bibitem [{\citenamefont {Hall}\ \emph {et~al.}(2012)\citenamefont {Hall},
  \citenamefont {Pinner},\ and\ \citenamefont {Ruderman}}]{Hall:2011aa}%
  \BibitemOpen
  \bibfield  {author} {\bibinfo {author} {\bibfnamefont {L.~J.}\ \bibnamefont
  {Hall}}, \bibinfo {author} {\bibfnamefont {D.}~\bibnamefont {Pinner}}, \ and\
  \bibinfo {author} {\bibfnamefont {J.~T.}\ \bibnamefont {Ruderman}},\
  }\href@noop {} {\bibfield  {journal} {\bibinfo  {journal} {JHEP},\ }\textbf
  {\bibinfo {volume} {1204}},\ \bibinfo {pages} {131} (\bibinfo {year}
  {2012})},\ \Eprint {http://arxiv.org/abs/1112.2703} {arXiv:1112.2703
  [hep-ph]} \BibitemShut {NoStop}%
%%CITATION = ARXIV:1112.2703;%%
\bibitem [{\citenamefont {Heinemeyer}\ \emph {et~al.}(2012)\citenamefont
  {Heinemeyer}, \citenamefont {Stal},\ and\ \citenamefont
  {Weiglein}}]{Heinemeyer:2011aa}%
  \BibitemOpen
  \bibfield  {author} {\bibinfo {author} {\bibfnamefont {S.}~\bibnamefont
  {Heinemeyer}}, \bibinfo {author} {\bibfnamefont {O.}~\bibnamefont {Stal}}, \
  and\ \bibinfo {author} {\bibfnamefont {G.}~\bibnamefont {Weiglein}},\ }\Doi
  {10.1016/j.physletb.2012.02.084} {\bibfield  {journal} {\bibinfo  {journal}
  {Phys.Lett.},\ }\textbf {\bibinfo {volume} {B710}},\ \bibinfo {pages} {201}
  (\bibinfo {year} {2012})},\ \Eprint {http://arxiv.org/abs/1112.3026}
  {arXiv:1112.3026 [hep-ph]} \BibitemShut {NoStop}%
%%CITATION = ARXIV:1112.3026;%%
\bibitem [{\citenamefont {Arbey}\ \emph {et~al.}(2012)\citenamefont {Arbey},
  \citenamefont {Battaglia}, \citenamefont {Djouadi}, \citenamefont
  {Mahmoudi},\ and\ \citenamefont {Quevillon}}]{Arbey:2011ab}%
  \BibitemOpen
  \bibfield  {author} {\bibinfo {author} {\bibfnamefont {A.}~\bibnamefont
  {Arbey}}, \bibinfo {author} {\bibfnamefont {M.}~\bibnamefont {Battaglia}},
  \bibinfo {author} {\bibfnamefont {A.}~\bibnamefont {Djouadi}}, \bibinfo
  {author} {\bibfnamefont {F.}~\bibnamefont {Mahmoudi}}, \ and\ \bibinfo
  {author} {\bibfnamefont {J.}~\bibnamefont {Quevillon}},\ }\Doi
  {10.1016/j.physletb.2012.01.053} {\bibfield  {journal} {\bibinfo  {journal}
  {Phys.Lett.},\ }\textbf {\bibinfo {volume} {B708}},\ \bibinfo {pages} {162}
  (\bibinfo {year} {2012})},\ \Eprint {http://arxiv.org/abs/1112.3028}
  {arXiv:1112.3028 [hep-ph]} \BibitemShut {NoStop}%
%%CITATION = ARXIV:1112.3028;%%
\bibitem [{\citenamefont {Carena}\ \emph {et~al.}(2012)\citenamefont {Carena},
  \citenamefont {Gori}, \citenamefont {Shah},\ and\ \citenamefont
  {Wagner}}]{Carena:2011aa}%
  \BibitemOpen
  \bibfield  {author} {\bibinfo {author} {\bibfnamefont {M.}~\bibnamefont
  {Carena}}, \bibinfo {author} {\bibfnamefont {S.}~\bibnamefont {Gori}},
  \bibinfo {author} {\bibfnamefont {N.~R.}\ \bibnamefont {Shah}}, \ and\
  \bibinfo {author} {\bibfnamefont {C.~E.}\ \bibnamefont {Wagner}},\
  }\href@noop {} {\bibfield  {journal} {\bibinfo  {journal} {JHEP},\ }\textbf
  {\bibinfo {volume} {1203}},\ \bibinfo {pages} {014} (\bibinfo {year}
  {2012})},\ \Eprint {http://arxiv.org/abs/1112.3336} {arXiv:1112.3336
  [hep-ph]} \BibitemShut {NoStop}%
%%CITATION = ARXIV:1112.3336;%%
\bibitem [{\citenamefont {Kadastik}\ \emph {et~al.}(2012)\citenamefont
  {Kadastik}, \citenamefont {Kannike}, \citenamefont {Racioppi},\ and\
  \citenamefont {Raidal}}]{Kadastik:2011aa}%
  \BibitemOpen
  \bibfield  {author} {\bibinfo {author} {\bibfnamefont {M.}~\bibnamefont
  {Kadastik}}, \bibinfo {author} {\bibfnamefont {K.}~\bibnamefont {Kannike}},
  \bibinfo {author} {\bibfnamefont {A.}~\bibnamefont {Racioppi}}, \ and\
  \bibinfo {author} {\bibfnamefont {M.}~\bibnamefont {Raidal}},\ }\href@noop {}
  {\bibfield  {journal} {\bibinfo  {journal} {JHEP},\ }\textbf {\bibinfo
  {volume} {1205}},\ \bibinfo {pages} {061} (\bibinfo {year} {2012})},\ \Eprint
  {http://arxiv.org/abs/1112.3647} {arXiv:1112.3647 [hep-ph]} \BibitemShut
  {NoStop}%
%%CITATION = ARXIV:1112.3647;%%
\bibitem [{\citenamefont {Ellwanger}(2012)}]{Ellwanger:2011aa}%
  \BibitemOpen
  \bibfield  {author} {\bibinfo {author} {\bibfnamefont {U.}~\bibnamefont
  {Ellwanger}},\ }\Doi {10.1007/JHEP03(2012)044} {\bibfield  {journal}
  {\bibinfo  {journal} {JHEP},\ }\textbf {\bibinfo {volume} {1203}},\ \bibinfo
  {pages} {044} (\bibinfo {year} {2012})},\ \Eprint
  {http://arxiv.org/abs/1112.3548} {arXiv:1112.3548 [hep-ph]} \BibitemShut
  {NoStop}%
%%CITATION = ARXIV:1112.3548;%%
\bibitem [{\citenamefont {Cao}\ \emph {et~al.}(2012){\natexlab{a}}\citenamefont
  {Cao}, \citenamefont {Heng}, \citenamefont {Li},\ and\ \citenamefont
  {Yang}}]{Cao:2011sn}%
  \BibitemOpen
  \bibfield  {author} {\bibinfo {author} {\bibfnamefont {J.}~\bibnamefont
  {Cao}}, \bibinfo {author} {\bibfnamefont {Z.}~\bibnamefont {Heng}}, \bibinfo
  {author} {\bibfnamefont {D.}~\bibnamefont {Li}}, \ and\ \bibinfo {author}
  {\bibfnamefont {J.~M.}\ \bibnamefont {Yang}},\ }\Doi
  {10.1016/j.physletb.2012.03.052} {\bibfield  {journal} {\bibinfo  {journal}
  {Phys.Lett.},\ }\textbf {\bibinfo {volume} {B710}},\ \bibinfo {pages} {665}
  (\bibinfo {year} {2012}{\natexlab{a}})},\ \Eprint
  {http://arxiv.org/abs/1112.4391} {arXiv:1112.4391 [hep-ph]} \BibitemShut
  {NoStop}%
%%CITATION = ARXIV:1112.4391;%%
\bibitem [{\citenamefont {Cao}\ \emph {et~al.}(2012){\natexlab{b}}\citenamefont
  {Cao}, \citenamefont {Heng}, \citenamefont {Yang}, \citenamefont {Zhang},\
  and\ \citenamefont {Zhu}}]{Cao:2012fz}%
  \BibitemOpen
  \bibfield  {author} {\bibinfo {author} {\bibfnamefont {J.}~\bibnamefont
  {Cao}}, \bibinfo {author} {\bibfnamefont {Z.}~\bibnamefont {Heng}}, \bibinfo
  {author} {\bibfnamefont {J.~M.}\ \bibnamefont {Yang}}, \bibinfo {author}
  {\bibfnamefont {Y.}~\bibnamefont {Zhang}}, \ and\ \bibinfo {author}
  {\bibfnamefont {J.}~\bibnamefont {Zhu}},\ }\Doi {10.1007/JHEP03(2012)086}
  {\bibfield  {journal} {\bibinfo  {journal} {JHEP},\ }\textbf {\bibinfo
  {volume} {1203}},\ \bibinfo {pages} {086} (\bibinfo {year}
  {2012}{\natexlab{b}})},\ \Eprint {http://arxiv.org/abs/1202.5821}
  {arXiv:1202.5821 [hep-ph]} \BibitemShut {NoStop}%
%%CITATION = ARXIV:1202.5821;%%
\bibitem [{\citenamefont {Ellwanger}\ and\ \citenamefont
  {Hugonie}(2012)}]{Ellwanger:2012ke}%
  \BibitemOpen
  \bibfield  {author} {\bibinfo {author} {\bibfnamefont {U.}~\bibnamefont
  {Ellwanger}}\ and\ \bibinfo {author} {\bibfnamefont {C.}~\bibnamefont
  {Hugonie}},\ }\href@noop {} { (\bibinfo {year} {2012})},\ \Eprint
  {http://arxiv.org/abs/1203.5048} {arXiv:1203.5048 [hep-ph]} \BibitemShut
  {NoStop}%
%%CITATION = ARXIV:1203.5048;%%
\end{thebibliography}%

\end{document}